\documentclass[12pt]{article}
\usepackage{latexsym}
\usepackage{epsfig,amssymb,euscript}
\usepackage{amsmath}
\usepackage{mathrsfs}
\numberwithin{equation}{section}  
\newsavebox{\ns}
\newsavebox{\dbrane}

\def\be{\begin{equation}}
\def\ee{\end{equation}}
\def\bea{\begin{eqnarray}}
\def\eea{\end{eqnarray}}

\newcommand{\nn}{\nonumber}

\def\Dslash{\,\,{\raise.15ex\hbox{/}\mkern-12mu D}}
\def\Dbarslash{\,\,{\raise.15ex\hbox{/}\mkern-12mu {\bar D}}}
\def\delslash{\,\,{\raise.15ex\hbox{/}\mkern-9mu \partial}}
\def\delbarslash{\,\,{\raise.15ex\hbox{/}\mkern-9mu {\bar\partial}}}
\def\pslash{\,\,{\raise.15ex\hbox{/}\mkern-9mu p}}
\def\calDslash{\,\,{\raise.15ex\hbox{/}\mkern-12mu {\cal D}}}

\newcommand\R{\mathbb{R}}
\newcommand\Z{\mathbb{Z}}

\newcommand\C{\mathbb{C}}

\newcommand\diff{\mathrm{d}}
\newcommand\ex{\mathrm{e}}

\newcommand{\vol}{\mathrm{vol}}
\newcommand{\dd}{\mathrm{d}}
\newcommand{\ii}{\mathrm{i}}

\newcommand{\me}{\mathrm{e}}
\newcommand{\comment}[1]{}
\newcommand{\secref}[1]{\S\ref{#1}}

\begin{document}
\begin{titlepage}
\begin{center}
\today

\vskip .5cm
{\Large \bf  Baryonic symmetries and M5 branes in \\ [3.4mm]

the AdS$_4$/CFT$_3$ correspondence} \\[3.4mm]

\vskip 1.3cm

{Nessi Benishti$^{1}$, ~~ Diego Rodr\'iguez-G\'omez$^{2}$,  ~~and~~ James Sparks$^{3}$}\\
\vskip 1cm

1: {\em Rudolf Peierls Centre for Theoretical Physics, \\ University of Oxford, \\
1 Keble Road, Oxford OX1 3NP, U.K.}\\

\vskip 0.5cm

2: {\em Queen Mary, University of London, \\
Mile End Road, London E1 4NS, U.K.}\\

\vskip 0.5cm

3: {\em Mathematical Institute, University of Oxford,\\
 24-29 St Giles', Oxford OX1 3LB, U.K.}\\

\vskip 1cm

\end{center}

\begin{abstract}
\noindent 

We study $U(1)$ symmetries dual to Betti multiplets in the $AdS_4/CFT_3$ correspondence for M2 branes at
Calabi-Yau four-fold singularities. Analysis of the boundary conditions for vector fields in $AdS_4$ allows for a choice where wrapped M5 brane states carrying non-zero charge under such symmetries can be considered. We begin by focusing on isolated toric singularities without vanishing six-cycles, and study in detail the cone over $Q^{111}$. The boundary conditions considered are dual to a CFT where the gauge group is $U(1)^2 \times SU(N)^4$. We find
agreement between the spectrum of gauge-invariant baryonic-type operators in
this theory and wrapped M5 brane states. Moreover, the physics of vacua in
which these symmetries are spontaneously broken precisely matches a dual
gravity analysis involving resolutions of the singularity, where we are able
to match condensates of the baryonic operators, Goldstone bosons and global
strings. We also argue more generally that theories where the 
resolutions have six-cycles are expected to receive non-perturbative corrections
from M5 brane instantons. We give a general formula relating the instanton 
action to normalizable harmonic two-forms, and compute it explicitly 
for the $Q^{222}$ example. The
holographic interpretation of such instantons is currently unclear.

\end{abstract}

\vskip .8cm

\vfill
\hrule width 5cm
\vskip 5mm

\end{titlepage}
\pagestyle{plain}
\setcounter{page}{1}
\newcounter{bean}
\baselineskip18pt

\tableofcontents


\section{Introduction}\label{sec:intro}

Over the last two years there have been major advances towards understanding the $AdS_4/CFT_3$ duality. Elaborating on \cite{Gustavsson:2007vu, Bagger:2007vi}, Aharony, Bergman, Jafferis and Maldacena \cite{Aharony:2008ug} proposed a theory conjectured to be dual to M2 branes probing a $\mathbb{C}^4/\mathbb{Z}_k$ singularity, where $\Z_k$ acts with weights $(1,1,-1,-1)$ on the coordinates of $\C^4$. This low energy theory on the worldvolume of $N$ coincident M2 branes is a $U(N)_k\times U(N)_{-k}$ quiver Chern-Simons (CS) theory, with a marginal quartic superpotential whose coefficient is related by the high degree of SUSY to the CS coupling. Indeed, for generic CS coupling $k$ the theory 
enjoys $\mathcal{N}=6$ SUSY, and as such 
possesses an $SO(6)_R$ symmetry which is manifest in the potential \cite{Benna:2008zy}. 
The theory is then 
automatically conformal at the quantum level.  For $k=1,\, 2$ the SUSY is enhanced to $\mathcal{N}=8$. In field theory it has been argued \cite{Aharony:2008ug}  that this enhancement is due to quantum effects where 't Hooft monopole operators play a key r\^ole. Indeed, the 
ABJM theory has just the right structure \cite{Benna:2009xd} for these monopole operators to have appropriate quantum numbers that then allow for such a symmetry enhancement.

Motivated by this progress in understanding the maximally SUSY case, it is natural to consider M2 branes moving in less symmetric spaces, leading to versions of the duality with reduced SUSY. Inspired by ABJM \cite{Aharony:2008ug}, the theories considered are $\prod_{a=1}^G U(N)_{k_a}$ quiver CS (QCS) theories with bifundamental matter. The r\^ole of the CS levels is far from trivial, and it has been argued in \cite{Gaiotto:2009mv, Gaiotto:2009yz} that the sum $\sum_{a=1}^G k_a$ corresponds to the Type IIA SUGRA Romans mass parameter. In this paper we will focus entirely on the case in which the CS levels sum to zero; the Romans mass then vanishes and the system admits an M-theory lift. 

Since the kinetic terms for the gauge fields are given by the CS action, the only classically dimensionful parameters are the superpotential couplings. Strong gauge dynamics is then conjectured to drive the theory to a superconformal IR fixed point. In  \cite{Jafferis:2008qz,Martelli:2008si,Hanany:2008cd}, a general analysis of the moduli spaces of such superconformal gauge theories was presented. In particular, it is crucial that the CS levels sum to zero  if there is to be a so-called geometric branch of the moduli space 
which is a Calabi-Yau four-fold cone, where the branes are interpreted as moving.
In parallel to the ABJM case, multiplying the vector of CS levels by an integer orbifolds this moduli space in a certain way. More precisely, the theory with $k=\mathrm{gcd}\{k_a\}$ has a (abelian) moduli space which is a $\Z_k$ 
quotient of the moduli space of the theory with CS levels $\{k_a/k\}$. Generically this group will not act freely away from the tip of the cone. In this case one might expect additional 
gauge symmetries at such fixed points.  This has recently motivated \cite{Benini:2009qs, Jafferis:2009th} the consideration of 
 dual field theories which involve fundamental, as well as bifundamental, matter. It is however fair to say that, at present, there is no comprehensive understanding of these constructions.

On general grounds, the presence of global symmetries is of great help in classifying the spectrum of a gauge theory. One particularly important such global symmetry is the R-symmetry. In three dimensions a theory preserving $\mathcal{N}$ supersymmetries admits the action of an $SO\left(\mathcal{N}\right)$ R-symmetry. Thus the existence of a non-trivial R-symmetry, which can then provide 
important constraints on the dynamics, requires that we focus on $\mathcal{N}\ge2$, implying there is at least a $U(1)_R$. In particular it then follows that, assuming the theory flows to an IR superconformal fixed point, the scaling dimensions of chiral primary operators coincide with their R-charges. We note that, generically, the $\mathcal{N}=2$ theories considered have classically irrelevant superpotentials. Strong gauge dynamics is 
required to give large anomalous dimensions, thus making it possible to reach a non-trivial IR fixed point. However, in three dimensions there 
are few independent field theory checks on the existence of such a fixed point. For example, there is 
no useful analogous version of $a$-maximization \cite{Intriligator:2003jj}, which for four-dimensional $\mathcal{N}=1$ theories allows one to determine the R-charge in the superconformal algebra at the IR fixed point. This places the conjectured dualities on a much weaker footing than their four-dimensional cousins in Type IIB string theory.

The $\mathcal{N}=2$ QCS theories that we consider are expected to be dual to M2 branes moving in a Calabi-Yau four-fold cone over a seven-dimensional Sasaki-Einstein base $Y$, thus giving rise to an $AdS_4\times Y$ near horizon dual geometry. Such Sasaki-Einstein manifolds $Y$ will typically have  non-trivial topology, implying the existence of Kaluza-Klein (KK) modes obtained by reduction of SUGRA fields along the corresponding homology
cycles. Of particular interest are five-cycles, on which one can reduce the M-theory six-form potential to obtain $b_2(Y)=\dim H_2(Y,\R)$ vector fields in $AdS_4$. These vector fields are part of short multiplets of the KK reduction on $Y$, known as \textit{Betti multiplets} \cite{D'Auria:1984vv, D'Auria:1984vy} (for a discussion relevant to the cases we will consider, see also \cite{Fabbri:1999hw, Merlatti:2000ed}). In analogy with the Type IIB case, where these symmetries are well-known to correspond to global baryonic symmetries \cite{Klebanov:1999tb}, we will sometimes employ the same terminology here and  refer to these as baryonic $U(1)$s. 

In this paper we set out to study the above symmetries in the $AdS_4/CFT_3$ correspondence. In the rather better-understood $AdS_5/CFT_4$ correspondence in Type IIB string theory, from the field theory point of view these  baryonic $U(1)$ symmetries appear as non-anomalous combinations of the diagonal $U(1)$ factors inside the $U(N)$ gauge groups.\footnote{For a complete discussion of the Type IIB case we refer to \textit{e.g.} \cite{Franco:2005sm, Martelli:2007mk,Martelli:2008cm}.} The key point is that, in four dimensions, abelian gauge fields are IR free and thus become global symmetries in the IR. However, this is no longer true in three dimensions, thus raising the question of the fate of these abelian symmetries. From the gravity perspective, in the dual $AdS_4$ the vector fields admit two admissible fall-offs at the boundary of $AdS_4$ \cite{Witten:2003ya, Marolf:2006nd}. This is in contrast to the $AdS_5$ case where only one of them, that which leads to the interpretation as dual to a global current, is allowed. That the two behaviours are permitted implies that the corresponding boundary symmetries remain either gauged or ungauged, respectively, defining in each case a different boundary CFT. This issue is closely related to the gauge groups being either $U(N)$ or $SU(N)$ in the case at hand. From the point of view of the QCS theory with $U(N)$ gauge groups, at lowest CS level $k=1$ there is no real distinction between $U(N)$ and $U(1)\times SU(N)$ gauge groups  \cite{Aharony:2008ug, Imamura:2008nn,  Lambert:2010ji} . Therefore the discussion in \cite{Witten:2003ya} can be applied to the abelian part of the symmetry. In this way it is possible to connect the $SU(N)$ and the $U(N)$ theories in a rather precise manner, while keeping track of the corresponding action on the gravity side, which amounts to selecting one particular fall-off for the vector fields in $AdS_4$. This provides motivation to look at the $SU(N)$ version of the theory as dual to a particular choice of boundary conditions in the dual gravity picture.

In the first part of this paper we focus on the simplest class of examples, namely isolated toric Calabi-Yau four-fold singularities with no vanishing six-cycles (no exceptional divisors in a crepant resolution). These are discussed in more detail in \secref{sec:classification}. In particular we study in detail the example of $\mathcal{C}(Q^{111})$. A dual $U(N)^4$ QCS field theory was proposed for this singularity in \cite{Franco:2008um}, and further studied in \cite{Franco:2009sp} where the non-abelian chiral ring of the theory (at large $k$) was shown to precisely match the coordinate ring of the variety. Motivated by the analysis of the behaviour of gauge fields in $AdS_4$, we will choose boundary conditions where the $b_2(Q^{111})=2$ Betti multiplets are dual to global symmetries. This amounts to focusing on a certain version of the theory with gauge group $U(1)^2\times SU(N)^4$. On the other hand, gauge fields in $AdS_4$ can have {\it a priori} both electric sources, corresponding to wrapped M5 branes, and magnetic sources, corresponding to wrapped M2 branes. It turns out that the boundary conditions necessary to define the $AdS/CFT$ correspondence allow for just one of the two types at a time  \cite{Witten:2003ya}. In particular, the chosen $U(1)^2\times SU(N)^4$ quantization allows only for electric sources; that is, wrapped supersymmetric M5 branes. In turn, these correspond to baryonic operators \cite{Imamura:2008ji} in the field theory that are charged under the global symmetries. We will analyse this correspondence in detail, finding the expected agreement.

On the other hand, magnetic sources correspond to M2 branes \cite{Imamura:2008ji}. While in the $AdS$ geometry these wrap non-supersymmetric cycles, we can also consider resolutions of the corresponding cone where there are supersymmetric wrapped M2 branes. Along the lines of \cite{Klebanov:2007us, Klebanov:2007cx}, we will identify the relevant operator, responsible for the resolution, which is acquiring a VEV. Very much as in reference \cite{Klebanov:2007cx}, it is possible to find an interpretation of these solutions as spontaneous symmetry breaking (SSB) through the explicit appearance of a Goldstone boson in the SUGRA dual. 

A natural next step is to enlarge the class of singularities under consideration by allowing dual geometries with exceptional six-cycles. One such example is a $\mathbb{Z}_2$ orbifold of $\mathcal{C}(Q^{111})$ known as $\mathcal{C}(Q^{222})$. A dual field theory candidate has been proposed in \cite{Franco:2009sp, Amariti:2009rb, Davey:2009sr}. Further tests of this theory were performed in \cite{Franco:2009sp}, where it was shown that its chiral ring matches the gravity computation at large $k$. The interpretation of such six-cycles is somewhat obscure holographically. Indeed, such six-cycles, when resolved, can support M5 brane instantons leading to non-perturbative corrections \cite{Witten:1996bn}. In the second part of this paper we set up the study of such corrections by finding a general expression for the Euclidean action of such branes in terms of normalizable harmonic two-forms, and compute this explicitly for $Q^{222}$. We leave a full understanding of such non-perturbative effects from the gauge theory point of view for future work.

The organization of this paper is as follows. In \secref{sec:2} we review the Freund-Rubin-type solutions which are eleven-dimensional $AdS_4\times Y$  backgrounds. We then turn to KK reduction of the SUGRA six-form potential on five-cycles in $Y$, leading to the Betti multiplets of interest. General analysis of gauge fields in $AdS_4$ shows that two possible fall-offs are admissible. We then review the construction in \cite{Witten:2003ya} relating these different boundary conditions for a single abelian gauge field in $AdS_4$ to the action of $SL(2,\mathbb{Z})$. In \secref{sec:3} we turn in more detail to the field theory description. We start by reviewing general aspects of $U(N)$ QCS theories that have appeared in the literature, before turning in \secref{sec:Q111} to the example of interest. We then propose a set of boundary conditions dual to the $U(1)^2\times SU(N)^4$ theory. We identify the ungauged $U(1)$s via the electric M5 branes wrapping holomorphic divisors in the geometry. In  \secref{sec:4} we turn to the spontaneous breaking of these baryonic symmetries. We compute on the gravity side the baryonic condensate and identify the Goldstone boson of the SSB. In \secref{sec:5} we initiate the study of exceptional six-cycles. We compute the warped volume of a Euclidean brane in the resolved $\mathcal{C}(Q^{222})$ geometry. By extending our results on warped volumes to arbitrary geometries, both for the baryonic condensate and the Euclidean brane, we find general formulae for such warped volumes. We end with some concluding comments in \secref{sec:6}. Finally, a number of relevant calculations and formulae are collected in the appendices.

\vspace{.5cm}
\textbf{Note added:} as this paper was being finalized the preprint \cite{Klebanov:2010tj} appeared, which has partial overlap with our results.


\section{$AdS_4$ backgrounds and abelian symmetries} \label{sec:2}

We begin by reviewing general properties of Freund-Rubin $AdS_4$ backgrounds, and also introduce the $Q^{111}$ and $Q^{222}=Q^{111}/\Z_2$ examples of main interest. KK reduction of the M-theory potentials on topologically non-trivial cycles leads to gauge symmetries in $AdS_4$. We review their dynamics in the $AdS/CFT$ context and the sources allowed, depending on the chosen quantization. Of central relevance for our purposes will be wrapped supersymmetric M5 branes.

\subsection{Freund-Rubin solutions}\label{sec:FR}

The $AdS_4$ backgrounds of interest are of Freund-Rubin type, with eleven-dimensional metric and 
four-form given by
\bea\label{AdSbackground} 
\diff s^2_{11} &= &R^2\left(\frac{1}{4}\diff s^2(AdS_4) + \diff s^2(Y)\right)~,\\ \nn 
G &=& \frac{3}{8}R^3 \mathrm{Vol}(AdS_4)~. 
\eea
Here the $AdS_4$ metric is normalized so that 
$R_{\mu\nu} = -3g_{\mu\nu}$. The Einstein equations 
imply that $Y$ is an Einstein 
manifold of positive Ricci curvature, with metric normalized so that 
$R_{ij} = 6g_{ij}$. The flux quantization condition
\bea
\frac{1}{(2\pi \ell_p)^6}\int_Y \star_{11}\, G = N\in \mathbb{Z}~,
\eea
then leads to the relation
\bea 
R = 2\pi \ell_p\left(\frac{N}{6\vol(Y)}\right)^{1/6}~,
 \eea
where $\ell_p$ denotes the eleven-dimensional Planck length.

As is well-known, such solutions arise as the near-horizon limit of $N$ M2 branes placed at the tip $r=0$ of the Ricci-flat cone
\bea\label{cone}
\diff s^2({\cal C}(Y)) = \diff r^2 + r^2 \diff s^2(Y)~. 
\eea
More precisely, the eleven-dimensional solution is
\bea\label{background} 
\dd s^2_{11} &=& h^{-2/3} \dd s^2(\R^{1,2}) + h^{1/3} \dd s^2(X)~, \\ \nn 
G &=& \dd^3 x \wedge \dd h^{-1}~, 
\eea
where in the case at hand we take the eight-manifold $X={\cal C}(Y)$ with conical metric (\ref{cone}).
Placing $N$ Minkowski space-filling M2 branes at $r=0$ leads, after including their gravitational back-reaction, to the warp factor
\bea \label{warping}
h = 1 + \frac{R^6}{r^{6}}~. 
\eea
In the near-horizon limit, near to $r=0$, the background (\ref{background}) approaches the $AdS_4$ background (\ref{AdSbackground}).
In fact the warp factor $h=R^6/r^6$ is precisely the $AdS_4$ background in a Poincar\'e slicing. More precisely, writing 
\bea
\label{z_coordinate}
z= \frac{R^2}{r^2} \ , \qquad \diff s^2(AdS_4) = z^{-2}\left(\diff z^2 + \diff s^2(\R^{1,2})\right)~,
\eea
leads to the metric (\ref{AdSbackground}).

\begin{table}[ht]
\centering
\begin{tabular}{|c|c|c|} 
\hline
Supersymmetries $\mathcal{N}$ & $Y$ & ${\cal C}(Y)$ \\
\hline
1 & weak $G_2$ holonomy & $Spin(7)$ holonomy \\
2 & Sasaki-Einstein & $SU(4)$ holonomy (Ricci-flat K\"ahler) \\
3 & 3-Sasakian & $Sp(2)$ holonomy (hyperK\"ahler) \\
\hline
\end{tabular}
\caption{Relation between the number of supersymmetries $\mathcal{N}$ in $AdS_4$ and the special Einstein geometry of $Y$ and its cone ${\cal C}(Y)$.}
\label{table}
\end{table}
We shall be interested in solutions of this form preserving supersymmetry in $AdS_4$. 
The well-known result \cite{Acharya:1998db} is summarized in Table \ref{table}. As mentioned in the introduction, in general $\mathcal{N}$ supersymmetries leads to the R-symmetry group $SO(\mathcal{N})$, and thus supersymmetry provides 
a strong constraint on the spectrum only for $\mathcal{N}\geq 2$. We hence restrict 
attention to the $\mathcal{N}=2$ Sasaki-Einstein case, which includes the $\mathcal{N}=3$ geometry as a special case. 
It is then equivalent to say that the cone metric on ${\cal C}(Y)$ is K\"ahler as well as as Ricci-flat, {\it i.e.} Calabi-Yau. 
Geometries with $\mathcal{N}\geq 4$ supersymmetries are necessarily quotients of $S^7$.

Only a decade ago the only known examples of such Sasaki-Einstein 
seven-manifolds were homogeneous spaces. Since then there has been dramatic progress. 
3-Sasakian manifolds, with $\mathcal{N}=3$, may be constructed via an analogue of the hyperK\"ahler quotient, leading 
to rich infinite classes of examples \cite{BG}. For $\mathcal{N}=2$ supersymmetry
one could take $Y$ to be one of the explicit $Y^{p,k}$  
manifolds constructed in \cite{Gauntlett:2004hh}, and further 
studied in \cite{Res, Martelli:2008rt}, or any of their subsequent generalizations. These $\mathcal{N}=2$ examples 
are all toric, meaning that the isometry group contains
$U(1)^4$ as a subgroup. In fact, toric Sasaki-Einstein manifolds are now
completely classified thanks to the general existence and uniqueness result in 
\cite{FOW}. At the other extreme, there are also non-explicit metrics in which $U(1)_R$ is the only 
isometry \cite{BG}. 

However, for our purposes it will be sufficient to focus on two specific homogeneous examples, namely 
$Q^{111}$ and $Q^{222}=Q^{111}/\Z_2$, with 
$\Z_2\subset U(1)_R$ being along the R-symmetry of $Q^{111}$. These will turn out to be simple enough so that 
 everything can be computed explicitly, and yet at the same time 
we shall argue that many of the features seen in these cases hold also for the more general geometries mentioned above.
In both cases the isometry group is
$SU(2)^3\times U(1)_R$, and in local coordinates the explicit metrics are
\bea\label{Qiiimetric}
\diff s^2 = \frac{1}{16}\left(\diff\psi+\sum_{i=1}^3 \cos\theta_i\diff\phi_i\right)^2 + \frac{1}{8}\sum_{i=1}^3 
\left(\diff \theta_i^2 + \sin^2\theta_i\diff\phi_i^2\right)~.
\eea
Here $(\theta_i,\phi_i)$ are standard coordinates on three copies of $S^2=\mathbb{CP}^1$, $i=1,2,3$, and $\psi$ has period $4\pi$ for $Q^{111}$ and period 
$2\pi$ for $Q^{222}$. The two Killing spinors are charged under $\partial_\psi$, which is dual to the $U(1)_R$ symmetry. The metric 
(\ref{Qiiimetric}) shows very explicitly the regular structure of a $U(1)$ bundle over the standard K\"ahler-Einstein metric on
$\mathbb{CP}^1\times\mathbb{CP}^1\times\mathbb{CP}^1$, where $\psi$ is the fibre coordinate and the Chern numbers are $(1,1,1)$ and $(2,2,2)$ respectively. These are hence 
natural generalizations\footnote{The other natural such generalization is the homogeneous space $V_{5,2}=SO(5)/SO(3)$, which has been studied in detail in \cite{Martelli:2009ga}.} to seven dimensions of the $T^{11}$ and $T^{22}$ manifolds. 

\subsection{$C$-field modes}\label{sec:Cfield}

One might wonder whether it is possible to turn on an internal $G$-flux $G_Y$ on $Y$, in addition to the $G$-field in (\ref{AdSbackground}), and still preserve supersymmetry, {\it i.e.}
\bea
G = \frac{3}{8}R^3 \mathrm{Vol}(AdS_4) + G_Y~.
\eea
In fact necessarily $G_Y=0$. This follows from the results of \cite{Becker:1996gj}: 
for any warped Calabi-Yau four-fold background with metric of the form (\ref{background}), one can turn on a 
$G$-field $G_X$ on $X$ without changing the Calabi-Yau metric on $X$ only if 
$G_X$ is self-dual. But for a cone, with $G_X=G_Y$ a pull-back from the base $Y$, this obviously implies that $G_X=0$. 

However, more precisely the $G$-field in M-theory determines a 
class\footnote{This is true since the membrane global anomaly described in \cite{Witten:1996md} is always zero on a seven-manifold $Y$ that is spin.} in $H^4(Y,\Z)$. The differential form part of $G$ captures 
only the image of this in $H^4(Y,\R)$, and so $G_Y=0$ still allows for a topologically non-trivial $G$-field classified by the torsion part $H^4_{\mathrm{tor}}(Y,\Z)$. This is also captured, up to gauge equivalence, by the holonomy of the corresponding flat $C$-field through dual torsion three-cycles in $Y$.
There are hence $|H^4_{\mathrm{tor}}(Y,\Z)|$ physically distinct $AdS_4$ Freund-Rubin backgrounds associated to the same geometry, which should 
thus correspond to physically inequivalent dual SCFTs. In a small number of examples with proposed Chern-Simons quiver duals, including the original ABJ(M) theory, different choices of this torsion $G$-flux have been 
argued to be dual to changing the \emph{ranks} in the quiver \cite{Martelli:2009ga, Aharony:2008gk}. 
However, the related Seiberg-like dualities are currently very poorly understood in examples without 
Hanany-Witten-type brane duals. In particular, for example, one can compute $H^4(Q^{111},\Z)\cong\Z_2$, implying there are two distinct M-theory backgrounds with the same $Q^{111}$ geometry but different $C$-fields. This is an important aspect of the $AdS_4/CFT_3$ duality that we shall not discuss any further in this paper.

More straightforwardly, if one has $b_3(Y)=\dim H_3(Y,\R)$ three-cycles in $Y$ then one can also turn on a closed three-form $C$ with non-zero periods through these 
cycles. Including large gauge transformations, this gives a space $U(1)^{b_3(Y)}$ of such flat $C$-fields. 
Since these are continuously connected to each other they would be dual to marginal deformations in the dual field theory. Indeed, the \emph{harmonic} three-forms on 
a Sasaki-Einstein seven-manifold are in fact paired by an almost complex structure \cite{Boyer:1998sf} and thus $b_3(Y)$ is always even, allowing these 
to pair naturally into complex parameters as required by $\mathcal{N}=2$ supersymmetry. However, for the class of toric singularities 
studied in this paper, including $Q^{111}$ and $Q^{222}$, it is 
straightforward\footnote{There are, however, examples: the Calabi-Yau four-fold hypersurfaces $\sum_{i=1}^5 z_i^d=0$, where 
$d=3, 4$, are known to have Calabi-Yau cone metrics, and these have $b_3(Y)=10$, $60$, respectively \cite{Boyer:1998sf}.} to show that $b_3(Y)=0$ and there are hence no such marginal deformations associated to the $C$-field. 

Finally, since $H_6(Y,\R)=0$ for any positively curved Einstein seven-manifold, there are never
periods of the dual potential $C_6$ through six-cycles in $Y$.

\subsection{Baryonic symmetries and wrapped branes}\label{sec:baryons}

Of central interest in this paper will be symmetries associated to the topology of $Y$, and the corresponding 
charged BPS states associated to wrapped M branes. By analogy with the corresponding situation in $AdS_5\times Y_5$ in Type IIB string theory, 
we shall refer to these symmetries as baryonic symmetries; the name will turn out to be justified.

Denote by $b_2(Y)=\dim H_2(Y,\R)$ the second Betti number of $Y$. By Poincare duality we have $\dim H_5(Y,\R)=\dim H_2(Y,\R)=b_2(Y)$. Let $\alpha_1,\ldots,\alpha_{b_2(Y)}$ be a set of dual harmonic five-forms with integer periods. Then for the $AdS_4 \times Y$ Freund-Rubin background we may  write the KK ansatz
\bea 
\delta C_6 = \frac{2\pi}{T_5} \sum_{I=1}^{b_2(Y)} \mathcal{A}_I\wedge\alpha_I~,
\label{3-form-to-global}
\eea
where $T_5={2\pi}/{(2\pi \ell_p)^6}$ is the M5 brane tension. 
This gives rise to $b_2(Y)$ massless $U(1)$ gauge fields $\mathcal{A}_I$ in $AdS_4$. 
For a supersymmetric theory these gauge fields of course sit in certain multiplets, known as 
\emph{Betti multiplets}. See, for example, \cite{D'Auria:1984vv, D'Auria:1984vy, Fabbri:1999hw, Merlatti:2000ed}. 

\subsubsection{Vector fields in $AdS_4$, boundary conditions and dual CFTs}

The $AdS/CFT$ duality requires specifying the boundary conditions for the fluctuating fields in $AdS$. In particular, vector fields in $AdS_4$ admit different sets of boundary conditions \cite{Witten:2003ya, Marolf:2006nd} leading to different boundary CFT´s. In order to see this, let us consider a vector field in $AdS_{d+1}$. Using the straightforward generalization to $AdS_{d+1}$ of the coordinates in (\ref{z_coordinate}), in the gauge $A_z=0$ the bulk equations of motion set
\begin{equation}
\label{gauge_field_In_AdS4}
A_{\mu}=a_{\mu}+j_{\mu}\, z^{d-2} \ ,
\end{equation}
where $a_{\mu},\, j_{\mu}$ satisfy the free Maxwell equation in Lorentz gauge in the Minkowski space. It is not hard to see that in $d<4$ both behaviours have finite action, and thus can be used to define a consistent $AdS/CFT$ duality.

Let us now concentrate on the case of interest $d=3$, where both quantizations are allowed. In order to have a well-defined variational problem for the gauge field in $AdS_4$ we should be careful with the boundary terms when varying the action. In general, we have
\begin{equation}
\delta S = \int \Big\{ \frac{\partial \sqrt{\det g}\,\mathcal{L}}{\partial A_M}-\partial_N\frac{\partial \sqrt{\det g}\,\mathcal{L}}{\partial\partial_NA_M}\Big\}\, \delta A_M +\partial_N\Big\{ \frac{\partial\sqrt{\det g}\, \mathcal{L}}{\partial\partial_NA_M}\, \delta A_M\Big\} \ .
\end{equation}
The bulk term gives the equations of motion whose solution behaves as (\ref{gauge_field_In_AdS4}). In turn, the boundary term can be seen to reduce to 
\begin{equation}
\delta S_B=- \frac{1}{2}\,\int_{\mathrm{Boundary}} \, j_{\mu}\delta a^{\mu} \, \diff^3 x~.
\end{equation}
Therefore, in order to have a well-posed variational problem, we need to demand $\delta a_{\mu}=0$; that is, we need to impose boundary conditions where $a_{\mu}$ is fixed in the boundary. 

On the other hand, since in $d=3$ both behaviours for the gauge field have finite action, we can consider adding suitable boundary terms such that the action becomes \cite{Marolf:2006nd}
\begin{equation}
\label{dynamical_a}
S=\frac{1}{4}\int \sqrt{\det g}\, F_{AB}\, F^{AB} +\frac{1}{2}\,\int_{\mathrm{Boundary}}\, \sqrt{\det g}\, A^{\mu} \, F_{r\mu}|_{\mathrm{Boundary}} \, \dd^3x .
\end{equation}
The boundary term is now 
\begin{equation}
\delta S_B=\frac{1}{2}\,\int_{\mathrm{Boundary}} \,  a_{\mu}\delta j^{\mu} \, \diff^3 x~,
\end{equation}
so that we need to impose the boundary condition $\delta j_{\mu}=0$; that is, fix the boundary value of $j_{\mu}$. 

Defining $\vec{B}=\frac{1}{2}\epsilon^{\mu\nu\rho}\, F_{\nu\rho}$ and $\vec{E}=F_{\mu r}$, we have
\begin{equation}
B^{\mu}=\epsilon^{\mu\nu\rho}\partial_{\nu}a_{\rho}+\epsilon^{\mu\nu\rho}\partial_{\nu}j_{\rho}\, z\,, \qquad E^{\mu}=j^{\mu}\, z^2 \ .
\end{equation}
The two sets of boundary conditions then correspond to either setting $E_{\mu}=0$ while leaving $a_{\mu}$ unrestricted, or setting $B_{\mu}=0$ while leaving $j_{\mu}$ unrestricted. 

At this point we note that $a_{\mu},\, j_{\mu}$ are naturally identified, respectively,  with a dynamical gauge field and a global current in the boundary. In accordance with this identification, eq. (\ref{gauge_field_In_AdS4}) and the usual $AdS/CFT$ prescription shows each field to have the correct scaling dimension for this interpretation: for a gauge field $\Delta(a_{\mu})=1$, while for a global current $\Delta(j_{\mu})=2$. Therefore, the quantization $E_{\mu}=0$ is dual to a boundary CFT where the $U(1)$ gauge field is dynamical; while the quantization $B_{\mu}=0$ is dual to a boundary CFT where the $U(1)$ is ungauged and is instead a global symmetry. Furthermore, as discussed in \cite{Klebanov:1999tb} for the scalar counterpart, once the improved action is taken into account the two quantizations are Legendre transformations of one another \cite{Klebanov:2010tj}, as can be seen by \textit{e.g.} computing the free energy in each case. 

One can consider electric-magnetic duality in the bulk theory, which exchanges $E_{\mu}\leftrightarrow B_{\mu}$ thus exchanging the two boundary conditions for the $AdS_4$ gauge field quantization. This action translates in the boundary theory into the so-called $\mathcal{S}$ \textit{operation} \cite{Witten:2003ya}. This is an operation on three-dimensional CFTs with a global $U(1)$ symmetry, taking one such CFT to another. In addition, it is possible to construct a $\mathcal{T}$ \textit{operation}, which amounts, from the bulk perspective, to a shift of the bulk $\theta$-angle by $2\pi$. Following \cite{Witten:2003ya}, we can be more precise in defining these actions in the boundary CFT. Starting with a three-dimensional CFT with a global $U(1)$ current $J^{\mu}$, one can couple this global current to a background gauge field $A$ resulting in the action $S[A]$. The $\mathcal{S}$ operation then promotes $A$ to a dynamical gauge field and adds a BF coupling of $A$ to a new background field $B$, while the $\mathcal{T}$ operation instead adds a CS term for the background gauge field $A$:
\begin{equation}
\mathcal{S}:\, S[A]\,\rightarrow\, S[A]+\frac{1}{2\pi} \int B\wedge \dd A~,\qquad \mathcal{T}:\,S[A]\,\rightarrow\, S[A]+\frac{1}{4\pi}\int A\wedge \dd A \ .
\end{equation}
As shown in \cite{Witten:2003ya}, these two operations generate the group $SL(2,\mathbb{Z})$.\footnote{Even though we are explicitly discussing the effect of $SL(2,\mathbb{Z})$ on the vector fields, since these are part of a whole Betti multiplet we expect a similar action on the other fields of the multiplet. We leave this investigation for future work.} In turn, as discussed above, the $\mathcal{S}$ and $\mathcal{T}$ operations have the bulk interpretation of exchanging $E_{\mu}\leftrightarrow B_{\mu}$ and shifting the bulk $\theta$-angle by $2\pi$, respectively. It is important to stress that these actions on the bulk theory change the boundary conditions. Because of this, the dual CFTs living on the boundary are different.

\subsubsection{Boundary conditions and sources for gauge fields: M5 branes in toric manifolds} \label{s:bc}

We are interested in gauge symmetries in $AdS_4$ associated to the topology of $Y$; that is, arising from KK reductions as in (\ref{3-form-to-global}). All Kaluza-Klein modes, and hence their dual operators, carry zero charge under these $b_2(Y)$ $U(1)$ symmetries. However, there are operators associated to wrapped M branes that do carry charge under this group. In particular, an M5 brane wrapped on a five-manifold $\Sigma_5\subset Y$, such that the cone ${\cal C}(\Sigma_5)$ is a complex divisor in the K\"ahler cone ${\cal C}(Y)$, is supersymmetric and leads to a BPS particle propagating in $AdS_4$. Since the M5 brane is a source for $G$, this particle is electrically charged under the $b_2(Y)$ massless $U(1)$ gauge fields $\mathcal{A}_I$. One might also consider M2 branes wrapped on two-cycles in $Y$. However, such wrapped M2 branes are supersymmetric only if the cone ${\cal C}(\Sigma_2)$ over the two-submanifold $\Sigma_2\subset Y$ is calibrated in the Calabi-Yau cone, 
and there are no such calibrating three-forms. Thus these particles, although topologically stable, are not BPS. They are magnetically charged under the $U(1)^{b_2(Y)}$ gauge fields in $AdS_4$ \cite{Imamura:2008ji}.

As discussed above, the $AdS/CFT$ duality instructs us to choose, for each $U(1)$ gauge field, a set of boundary conditions where either $E_{\mu}$ or $B_{\mu}$ vanishes. Clearly, only the latter possibility allows for the existence of the SUSY electric M5 branes, otherwise forbidden by the boundary conditions. In turn, this quantization leaves, in the boundary theory, the $U(1)$ symmetry as a global symmetry. Therefore, in this case we should expect to find operators in the field theory that are charged under the global baryonic symmetries and dual to the M5 brane states. We turn to this point in the next section. We note that, with this choice of boundary condition, the r\^ole of the Betti multiplets is very similar to their $AdS_5$ counterparts, giving rise to global baryonic symmetries in the boundary theory, and hence motivating the use of the same name in the case at hand. 

For toric manifolds there is a canonical set of such wrapped M5 brane states, where ${\cal C}(\Sigma_5)$ are taken to be the toric 
divisors. Each such state leads to a 
corresponding dual chiral primary operator that is charged under the $U(1)^{b_2(Y)}$ global symmetries and will 
also have definite charge under the $U(1)^4$ flavour group dual to the isometries of $Y$. We refer the reader to the standard literature for a thorough introduction to toric geometry. 
However, the basic idea is simple to state. The cone ${\cal C}(Y)$ fibres over a  polyhedral cone in $\R^4$ with generic fibre $U(1)^4$. 
This polyhedral cone is by definition a convex set of the form $\bigcap\{\mathbf{x}\cdot \mathbf{v}_\alpha\geq 0\}\subset\R^4$, where $\mathbf{v}_\alpha\in\Z^4$ are integer vectors. 
This set of vectors is precisely the set of charge vectors specifying the $U(1)$ subgroups of $U(1)^4$ that have complex codimension one fixed point sets. 
These fixed point sets are, by definition, the toric divisors referred to above. 
The Calabi-Yau condition implies that, with a suitable choice of basis, we can write $\mathbf{v}_\alpha=(1,\mathbf{w}_\alpha)$, with 
$\mathbf{w}_\alpha\in\Z^3$. If we plot these latter points in $\R^3$ and take their convex hull, we obtain the \emph{toric diagram}. 

\begin{figure}[ht]
\begin{center}
\includegraphics[scale=.40,angle=-90]{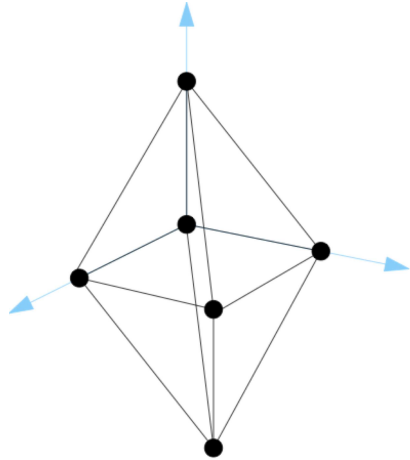}
\end{center}
\caption{The toric diagram for $\mathcal{C}(Q^{111})$.}
\label{fig:toricdiagramQ111}
\end{figure}

For the $Q^{111}$ example the toric divisors are given by taking $\Sigma_5=\{\theta_i=0\}$ or 
$\Sigma_5=\{\theta_i=\pi\}$, for any $i=1,2,3$, which are 6 five-manifolds in $Y$. The toric diagram for $Q^{111}$ is shown in 
Figure \ref{fig:toricdiagramQ111}, where one sees clearly these 6 toric divisors as the 6 external vertices. 
Notice that for $Q^{111}$ the full isometry group may be used to rotate $\{\theta_i=0\}$ into $\{\theta_i=\pi\}$, specifically using the $i$th
copy of $SU(2)$ in the $SU(2)^3\times U(1)_R$ isometry group. In fact these two five-manifolds are two points in an $S^2$ family of such five-manifolds related via the isometry group. 
Similar comments apply also to $Q^{222}$. 


\section{Baryonic symmetries in QCS theories}\label{sec:3}

In the previous section we discussed the r\^ole of vector fields in $AdS_4$. In particular, we have shown that there is a choice of boundary conditions where the Betti multiplets corresponding to (\ref{3-form-to-global}) are dual to global currents in the boundary theory. From the bulk perspective, this translates into the possibility of having electric M5 brane states in the theory, in a consistent manner. On general grounds, we expect these states to be dual to certain operators in the boundary theory charged under the global $U(1)^{b_2(Y)}$. In this section we turn to a more precise field theoretic description of this. We begin with a brief review of the $U(N)$ theories considered in the literature, before turning to our $\mathcal{C}(Q^{111})$ example and considering the r\^ole of the abelian symmetries in this case.

\subsection{$U(N)$ QCS theories}

Let us start by considering the $\prod_{a=1}^G U(N_a)$ theories. The Lagrangian, in $\mathcal{N}=2$ superspace notation, for a theory containing an arbitrary number of bifundamentals $X_{ab}$ in the representation $(\square_a,\, \bar{\square}_b)$ under the $(a,\, b)$-th gauge groups and a choice of superpotential $W$, reads 
\bea\label{action}
\mathcal{L}= &&\int \dd^4 \theta\, \mathrm{Tr}\left[ \sum\limits_{X_{ab}} X_{ab}^\dagger \, \ex^{-V_{a}} X_{ab} \, \ex^{V_{b}}
+ \sum\limits_{a=1}^G \frac{k_a}{2\pi} \int\limits_0^1 \dd t V_a \bar{D}^{\alpha}(\ex^{t V_a} D_{\alpha} \ex^{-tV_a})
\right] \nonumber \\
&& + 
\int \dd^2 \theta \, W(X_{ab}) \, + \, \mathrm{c.c.}~.
\eea
Here $k_a\in\Z$ are the CS levels for the vector multiplet $V_a$. For future convenience we define $k={\rm gcd}\{k_a\}$.

The classical vacuum moduli space (VMS) is determined in general by the following equations 
\cite{Martelli:2008si, Hanany:2008cd}
\begin{eqnarray}\label{VMSeqns}
\nn \partial_{X_{ab}} W &=& 0~,\\
\nn \mu_a := -\sum\limits_{b=1}^G {X_{ba}}^{\dagger} {X_{ba}} + 
\sum\limits_{c=1}^G  {X_{ac}} {X_{ac}}^{\dagger} &=& 
\frac{k_a\sigma_a}{2\pi}~, \\
\label{DF} \sigma_a X_{ab} - X_{ab} \sigma_b &=& 0~,
\end{eqnarray}
where $\sigma_a$ is the scalar component of $V_a$. Following \cite{Martelli:2008si}, upon diagonalization of the fields using $SU(N)$ rotations, one can focus on the branch where $\sigma_a=\sigma$, $\forall a$, so that the last equation is immediately satisfied.\footnote{We stress that there might be, and indeed even in the $Q^{111}$ example there are, other branches of the moduli space where the condition $\sigma_a=\sigma$ for all $a$ is not met, and yet still the bosonic potential is minimized.} Under the assumption that $\sum_{a=1}^G k_a=0$, the equations for the moment maps $\mu_a$ boil down to a system of $G-2$ independent equations for the bifundamental fields, analogous to D-term equations. Since for toric superpotentials the set of F-flat configurations, determining the so-called master space, is of dimension $G+2$, upon imposing the $G-2$ D-terms and dividing by the associated gauge symmetries we have a ${\rm dim}_\mathbb{C}\mathcal{M}=4$ moduli space $\mathcal{M}$ where the M2 branes move. 

However, due to the peculiarities of the CS kinetic terms, extra care has to be taken with the diagonal part of the gauge symmetry. At a generic point of the moduli space the gauge group is broken to $N$ copies of $U(1)^{G}$.  The diagonal gauge field $\mathcal{B}_G=\sum_{a=1}^G\mathcal{A}_a$ is completely decoupled from the matter fields, and only appears coupled to $\mathcal{B}_{G-1}=k^{-1}\,\sum_{a=1}^Gk_a\,\mathcal{A}_a$ through 
\begin{equation}\label{SBG}
S(\mathcal{B}_G)=\frac{k}{4\pi\, G}\int (\mathcal{B}_{G-1})_{\mu}\,\epsilon^{\mu\nu\rho}\, (\mathcal{G}_G)_{\nu\rho}~.
\end{equation}
Since $\mathcal{B}_G$ appears only through its field strength, it can be dualized into a scalar $\tau$. Following the standard procedure, it is easy to see that integrating out $\mathcal{G}_G=\dd\mathcal{B}_G$ sets
\begin{equation}
\label{identification}
\mathcal{B}_{G-1}=\frac{G}{k}\, \dd\tau~,
\end{equation}
such that the relevant part of the action becomes a total derivative
\begin{equation}
S(\mathcal{B}_G)=\int \dd\Big(\frac{\tau}{2\pi}\,\mathcal{G}_G\Big)~.
\end{equation}
Around a charge $n\in\Z$ monopole in the diagonal $U(1)$ gauge field $\mathcal{B}_G$ we then have $\int \mathcal{G}_G=2\pi\,G\, n$, so that $\tau$ must have period $2\pi/G$ in order for the above phase to be unobservable \cite{Martelli:2008si}. Gauge transformations of  $\mathcal{B}_{G-1}$ then allow one to gauge-fix $\tau$ to a particular value via (\ref{identification}), but this still leaves a residual discrete set of $\Z_k$ gauge symmetries that leave this gauge choice invariant. The space of solutions to (\ref{DF}) is then quotiented by gauge transformations where the parameters $\theta_a$ satisfy $\sum_{a=1}^G k_a\, \theta_a =0$, together 
with the residual discrete $\Z_k$ gauge transformations generated by $\theta_a = 2\pi/k$ for all $a$. Altogether this leads to a $U(1)^{G-2}\times \Z_k$ quotient. We refer to \cite{Martelli:2008si} for further discussion, and to  \cite{Franco:2009sp} for a discussion in the context of the $Q^{111}$ theory in particular.

An alternative point of view has recently appeared in the literature \cite{Benini:2009qs, Jafferis:2009th}, in which the existence of two special monopole operators $T, \tilde{T}$ is noted. These monopole operators, which have charges $\pm (k_1,\cdots, k_G)$ respectively under each gauge group, are conjectured to satisfy a relation in the chiral ring of the form $T\, \tilde{T}=1~$. In this approach the moduli space is defined as the chiral ring of the abelian theory enhanced by the operators $T, \tilde{T}$, together with the constraint.

\subsection{The $\mathcal{C}(Q^{111})$ theory}\label{sec:Q111}

\subsubsection{The theory and its moduli space}

\begin{figure}[ht]
\begin{center}
\includegraphics[scale=1.1]{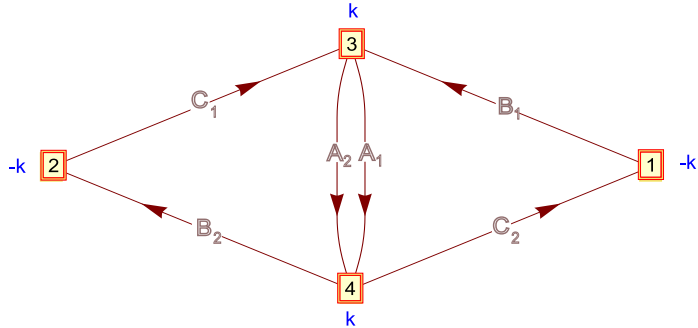}
\end{center}
\caption{The quiver diagram for a conjectured dual of $\mathcal{C}(Q^{111})$.}
\label{fig:quiverdiagramQ111}
\end{figure}

A field theory candidate dual to M2 branes probing $\mathcal{C}(Q^{111})/\Z_k$ was proposed in \cite{Franco:2008um} and further studied in \cite{Franco:2009sp}. The proposal in those references is a $U(N)^4$ Chern-Simons gauge theory with CS levels $(k,\, k,\, -k,\, -k)$, with matter content summarized by the quiver in Figure \ref{fig:quiverdiagramQ111}. 

In addition, there is a superpotential given by
\begin{equation}
\label{WQ111}
W={\rm Tr}\, \Big(\, C_2\, B_1\, A_1\, B_2\, C_1\, A_2\,-\,C_2\,B_1\, A_2\, B_2\, C_1\, A_1\, \Big)\ .
\end{equation}
As expected for a field theory dual to $N$ point-like branes moving in $\mathcal{C}(Q^{111})/\Z_k$, the moduli space contains a branch which is the symmetric product of $N$ copies of this conical singularity. To see this, let us begin with the abelian theory in which all the gauge groups are 
$U(1)$. As shown in  \cite{Franco:2009sp}, after integrating out the auxiliary $\sigma$ scalar the geometric 
branch of the moduli space with $k=1$ is described by $G-2=2$ D-term equations. Recalling the special r\^ole played by $\mathcal{B}_{G-1}=\mathcal{B}_3,\, \mathcal{B}_G=\mathcal{B}_4$, it is useful to introduce the following basis for the $U(1)$ gauge fields:
\bea
\label{redefinedGF}
\nn
&&\mathcal{A}_{I}=\frac{1}{2}(\mathcal{A}_{1}-\mathcal{A}_{2}+\mathcal{A}_{3}-\mathcal{A}_{4})~, \quad
\mathcal{A}_{II}=\frac{1}{2}(\mathcal{A}_{1}-\mathcal{A}_{2}-\mathcal{A}_{3}+\mathcal{A}_{4})~,\\ \nn
&&\mathcal{B}_3=\mathcal{A}_{1}+\mathcal{A}_{2}-\mathcal{A}_{3}-\mathcal{A}_{4}~,\quad
\mathcal{B}_4=\mathcal{A}_{1}+\mathcal{A}_{2}+\mathcal{A}_{3}+\mathcal{A}_{4}~.
\eea
Then the two D-terms to impose are those for $\mathcal{A}_I,\,\mathcal{A}_{II}$. In turn, the charge matrix is
\bea
\label{gaugedU(1)charges}
\begin{array}{l | c c c c c c} 
 & A_1 & A_2 & B_1 & B_2 & C_1 & C_2 \\ \hline
 U(1)_{I} & 1 & 1 & 0 & 0 & -1 & -1 \\ 
 U(1)_{II} & -1 & -1 & 1 & 1 & 0 & 0 \\
 U(1)_{\mathcal{B}_3} & 0 & 0 & -2 & 2 & -2 & 2 \\ 
 U(1)_{\mathcal{B}_4} & 0 & 0 & 0 & 0 & 0 & 0 
\end{array}~.
\eea
Notice the appeareance of the $SU(2)^3$ global symmetry, under which the pairs $A_i$, $B_i$, $C_i$ transform as doublets under each of the 
respective factors. 

Since for the abelian theory the superpotential is identically zero, one can determine the abelian moduli space 
by constructing the gauge-invariants with respect to the gauge transformations for $\mathcal{A}_I,\, \mathcal{A}_{II}$. Borrowing the results from \cite{Franco:2009sp}, for CS level $k=1$ these are
\begin{equation}
\begin{array}{lclclcl}
w_1 = A_1\,B_2\,C_1~, &  \  & w_2 = A_2\,B_1\,C_2~, & \ &  w_3 = A_1\,B_1\,C_2~, &  \ & w_4 = A_2\,B_2\,C_1~, \\
w_5 = A_1\,B_1\,C_1~, &  \ & w_6 = A_2\,B_1\,C_1~, & \  & w_7 = A_1\,B_2\,C_2~, & \ & w_8 = A_2\,B_2\,C_2  \, .
\end{array}
\label{ws}
\end{equation}
One can then check explicitly that these satisfy the 9 relations defining $\mathcal{C}(Q^{111})$ as an affine variety:
\begin{equation}
\begin{array}{cccccccc}
w_1\,w_2 - w_3\,w_4 & = & w_1\,w_2 - w_5\,w_8 & = & w_1\,w_2 - w_6\,w_7 & = & 0~, & \\
w_1\,w_3 - w_5\,w_7 & = & w_1\,w_6 - w_4\,w_5 & = & w_1\,w_8 - w_4\,w_7 & = & 0~, & \\
w_2\,w_4 - w_6\,w_8 & = & w_2\,w_5 - w_3\,w_6 & = & w_2\,w_7 - w_3\,w_8 & = & 0~.
\label{eq_Q11}
\end{array}
\end{equation}
This is an affine toric variety, with toric diagram given by Figure \ref{fig:toricdiagramQ111}. Indeed, we also notice 
that for the abelian theory the description of the moduli space as a $U(1)^2$ K\"ahler quotient of $\C^6$ with 
coordinates $\{A_i,B_i,C_i\}$ is precisely the minimal gauged linear sigma model (GLSM) description. 
Thus the 6 toric divisors in Figure \ref{fig:toricdiagramQ111}, discussed in \secref{sec:baryons}, 
are defined by $\{A_i=0\}$, $\{B_i=0\}$, $\{C_i=0\}$, $i=1,2$.

For CS level $k>1$ one obtains an $\mathcal{N}=2$ supersymmetric $\Z_k\subset U(1)_{\mathcal{B}_{G-1}}$ orbifold of $\mathcal{C}(Q^{111})$. 
Notice that $\{w_i\mid i=1,\ldots,4\}$ are invariant under this action, while $\{w_5,w_6\}$ and $\{w_7,w_8\}$ are rotated with equal and opposite 
phase. On the other hand, for the non-abelian theory with $N>1$ it was shown in \cite{Franco:2009sp} that for large $k$, where 
the use of still poorly-understood monopole operators is evaded, upon using the F-terms of the full non-abelian superpotential 
(\ref{WQ111}) the chiral ring matches that expected for the corresponding orbifold. 
In this case the chiral primaries at the non-abelian level are just the usual gauge-invariants given by
\begin{equation}
\label{nonabelianchirals}
{\rm Tr}\, \Big( \prod_{a=1}^r\, X^{\pm}_{i_a}\Big)\ ,\quad \mbox{where}\quad X^+_i=A_i\, C_2\, B_1\ , \quad X^-_i=A_i\, B_2\, C_1~.
\end{equation}
An important subtlety in this theory is that 
 $U(1)_{\mathcal{B}_{G-1}}$ does not act freely on $Q^{111}$: it fixes two disjoint copies of $S^3$ inside $Q^{111}$, as explained in \cite{Benini:2009qs}. Indeed, using (\ref{gaugedU(1)charges}) one sees that the corresponding two cones $\C^2=\mathcal{C}(S^3)$ are parametrized 
respectively by $\{w_1,w_4\}$ and $\{w_2,w_3\}$, with in each case all other $w_i=0$. Thus for $k>1$ the 
horizon $Y=Q^{111}/\Z_k$ has orbifold singularities in codimension four. This means that the SUGRA approximation cannot be trusted for $k>1$. In fact these are $A_{k-1}$ singularities which can support ``fractional'' M2 branes wrapping the collapsed cycles, and one expects an $SU(k)$ gauge theory to be supported on these $S^3$s. A different perspective can be obtained by interpreting $U(1)_{\mathcal{B}_{G-1}}$ as the M-theory circle and reducing to Type IIA. This results in $k$ D6 branes wrapping these two $S^3$ submanifolds. From now on we will therefore assume that $k=1$. 

\subsubsection{Gauged versus global abelian subgroups and $SL(2,\mathbb{Z})$}

At $k=1$ the orbifold identification due to the CS terms is trivial. Indeed, in this case there is no real distinction between $U(N)$ and $SU(N)\times U(1)$ gauge groups, as discussed in \cite{Aharony:2008ug, Imamura:2008nn,  Lambert:2010ji} for the ABJM theory and orbifolds of it. We shall argue that ungauging some of the $U(1)$s is dual to a particular choice of boundary conditions on the gravity side. That is, we apply the general discussion in \secref{sec:baryons} to the $b_2(Q^{111})=2$ $U(1)$ gauge fields, and argue that the associated $U(1)$ symmetries are those in $SU(N)\times U(1)$, for appropriate gauge group factors. This raises the important problem of how to identify the relevant two $U(1)$ symmetries dual to the Betti multiplets in the QCS theory proposed above. The key is to recall that the boundary conditions which amount to ungauging these $U(1)$s in turn allow for the existence of supersymmetric M5 branes on the gravity side. As discussed in \secref{sec:baryons}, from an algebro-geometric point of view the corresponding divisors are easy to identify. In turn we notice that, for the abelian theory, the fields $\{A_i, B_i, C_i\}$ are also the minimal GLSM coordinates. Setting each to zero therefore gives one of the 6 toric divisors that may be wrapped by an M5 brane. The charges of the resulting M5 brane states under $U(1)^{b_2(Y)}$ are then the \emph{same} as the charges of these fields under the $U(1)_I\times U(1)_{II}$ we quotient by in forming the abelian moduli space -- this was shown for the D3 brane case in \cite{Franco:2005sm}, and the same argument applies here also. This strongly suggests that the gauge symmetries $U(1)_I$, $U(1)_{II}$ should in fact be dual to the Betti multiplets discussed in \secref{sec:baryons}. 

Once we have identified the relevant abelian symmetries, we can consider acting with the $\mathcal{S}$ and $\mathcal{T}$ operations. We schematically write the action of the $U(N)^4$ $Q^{111}$ theory (which we will denote as $S_U$), separating the $U(1)$ sector from the rest, as
\begin{equation}
\label{S_gauged}
S_U\sim \int  \mathcal{B}_3\wedge \dd \mathcal{B}_4+ \mathcal{A}_I\wedge \dd \mathcal{A}_{II}+\int \mathcal{L}_R \ ,
\end{equation}
where $\int \mathcal{L}_R$ stands for the remaining terms. We can then consider a theory without the gauge fields $\mathcal{A}_I$, $\mathcal{A}_{II}$, constructed schematically as $S_{SU}=\int \mathcal{B}_3\wedge \dd \mathcal{B}_4+\int \mathcal{L}_R$. By construction, this theory has exactly 2 global symmetries satisfying all the properties  expected as dual to Betti multiplets. Following \cite{Witten:2003ya}, we can introduce a background gauge field for one of them, which we can call $\mathcal{A}_I$. Then, as reviewed in \secref{sec:baryons}, the $\mathcal{S}$-operation amounts to regarding this field as dynamical, while at the same time introducing a coupling to another background field $\mathcal{C}_I$ as
\begin{equation}
S_{SU}\rightarrow S_{SU}[\mathcal{A}_I]+\int \mathcal{C}_I\wedge \dd\mathcal{A}_I \ .
\end{equation}
We can introduce yet another background gauge field $\mathcal{A}_{II}$ for the second global symmetry and perform yet another  $\mathcal{S}$-operation. However, this time we will choose to regard $\mathcal{C}_I-\mathcal{A}_{II}$ as the background gauge field on which to act with the $\mathcal{S}$-generator. This results in
\begin{equation}
 S_{SU}[\mathcal{A}_I]+\int \mathcal{C}_I\wedge \dd\mathcal{A}_I \rightarrow S_{SU}[\mathcal{A}_I,\, \mathcal{A}_{II}]+\int \mathcal{C}_I\wedge \dd\mathcal{A}_I + \int \mathcal{C}_{II}\wedge \dd(\mathcal{C}_I-\mathcal{A}_{II}) \ .
\end{equation}
Integrating by parts yields
\begin{equation}
S_{SU}[\mathcal{A}_I,\, \mathcal{A}_{II}]+\int \mathcal{C}_I\wedge \dd(\mathcal{C}_{II}+\mathcal{A}_I) - \int \mathcal{C}_{II}\wedge \dd\mathcal{A}_{II} \ .
\end{equation}
Since $\mathcal{C}_I$ only appears linearly, its functional integral gives rise to a delta functional setting $\mathcal{C}_{II}=-\mathcal{A}_I$, which leads to an action of the precise form (\ref{S_gauged}). We have therefore been able to establish a connection between a theory where the gauge group is $U(1)^2\times SU(N)^4$, and whose action is $S_{SU}$, with the original $U(N)^4$ theory, whose action is given by $S_U$, via repeated action with the $\mathcal{S}$-operation. 

More generally, the whole of $SL(2,\Z)$ will act on the boundary conditions for the bulk gauge fields, leading in general to an infinite orbit of CFTs for each $U(1)$ gauge symmetry in $AdS_4$. This is a rich structure that deserves considerable further investigation. In this paper, however, we will content ourselves to study the particular choice of boundary conditions described by the $S_{SU}$ theory. Since the dual to the $\mathcal{S}$ operation is the exchange of the $E_{\mu}\leftrightarrow B_{\mu}$ boundary conditions, we expect the gravity dual to the $S_{SU}$ theory to still be $AdS_4\times Q^{111}$, but with an appropriate choice of boundary conditions. In turn, these boundary conditions allow for the existence of the electrically charged M5 branes which we used to identify the symmetries. These M5 branes would not be allowed in the quantization $E_{\mu}=0$, which in turn would be dual to a CFT where the corresponding $U(1)$ factors would remain gauged. In agreement, the dual operators which we will propose below would not be gauge-invariant in that case.

Let us now consider the effect of the $U(1)^2\times SU(N)$ gauge group on the construction of the moduli space. The diagonalization of the $\sigma_a$ auxiliary fields in the equations defining the moduli space (\ref{DF}) relies on the non-abelian part of the gauge symmetry, and therefore it applies even if we consider ungauging some of the diagonal $U(1)$ factors. More crucially, in order to obtain the correct four-fold moduli space we needed the $S(\mathcal{B}_4)$ piece (\ref{SBG}) of the CS action so that, upon dualizing the $\mathcal{B}_4$ field, the dual scalar $\tau$ is gauge-fixed via gauge transformations of $\mathcal{B}_{3}$. Thus provided we leave 
$\mathcal{B}_{4}$ and $\mathcal{B}_{3}$ gauged, with the same CS action, all of this discussion is unaffected if we ungauge the remaining $U(1)_I$, $U(1)_{II}$.
Correspondingly, we will still have the 8 gauge-invariants (\ref{ws}), which will give rise to the same 9 equations defining 
$\mathcal{C}(Q^{111})$ as a non-complete intersection as ``mesonic'' moduli space. 
The remarks on the non-abelian chiral ring elements spanned by (\ref{nonabelianchirals}) are also unchanged. 
However, with only a  $U(1)_{\mathcal{B}_3}\times U(1)_{\mathcal{B}_4}\times SU(N)^4$ gauge symmetry we also have additional 
chiral primary operators, charged under the now global $U(1)_I$, $U(1)_{II}$. Indeed, we have the following ``baryonic'' type operators:
\bea
\nn
\mathscr{B}_{A_{I_1...I_N}}&=&\frac{1}{N!}\, \epsilon^{i_1\cdots i_N}\, \epsilon_{j_1\cdots j_N}\, (A_{I_1})^{j_1}_{i_1}\cdots  (A_{I_N})^{j_N}_{i_N}~,\\ \nn
\mathscr{B}_{B_i}&=&\frac{1}{N!}\, \epsilon^{i_1\cdots i_N}\, \epsilon_{j_1\cdots j_N}\, (B_i)^{j_1}_{i_1}\cdots  (B_i)^{j_N}_{i_N}\,\me^{\ii\,(-1)^{i-1}\,N\,\tau}~,\\
\mathscr{B}_{C_i}&=&\frac{1}{N!}\, \epsilon^{i_1\cdots i_N}\, \epsilon_{j_1\cdots j_N}\, (C_i)^{j_1}_{i_1}\cdots  (C_i)^{j_N}_{i_N}\,\me^{\ii\,(-1)^{i-1}\,N\,\tau}~.
\label{b-operators}
\eea
In particular, for the 6 fields in the quiver there is a canonical set of 6 baryonic operators given by determinants of these fields, 
dressed by appropriate powers of the disorder operators $\me^{\ii \tau}$ to obtain gauge-invariants under $\mathcal{B}_3$. 
These operators are in 1-1 correspondence with the toric divisors in the geometry. This is precisely the desired mapping between 
baryonic operators in the field theory and M5 branes wrapping such toric submanifolds, with one M5 brane 
state for each divisor. Indeed, the charges of these operators under the two baryonic $U(1)$s are
\begin{displaymath}
\begin{array}{c | c c c} 
 & \mathscr{B}_{A_{I_1..I_N}} & \mathscr{B}_{B_i} & \mathscr{B}_{C_i} \\ \hline
 U(1)_{I} & N & 0 & -N \\ 
 U(1)_{II} & -N & N & 0 \\
 \end{array}~.
 \end{displaymath}
These are precisely the charges of M5 branes, wrapped on the five-manifolds corresponding to the divisors $\{A_i=0\}$, $\{B_i=0\}$, $\{C_i=0\}$, 
under the two $U(1)^{b_2(Y)}$ symmetries in $AdS_4$. Indeed, recall that the two two-cycles in $Q^{111}$ may be taken to be 
the anti-diagonal $S^2$s in two factors of $\mathbb{CP}^1\times \mathbb{CP}^1\times \mathbb{CP}^1$, at $\psi=0$. Let us choose these to be the anti-diagonal in the
first and third factor, and second and first factor, respectively. 
The charge of an M5 brane wrapped 
on a five-cycle $\Sigma_5\subset Y$ under each $U(1)$ is then the intersection number of $\Sigma_5$ with each corresponding two-cycle. 
Thus with this basis choice, the charges of the operator associated to an M5 brane wrapped on the base of one of the 
 6 toric divisors $\{A_i=0\}$, $\{B_i=0\}$, $\{C_i=0\}$ are precisely those listed in the above table. 

Being chiral primary, the conformal dimensions of these operators are given by 
 $N\, \Delta[X]=N\, R[X]$, $R[X]$ being the R-charge of the field $X$. The conformal dimension 
of an M5 brane wrapping a supersymmetric five-cycle $\Sigma_5\subset  Y$ is given by the general formula 
\cite{Gubser:1998fp}
\begin{equation}
 \Delta[\Sigma_5] = \frac{N\pi \vol(\Sigma_5)}{6\vol(Y)}~.
\end{equation}
These volumes are easily computed for the $Q^{111}$ metric (\ref{Qiiimetric}): 
$\vol(Q^{111})=\frac{\pi^4}{8}$, $\vol(\Sigma_5)=\frac{\pi^3}{4}$, where 
$\Sigma_5$ is any of the 6 toric five-cycles. From this one obtains
$\Delta[\Sigma_5]=\frac{N}{3}$ in each case, giving conformal dimensions 
$\Delta[X]=\frac{1}{3}$ for each field. 
With this R-charge assignment we see that the superpotential 
(\ref{WQ111}) has R-charge 2, precisely as it must at a superconformal fixed point. Indeed, the 
converse argument was applied in \cite{Hanany:2008fj} to obtain this R-charge assignment. 
We thus regard 
this as further evidence in support of our claim in this section, as well as further support for 
these theories as candidate SCFT duals to $AdS_4\times Q^{111}$.

\subsection{QCS theories dual to isolated toric Calabi-Yau four-fold singularities}\label{sec:gen}

We would like to apply the preceding discussion to more general $\mathcal{N}=2$ CS-matter theories dual to M2 branes probing Calabi-Yau four-fold cones. With the exception of $\C^4$, the apex of the cone always corresponds to a \emph{singular point} $p$ in the toric variety. The simplest class of such singularities occurs when this singular point $p$ is isolated; that is, when no other singular loci intersect it. In this case the base of the cone $Y$ is a smooth Sasakian seven-manifold. The condition for the singular point $p$ to be isolated was given in \cite{Lerman2} and interpreted in terms of the toric diagram in \cite{Benishti:2009ky}. 

An additional ingredient is the possible presence of vanishing six-cycles at the tip of the cone. In terms of the toric data, these six-cycles are signalled by internal lattice points in the toric diagram. These codimension 2 cycles, in very much the same spirit as their four-cycle Type IIB counterparts (see \textit{e.g.} \cite{Benvenuti:2005qb}), represent a further degree of complexity. We postpone the analysis of geometries with exceptional six-cycles to \secref{sec:5}.

We study such isolated Calabi-Yau singularities without vanishing six-cycles in more detail in 
 \secref{sec:classification}, in particular classifying the singularities with 4 or 5 external vertices in the toric diagram. In the cases where a Lagrangian description of the M2 brane theory exists, it turns out that for all these cases one can construct an appropriate toric superpotential, so that there is a toric gauge theory which realizes at the abelian level the minimal GLSM.  This toric gauge theory has $b_2(Y)+2$ gauge group factors, and can be promoted to have $U(N)$ gauge groups.  Such quiver Chern-Simons theories have been considered in the past in \cite{Aharony:2008ug, Franco:2008um, Benishti:2009ky}. 

We would like to generalize our proposal to this simplest class of isolated singularities with no vanishing six-cycles. Indeed, we expect that a similar sequence of $\mathcal{T}$ and $\mathcal{S}$ operations amounts to ungauging of precisely $b_2(Y)$ $U(1)$ factors. In very much the same spirit as in the $\mathcal{C}(Q^{111})$ example, this should correspond to a particular choice of boundary conditions in the dual $AdS_4$. Furthermore, we conjecture the gauge group to be $U(1)^2\times SU(N)^{b_2(Y)+2}$, the two $U(1)$ factors being those corresponding to the $\mathcal{B}_{G}$ and $\mathcal{B}_{G-1}$ gauge fields. This way we are naturally left with $b_2(Y)$ global $U(1)$ symmetries which exactly correspond to the $b_2(Y)$ expected $U(1)$ baryonic symmetries. Furthermore, the M5 branes would be naturally identified with the corresponding baryonic operators, constructed in a similar manner as in the $\mathcal{C}(Q^{111})$ example.
 
As a final remark, for the $\mathcal{Q}^{111}$ case one can see that the total moduli space can be described as two extra complex directions that are fibred over the mesonic moduli space. The resulting variety is often called the \textit{master space}. More generally one can see that the moduli spaces of the class of theories that we described above can be described as $b_2(Y)$ baryonic directions fibred over the mesonic moduli space. This is implied by the fact that the number of gauge nodes in these theories is $b_2(Y)+2$.

\section{Gravity duals of baryonic symmetry breaking}\label{sec:4}

In \secref{sec:baryons} we explained that for each $U(1)$ gauge field in $AdS_4$, arising from reduction of 
$C_6$ along five-cycles in $Y$, there are two different $AdS$ quantizations: one of these gives rise to a
conserved $U(1)$ current in the dual CFT, while the other instead gives rise to a \emph{dynamical} $U(1)$ gauge field. As discussed in \secref{sec:baryons}, in this paper we content ourselves with studying the 
quantization which is more closely analogous to the case in Type IIB string theory, in which the 
$b_2(Y)$ baryonic $U(1)$ gauge fields (\ref{3-form-to-global}) in $AdS_4$ are dual to conserved currents in the dual SCFT. As we have argued, 
in this theory M5 branes wrapped on supersymmetric cycles in $Y$ should appear as chiral primary baryonic-type operators 
in the dual SCFT. Indeed, at least for toric theories with appropriate smooth supergravity horizons $Y$ we expect the dual 
SCFT to be described by a QCS theory with gauge group $U(1)^2\times SU(N)^{b_2(Y)+2}$. The M5 brane states  
are then the usual gauge-invariant determinant-like operators in these theories, as we discussed in detail for the 
$Q^{111}$ theory in the previous section. 

We may then study the gravity duals to vacua in which the 
$b_2(Y)$ global $U(1)$s are (spontaneously) broken. On general grounds, these should correspond to 
supergravity solutions constructed from resolutions of the corresponding cone over $Y$. The baryonic operators are charged under the global 
baryonic symmetries, and vacua in which these operators obtain a VEV lead to 
spontaneous symmetry breaking. By giving this VEV we pick a point in the moduli space of the theory, which at the same time introduces a scale and thus an RG flow, whose endpoint will be a different SCFT. The supergravity dual of this RG flow was first discussed in the Type IIB context by Klebanov-Witten \cite{Klebanov:1999tb}, and to some extent in the M-theory context in \cite{Benishti:2009ky}. 
In this section we begin by discussing in detail these gravity solutions for the case of $Q^{111}$. The physics in fact very closely 
resembles the physics in the Type IIB context. We then describe how to generalize this discussion for general Calabi-Yau 
four-fold singularities. In particular, we will obtain a general formula for M5 brane condensates, or indeed more generally still 
a formula for the on-shell action of a wrapped brane in a warped Calabi-Yau background. Essentially this formula 
appeared in \cite{Baumann:2006th}, where it was checked in some explicit examples. Here we provide a general proof of this formula.

We emphasize that the interpretation of the gravity backgrounds considered in this section is only for the special choice of dual CFT in which 
the $b_2(Y)$ baryonic $U(1)$ gauge fields in $AdS_4$ are dual to global symmetries in the dual SCFT. More generally, different choices of 
boundary conditions will imply that some, or all, of the M5 brane states considered here are absent. It is then clearly very interesting 
to ask what is the dual field theory interpretation of these gravity backgrounds in such situations. Again, we leave this for future work.

\subsection{Resolutions of $\mathcal{C}(Q^{111})$}\label{sec:resQ111}

In this section we consider the warped resolved gravity backgrounds for $Q^{111}$. We begin by discussing this 
in the context of the GLSM, and then proceed to construct corresponding explicit supergravity solutions. 

\subsubsection{Algebraic analysis}\label{sec:GKZQ111}

The toric singularity $\mathcal{C}(Q^{111})$ may be described by a GLSM with 6 fields, $a_i$, $b_i$, $c_i$, $i=1,2$, 
and gauge group $U(1)^2$. This is also the same as the abelian QCS theory presented in \secref{sec:3}, but without the CS terms. 
The charge matrix is
\bea
\begin{array}{c|cccccc} & a_1 & a_2 & b_1 & b_2 & c_1 & c_2 \\ \hline U(1)_1 & -1 & -1 & 1 & 1 & 0 & 0 \\ U(1)_2 & -1 & -1 & 0 & 0 & 1 & 1 \end{array}~.
\eea
The singular cone $\mathcal{C}(Q^{111})$ is the moduli space of this GLSM where the FI parameters 
$\zeta_1=\zeta_2=0$ are both zero. However, more generally we may allow 
$\zeta_1, \zeta_2\in \R$, leading to different (partial) resolutions of the singularity. In fact 
since there are no internal points in the toric diagram in Figure
\ref{fig:toricdiagramQ111}, this 
GLSM in fact describes \emph{all} possible (partial) resolutions of the singular cone.

It is straightforward to analyse the various cases. 
Suppose first that $\zeta_1,\zeta_2>0$ are both positive. We may write the two D-terms of the GLSM as
\bea\label{Dpos}
|b_1|^2 + |b_2|^2 &=& \zeta_1 + |a_1|^2 + |a_2|^2>0~,\nonumber\\
|c_1|^2 + |c_2|^2 &=& \zeta_2 + |a_1|^2 + |a_2|^2>0~.
\eea
In particular, for $a_1=a_2=0$ we obtain $\mathbb{CP}^1\times\mathbb{CP}^1$ where the K\"ahler class of each factor is 
proportional to $\zeta_1$ and $\zeta_2$, respectively. Here $b_i$ and $c_i$ may be thought of as homogeneous coordinates on the $\mathbb{CP}^1$s. 
Altogether, this describes the total space of the bundle $\mathcal{O}(-1,-1)\oplus \mathcal{O}(-1,-1)\rightarrow\mathbb{CP}^1\times\mathbb{CP}^1$, with $a_1, a_2$ the two fibre coordinates on the $\C^2$ fibres. 

Suppose instead that $\zeta_1<0$. We may then rewrite the D-terms as
\bea
|a_1|^2 + |a_2|^2 &=& -\zeta_1 + |b_1|^2 + |b_2|^2>0~,\nonumber\\
|c_1|^2 + |c_2|^2 &=& \zeta_2-\zeta_1 + |b_1|^2 + |b_2|^2~.
\eea
Provided also $\zeta_2-\zeta_1>0$, we hence obtain precisely the same geometry as when $\zeta_1,\zeta_2>0$, but with the $\mathbb{CP}^1\times\mathbb{CP}^1$ zero section now parametrized by $a_i$ and $c_i$ and with K\"ahler classes proportional to $-\zeta_1$ and $\zeta_2-\zeta_1$, respectively. There is a similar situation with $\zeta_2<0$ and $\zeta_1-\zeta_2>0$.

\begin{figure}[ht]
\begin{center}
\includegraphics[scale=0.8]{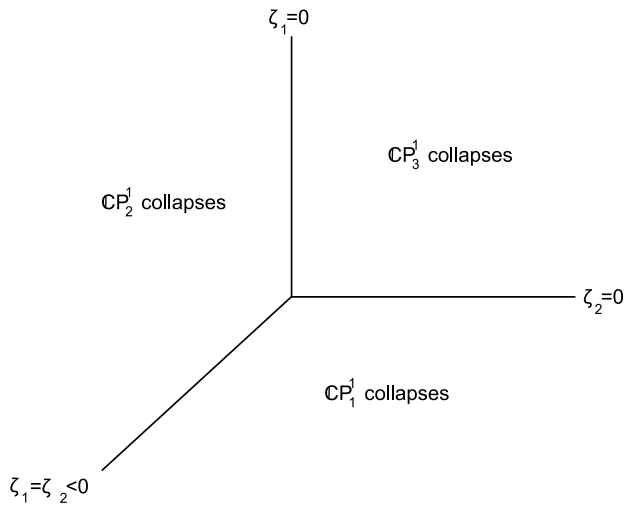}
\end{center}
\caption{The GKZ fan for $Q^{111}$ is $\R^2$, divided into three cones.}
\label{fig:GKZ1}
\end{figure}
Hence in total there are three different resolutions of $\mathcal{C}(Q^{111})$, corresponding to choosing which of the three $\mathbb{CP}^1$s in $Q^{111}$ collapses at the zero section in $\mathcal{O}(-1,-1)\oplus \mathcal{O}(-1,-1)\rightarrow\mathbb{CP}^1\times\mathbb{CP}^1$. We label these three $\mathbb{CP}^1$s as $\mathbb{CP}^1_a$, $a=1,2,3$, which in $Q^{111}$ are parametrized by $c_i$, $b_i$, $a_i$, respectively. This is shown in Figure \ref{fig:GKZ1}, which is known more generally as the \emph{GKZ fan}. Notice there is a $\Sigma_3$ permutation symmetry of the three $\mathbb{CP}^1$s in $Q^{111}$ and the three different resolutions are permuted by this symmetry.

The boundary edges between the regions correspond to collapsing another of the $\mathbb{CP}^1$s, leading only to a partial resolution 
of the singularity. Thus, for example, take $\zeta_1>0$ but $\zeta_2=0$. The D-terms are now
\bea
|b_1|^2 + |b_2|^2 &=& \zeta_1 + |a_1|^2 + |a_2|^2>0~,\nonumber\\
|c_1|^2 + |c_2|^2 &=& |a_1|^2 + |a_2|^2~.
\eea
The second line describes the conifold singularity, which is then fibred over a $\mathbb{CP}^1$, parametrized by the $b_i$, of K\"ahler class $\zeta_1$. 

\subsubsection{Supergravity analysis}\label{sec:sugraQ111}

For each of the resolutions of $\mathcal{C}(Q^{111})$ described above there is a corresponding Ricci-flat K\"ahler metric
that is asymptotic to the cone metric over $Q^{111}$. More precisely, there is a unique such metric for each choice of 
K\"ahler class, or equivalently FI parameter $\zeta_1,\zeta_2\in \R$. As we shall discuss later in this section, this is guaranteed
by a general theorem that has only just been proven in the mathematics literature. However, for $Q^{111}$ these 
metrics may in fact be written down explicitly. Denoting the (partially) resolved Calabi-Yau generically by $X$, the 
Calabi-Yau metrics are given by
\begin{eqnarray}\label{resolvedQ111}
\diff s^2(X) &=&\kappa(r)^{-1}\dd r^2+\kappa(r)\frac{r^2}{16}\Big(\dd\psi+\sum_{i=1}^3 \cos\theta_i \dd\phi_i\Big)^2+\frac{(2a+r^2)}{8}\Big(\dd\theta_2^2+\sin^2\theta_2 \dd\phi_2^2\Big)\nonumber \\ &&+\frac{(2b+r^2)}{8}\Big(\dd\theta_3^2+\sin^2\theta_3\dd\phi_3^2\Big)+\frac{r^2}{8}\Big(\dd\theta_1^2+\sin^2\theta_1\dd\phi_1^2\Big)~,
\end{eqnarray}
where 
\begin{equation}\label{kappa}
\kappa(r)=\frac{(2A_-+r^2)(2A_++r^2)}{(2a+r^2)(2b+r^2)}~,
\end{equation}
$a$ and $b$ are arbitrary constants, and we have also defined 
\begin{equation}
A_\pm=\frac{1}{3}\Big(2a+2b\pm\sqrt{4a^2-10ab+4b^2}\Big)~.
\end{equation}
One easily sees that at large $r$ the metric approaches the cone over the $Q^{111}$ metric (\ref{Qiiimetric}). This 
way of writing the resolved metric breaks the $\Sigma_3$ symmetry, since it singles out the $\mathbb{CP}^1$ parametrized by 
$(\theta_1,\phi_1)$ as that collapsing at $r=0$. Here we have an exceptional $\mathbb{CP}^1\times\mathbb{CP}^1$, parametrized 
by $(\theta_2,\phi_2)$, $(\theta_3,\phi_3)$, with K\"ahler 
classes proportional to $a>0$ and $b>0$, respectively. Thus setting $\{a=0, b>0\}$, or $\{b=0,a>0\}$, leads 
to a partial resolution with a residual $\mathbb{CP}^1$ family of conifold singularities at $r=0$.  We shall examine this 
in more detail below.
For further details of this solution we refer the reader to \secref{sec:Q111-geometry}.

We are interested in studying supergravity backgrounds corresponding to M2 branes localized on one of these resolutions 
of $\mathcal{C}(Q^{111})$. We thus consider the following ansatz for the background sourced by such M2 branes
\begin{eqnarray}\label{metricagain}
\dd s^2_{11}&=&h^{-2/3}\, \dd s^2\left(\R^{1,2}\right) +h^{1/3}\dd s^2(X)~, \nonumber\\ G&=&\dd^3 x \wedge \dd h^{-1}~,
\end{eqnarray}
where $\dd s^2(X)$ is the Calabi-Yau metric (\ref{resolvedQ111}).
If we place $N$ spacetime-filling M2 branes at a point $y\in X$, we must then also solve the equation
\bea
\Delta_x h[y] = \frac{(2\pi \ell_p)^6N}{\sqrt{\det g_X}} \delta^8(x-y) \ ,
\eea
for the warp factor $h=h[y]$. Here $\Delta$ is the scalar Laplacian on $X$. 
Having the explicit form of the metric we can compute this Laplacian and solve for the warp factor to obtain the full supergravity solution. 
This is studied in detail in \secref{sec:D}. 

In the remainder of this subsection let us analyse the simplified case 
in which we partially resolve the cone, 
setting $a=0$ and $b>0$. This corresponds to one of the boundary lines in the GKZ fan in Figure \ref{fig:GKZ1}, with the point on the boundary 
$\R_{> 0}$ labelled by the metric parameter $b>0$.
Here one can solve explicitly for the warp factor in the case where we put the $N$ M2 branes 
at the north pole of the exceptional $\mathbb{CP}^1$ parametrized by $(\theta_3,\phi_3)$; this is the point with coordinates $y=\{r=0, \theta_3=0\}$. 
Notice the choice of north pole is here without loss of generality, due to the $SU(2)$ isometry acting 
on the third copy of $\mathbb{CP}^1$. We denote the corresponding warp factor in this case as simply $h[y=\{r=0,\theta_3=0\}]\equiv h$. 
As shown in \secref{sec:Q111-warp}, $h=h(r,\theta_3)$ is then given explicitly in terms of hypergeometric functions by
\begin{eqnarray}
h(r,\theta_3) &=&\sum_{l=0}^\infty\,H_{l}(r)\, P_{l}(\cos\theta_3)~,\nonumber \\
H_{l}(r)&=& \mathcal{C}_{l}\, \Big(\frac{8b}{3r^2}\Big)^{3(1+\beta)/2}\, _2F_1\left(-\tfrac{1}{2}+\tfrac{3}{2}\beta,\tfrac{3}{2}+\tfrac{3}{2}\beta,1+3\beta,-\tfrac{8b}{3r^2}\right)~,
\label{Q111-warp-factor}
\end{eqnarray}
where $P_{l}$ denotes the $l$th Legendre polynomial, 
\begin{equation}
 \beta=\beta(l)=\sqrt{1+\frac{8}{9}l(l+1)}~,
\end{equation}
and the normalization factor $\mathcal{C}_{l}$ is given by
\bea
\mathcal{C}_{l}&=&\frac{3\Gamma(\frac{3}{2}+\frac{3}{2}\beta)^2}{2\Gamma(1+3\beta)}\left(\frac{3}{8b}\right)^3\, (2l+1)\, R^6~,\\\label{anothereqn}
 R^6&=&\frac{(2\pi\ell_p)^6 N}{6\text{vol}(Q^{111})}=\frac{256}{3}\pi^2 N \ell_p^6~.
\eea
In the field theory this solution corresponds to breaking one combination of the two global $U(1)$ baryonic symmetries, rather than both of them. 
This will become clear in the next subsection.
The resolution of the cone can be interpreted in terms of giving an expectation value to a certain operator $\mathcal{U}$ in the field theory. This operator is contained in the same multiplet as the current that generates the broken baryonic symmetry, and couples to the corresponding $U(1)$ gauge field in $AdS_4$. Since a conserved current has no anomalous dimension, the dimension of $\mathcal{U}$ is uncorrected in going from the classical description to supergravity \cite{Klebanov:1999tb}. According to the general $AdS/CFT$ prescription \cite{Klebanov:1999tb}, the VEV of the operator $\mathcal{U}$ is dual to the subleading correction to the warp factor. For large $r$ we can write
\begin{equation}
h(r,\theta_3)\sim\sum_{l=0}^\infty\, \mathcal{C}_{l}\, \Big(\frac{8b}{3r^2}\Big)^{3(1+\beta)/2}\, P_{l}(\cos\theta_3)~.
\end{equation}
Expanding the sum we then have
\begin{equation}
h(r,\theta_3)\sim \frac{R^6}{r^6}\left(1+\frac{18b\, \cos\theta_3}{5r^2}+\cdots\right)~.
\end{equation}
In terms of the $AdS_4$ coordinate $z= r^{-2}$ we have that the leading correction is of order $z$, which indicates that the dual operator $\mathcal{U}$ is dimension 1. This is precisely the expected result, since this operator sits in the same supermultiplet as the broken baryonic current, and thus has a protected dimension of 1. Furthermore, its VEV is proportional to $b$, the metric resolution parameter, which reflects the fact that in the conical ($AdS$) limit in which $b=0$ this baryonic current is not broken, and as such $\langle \mathcal{U}\rangle=0$.

The moduli space of the field theory in the new IR is equivalent to the geometry close to the branes. 
Recall that we placed the $N$  M2 branes at the north pole $\{\theta_3=0\}$ of the exceptional
$(\theta_3,\phi_3)$ sphere at $r=0$. Defining $\tilde{\psi}=\psi+\phi_3$ and introducing the new radial variables $\tilde{r}=\frac{b}{2}\,\theta_3$, 
$\rho=\frac{\sqrt{3}}{2}\, r$,  the geometry close to the branes becomes to leading order
\begin{eqnarray}
\dd s^2&=&\dd\rho^2+\frac{\rho^2}{9}\Big(\dd\tilde{\psi}+\sum_{i=1}^2 \cos\theta_i\dd\phi_i\Big)^2+\frac{\rho^2}{6}\Big(\dd\theta_2^2+\sin^2\theta_2\dd\phi_2^2\Big)\nonumber \\ &&+\frac{\rho^2}{6}\Big(\dd\theta_1^2+\sin^2\theta_1\dd\phi_1^2\Big)+\Big(\dd\tilde{r}^2+\tilde{r}^2\dd\phi_3^2\Big)~,
\label{c_x_conifold}
\end{eqnarray}
which is precisely the Ricci-flat K\"ahler metric of $\mathcal{C}(T^{11})\times \mathbb{C}$, in accordance with the discussion in the previous subsection.

\subsection{Higgsing the $Q^{111}$ field theory}\label{sec:HiggsQ111}

We have argued that the warped resolved supergravity solutions described in the previous section are dual to 
spontaneous symmetry breaking in the SCFT in which the M5 brane states appear as baryonic-type operators. 
Let us study this in more detail in the field theory described in \secref{sec:3}. 
In this SCFT the symmetries $U(1)_I$ and $U(1)_{II}$ in (\ref{gaugedU(1)charges}) are global, rather than gauge, symmetries, 
with the corresponding conserved currents coupling to the baryonic $U(1)$ gauge fields in $AdS_4$. By inspection 
of this charge matrix we conclude that it is possible to give a VEV to the $A_i$, $B_i$ and $C_i$ fields. 
These VEVs then break the corresponding baryonic $U(1)$ symmetries. In particular, by giving a VEV to 
any pair of fields ($A$s, $B$s or $C$s) we break only one particular $U(1)$ baryonic symmetry, leaving another combination
unbroken. In this section we will examine the resulting Higgsings of the gauge theory obtained by giving VEVs to 
different pairs of fields, and compare with the gravity results of the previous section. 

As explained in \secref{sec:2}, at each of the two poles for each copy of $\mathbb{CP}^1$ in the K\"ahler-Einstein base of $Q^{111}$, 
there is a supersymmetric five-cycle that may be wrapped by an M5 brane. Altogether these are 6 M5 brane states, corresponding to the 
toric divisors of $\mathcal{C}(Q^{111})$. Each pair are acted on by one of the $SU(2)$ factors in the isometry group 
$SU(2)^3\times U(1)_R$, rotating one into the other. Quantizing the BPS particles in $AdS_4$ one obtains dual baryonic-type operators
given by (\ref{b-operators}). In particular, consider the M5 branes that sit at a point on the copy of $\mathbb{CP}^1=S^2$ with coordinates 
$(\theta_3,\phi_3)$. In the next section we will compute the VEV of these M5 brane operators in the partially resolved gravity background 
described by (\ref{Q111-warp-factor}), 
showing that the baryonic operator dual to the M5 brane at $\theta_3=0$ vanishes, while that at the opposite pole $\theta_3=\pi$ 
is non-zero and proportional to the resolution parameter $b$ (see equation (\ref{vev})). Considering the $A$ fields in the field theory 
this corresponds to the fact that, after breaking the baryonic symmetry by giving diagonal VEVs to these fields, it is possilbe to use the $SU(2)$ flavour symmetry to find one combination of $A$ fields with zero VEV, and an orthogonal combination with non-vanishing VEV. Let us assume for example that $\langle A_1\rangle=0$ and $\langle A_2\rangle=b\, I_{N\times N}$. Thus only one baryonic operator in (\ref{b-operators}) has non-vanishing VEV, 
namely $\langle\mathscr{B}_{A_2}\rangle = b^{N\, \Delta_{A_2}}$.\footnote{As anticipated in \secref{sec:3}, at the IR superconformal fixed point the dimensions of the chiral fields are expected to be different from the free field fixed point. That is why generically the VEV of the baryonic operator is $\langle\mathscr{B}_{A_2}\rangle = b^{N\, \Delta_{A_2}}$.} This situation was analyzed in \cite{Franco:2009sp}, where it was shown that the effective field theory in the IR has CS quiver
\begin{center}
\includegraphics[scale=1.0]{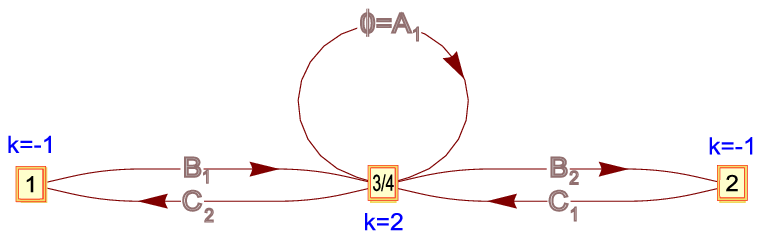}
\end{center}
with superpotential
\begin{equation}\label{con1}
W\,=\, {\rm Tr}\,\Phi\, \Big(\,C_2\, B_1\, B_2\, C_1\,-\,B_2\, C_1\, C_2\, B_1\,\Big)~.
\end{equation}
As shown in \cite{Franco:2009sp}, the moduli space of this theory is indeed $\mathcal{C}(T^{11})\times\C$. This is of course expected from the gravity analysis of equation (\ref{c_x_conifold}). Any other VEV for the $A$ fields corresponds to placing the M2 branes on $SU(2)$-equivalent points on the blown-up $\mathbb{CP}^1$, and therefore results in the same near horizon geometry.

The manifest symmetry exhibited by the Lagrangian of the $Q^{111}$ field theory is $SU(2) \times U(1)_R$, which is only a subgroup of the full $SU(2)^3 \times U(1)_R$ symmetry which is expected from the isometry of the $Q^{111}$ manifold. Therefore, in contrast to the situation with the 
$A$ fields, we see that different VEVs for the pairs of $B$ or $C$ fields result in different theories in the IR. Giving a non-vanishing VEV to $C_1$ and $C_2$, for example, results in the CS quiver
\begin{center}
\includegraphics[scale=.7]{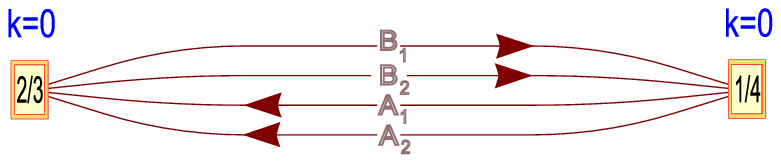}
\end{center}
with superpotential
\begin{equation}\label{zeroCS}
W\,=\, {\rm Tr}\,\Big(\,A_2\, B_1\, B_2\, A_1\,-\,B_2\, A_2\, A_1\, B_1\,\Big)~,
\end{equation}
while when the VEV of only one field is non-vanishing we instead obtain the CS quiver
\begin{center}
\includegraphics[scale=1]{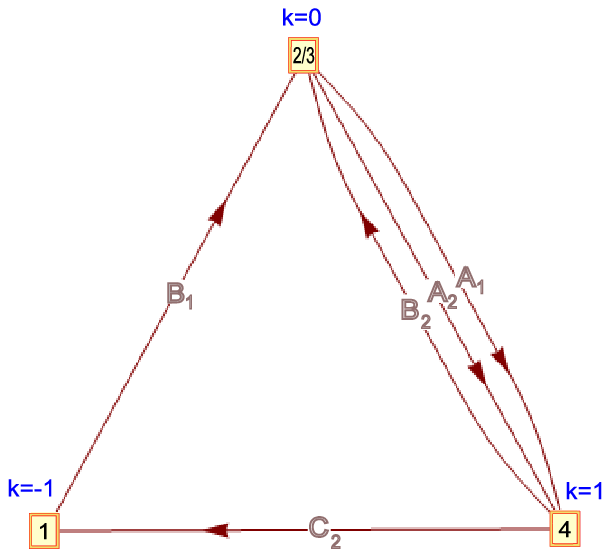}
\end{center}
with superpotential
\begin{equation}\label{con2}
W\,=\, {\rm Tr}\,C_2\, \Big(\,A_2\, B_1\, B_2\, A_1\,-\, A_2\, A_1\, B_1\,B_2\,\Big)~.
\end{equation}
Of course, both cases correspond geometrically to blowing up the same $\mathbb{CP}^1$, as can be seen from the explicit construction of the field 
theory moduli space. However, recall that the position of the M2 branes depends on the VEVs of $C_1$ and $C_2$. While in the gravity picture all  locations on the exceptional $\mathbb{CP}^1$ are $SU(2)$-equivalent, in the field theory since only part of the global symmetery is manifest we obtain different theories for different VEVs. The supergravity analysis hence suggests that the theories obtained in the IR above are dual. Indeed, one can check that the moduli spaces of these theories are the same, the details appearing in \cite{Davey:2009qx}, for example.

The QCS theory for $\mathcal{C}(T^{11})\times\C$ in (\ref{zeroCS}) has zero CS levels, and there is therefore no tunable coupling parameter in this theory. 
This may be understood as follows. After blowing up the cone to the partial resolution and placing the stack of M2 branes 
at a residual singular point on the exceptional $\mathbb{CP}^1$, the tangent cone geometry at this point is 
$\mathcal{C}(T^{11})\times\C$. In the field theory the $S^1$ that rotates the copy of $\C$ corresponds to the M-theory circle. 
Shrinking the size of this circle in the M-theory supergravity solution corresponds to a Type IIA limit describing 
a black D2 brane solution with no smooth near horizon. We should therefore not expect to find a dual field theory with a weak 
coupling limit. On the other hand, in the remaining two field theories (\ref{con1}), (\ref{con2}) for $\mathcal{C}(T^{11})\times\C$ described in this section 
the M-theory circle involves a $U(1)$ that acts also on $\mathcal{C}(T^{11})$.\footnote{The M-theory circle can be deduced by computing the $\Z_k$ orbifold action on the moduli space in each case.}

As a final comment in this section, notice that in general we will have an infinite set of CFTs dual to the $Q^{111}$ geometry, 
with different boundary conditions on the baryonic gauge fields (\ref{3-form-to-global}) in $AdS_4$. From our earlier discussion, 
these will have the same QCS theories in the $SU(N)$ sector, but different $U(1)$ sectors. In particular, in general 
different combinations of the diagonal $U(1)$s in $U(N)$ may be gauged, and with different $U(1)$ CS levels 
from the levels $k_a$ of the $SU(N)$ factors. These $U(1)$ sectors will in general behave differently under Higgsing. 
In particular, a global $U(1)$ can be sponteneously broken, while a gauged $U(1)$ can be Higgsed. It will be interesting 
to investigate this general structure in both the field theory and dual supergravity solutions, although we leave this 
for future work, focusing instead on the $U(1)^2\times SU(N)^4$ theory. 

\subsection{Baryonic condensates and M5 branes in the $Q^{111}$ solution}\label{sec:M5Q111}

In the previous section we discussed the RG flow triggered by giving a VEV to one of the fields, and hence baryonic operators,
in the $Q^{111}$ theory with gauge group $U(1)^2\times SU(N)^4$. In this section we calculate the VEV of this baryonic operator
in the corresponding gravity solution described by (\ref{Q111-warp-factor}). In order to do this we follow the analogous calculation in 
the Type IIB context, discussed in \cite{Martelli:2008cm, Klebanov:2007us}. In this prescription, to determine the VEV of a 
baryonic operator dual to an M5 brane wrapped on a five-submanifold $\Sigma_5\subset Y$ in the (partially) resolved 
supergravity background, we compute the Euclidean action of an M5 brane which wraps a minimal six-dimensional 
manifold in $X$ which at large $r$ asymptotes to the cone over $\Sigma_5$. In this section we 
present an explicit example, although later we will present a general formula for such VEVs.

Again, we focus on the partially resolved background for $Q^{111}$ described by (\ref{Q111-warp-factor}). We are interested in computing the 
VEV of the operator that was carrying charge under the baryonic symmetry before it was broken. As described in \secref{sec:2}, 
this symmetry originates on the supergravity side to a mode (\ref{3-form-to-global}) of the six-form potential $C_6$ along 
a five-cycle in the Sasaki-Einstein manifold $Q^{111}$. Consider a Euclidean M5 brane that is wrapped on the six-manifold 
at a fixed point in the $(\theta_3,\phi_3)$ copy of $\mathbb{CP}^1$, and is coordinatized by 
 $\{r,\psi,\theta_1,\theta_2,\phi_1,\phi_2\}$ in the partial resolution of $\mathcal{C}(Q^{111})$. This six-manifold is 
a divisor in the partially resolved background, and hence this wrapped M5 brane worldvolume is a calibrated submanifold. 
The M5 brane carries charge 
under the $U(1)$ gauge field in $AdS_4$ that descends from the corresponding harmonic five-form in $Q^{111}$. 
We calculate the Euclidean action of this wrapped M5 brane 
 up to a radial cut-off $r=r_c$, identifying 
 $\me^{-S(r_c)}$ with the classical field dual to the baryonic operator, as in \cite{Klebanov:2007us}. Explicitly, the action is given by
\bea
S(r_c)=T_{5}\int_{r\leq r_c} h\,\sqrt{\det g_6}\, \dd^6x~,
\label{m5-action}
\eea
where $T_{5}=2\pi/(2\pi\ell_p)^6$ is the M5 brane tension, the warp factor $h$ is given by (\ref{Q111-warp-factor}), and $\det g_6$ is the determinant of the metric induced on the M5 brane worldvolume. This induced metric is
\begin{eqnarray}
 \dd s_6^2&=&\kappa^{-1}\dd r^2+\kappa\frac{r^2}{16}\Big(\dd\psi+\sum_{i=1}^2 \cos\theta_i\dd\phi_i\Big)^2+\frac{r^2}{8}\Big(\dd\theta_2^2+\sin^2\theta_2\dd\phi_2^2\Big)\nonumber \\ &&+\frac{r^2}{8}\Big(\dd\theta_1^2+\sin^2\theta_1\dd\phi_1^2\Big)~,
\end{eqnarray}
where recall that $\kappa(r)$ is given by (\ref{kappa}).
A straightforward calculation hence gives
\bea
\sqrt{\det g_6}=\frac{\,r^5 \sin\theta_1 \sin\theta_2}{256}~.
\label{detg6}
\eea
Subtituting these results into (\ref{m5-action}) then gives
\bea
\nn
S(r_c)&=&\frac{T_{5}}{256}\int \dd\phi_1\, \dd\phi_2\, \dd\theta_1\, \dd\theta_2\, \dd\psi\, \sin \theta_1\, \sin\theta_2\, \int_0^{r_c} \dd r\, r^5 \sum_{l=0}^\infty H_{l}(r) P_{l}(\cos\theta_3) \nn \\
&=&\frac{\pi^3 T_{5}}{4} \left[ \int_0^{r_c} \dd r\, r^5 H_0 (r) + \sum_{l=1}^{\infty} \int_0^{r_c} \dd r\, r^5 H_{l}(r) P_{l}(\cos\theta_3) \right]~.
\label{mem-action}
\eea
Let us evaluate the integrals separately. The first diverges in the absence of the cut-off $r_c$, since
\bea
\int_0^{r_c} \dd r\, r^5 H_0 (r) \simeq \frac{R^6}{2} \Big[\frac{1}{3}+\log\left(1+\frac{3r_c^2}{8b}\right)\Big]~.
\eea
The second integral is finite and can be calculated straightforwardly:
\bea
\nn
\int_0^{\infty} \dd r\, r^5 \sum_{l=1}^{\infty}H_{l}(r) P_{l}(\cos\theta_3)&=&\frac{3\,R^6}{4}\sum_{l=1}^{\infty}\frac{(2l+1)}{l(l+1)}P_{l}(\cos\theta_3) \nn \\
&=& -\frac{3\,R^6}{2}\Big(\frac{1}{2}+\log \sin\frac{\theta_3}{2}\Big)~.
\eea
Recall here that $R$ is given by (\ref{anothereqn}). Subtituting these results into \eqref{mem-action} then gives
\bea
\me^{-S(r_c)}=\me^{{7N}/{18}} \left(\frac{8b}{3\,r_c^2}\right)^{\frac{N}{3}}\left(\sin \frac{\theta_3}{2}\right)^{N}~.
\label{vev}
\eea
The interpertation of this result is along the same lines as the discussion in the case of the conifold  in \cite{Klebanov:2007us}. We 
will therefore keep our discussion brief and refer the reader to \cite{Klebanov:2007us} for further details. 
Since the $AdS_4$ radial coordinate is related to $r$ as $z=r^{-2}$, we see that the operator dimension is $\Delta = \frac{N}{3}$, as anticipated by our  field theory discussion. Indeed, this provides a non-trivial check of the R-charge assignment required for the theory to have an IR superconformal fixed point, supporting the conjecture that the theory is indeed dual to M2 branes moving in $\mathcal{C}(Q^{111})$.

For the remaining M5 branes, wrapped on Euclidean six-submanifolds at a point in either the $(\theta_1,\phi_1)$ or $(\theta_2,\phi_2)$ 
copies of $\mathbb{CP}^1$, we note that $\me^{-S(r_c)}=0$. This is simply because the M5 brane worldvolumes intersect 
the M2 brane stack on the exceptional $\mathbb{CP}^1$ parametrized by $(\theta_3,\phi_3)$ in these cases, and hence 
the worldvolume action (\ref{m5-action}) is logarithmically divergent near to the M2 branes -- for
further discussion of this in the D3 brane case, see \cite{Martelli:2008cm}. Thus the dual gravity analysis of the 
partially resolved supergravity solution is in perfect agreement with the proposed $Q^{111}$ field theory, at least 
with the boundary conditions we study in this section.

\subsection{Wrapped branes and the phase of the condensate}\label{sec:wrapped}

In the resolved, or partially resolved, geometry one can consider various different kinds of stable wrapped branes in the $r\sim 0$ region. 
These shed further light on the physical interpretation of the supergravity solutions. 

In the fully resolved case with $a>0$ and $b>0$ one could consider an M5 brane wrapping the 
exceptional $\mathbb{CP}^1\times\mathbb{CP}^1$ at $r=0$ and filling the spacetime directions $x_0$, $x_1$. This is a domain wall in the Minkowski 
three-space in (\ref{metricagain}). Its tension, given by the energy of the probe brane, is 
\begin{equation}
E_{\mathrm{M5}}=\mathcal{T}_{\mathrm{wall}}=\frac{abT_5\pi^2}{16}~.
\end{equation}
Notice that the warp factor cancels out in this computation, and the brane remains of finite tension even if the stack of M2 
branes is placed at $r=0$.

Of more interest for us is to consider an M2 brane wrapping an exceptional $\mathbb{CP}^1$, either in the resolved or 
partially resolved background. In the former case notice there are homologically two choices of such $\mathbb{CP}^1$ inside $\mathbb{CP}^1\times\mathbb{CP}^1$ 
at $r=0$. Again, in either case this is a calibrated cycle. This leads to 
a point particle in the Minkowski three-space, moving along the time direction $x_0$, whose mass is
\begin{equation}
E_{\mathrm{M2}}=m_{\mathrm{point}}=\frac{bT_2\pi}{4}~.
\end{equation}
Again, its energy remains finite even when the M2 brane stack is at $r=0$. This wrapped M2 brane is the analogue of the global string that was discussed in \cite{Klebanov:2007cx} for the conifold theory in Type IIB. In our case the worldline of this point particle is linked by a circle, as it lives in a three dimensional spacetime. As we shall explain below, there are certain light fields/particles that have 
monodromies around this circle.

Consider three-form fluctuations of the form
\begin{equation} \label{fluctuateC3}
\delta C_3=A\wedge \beta~,
\end{equation}
where $\beta$ is a two-form on the Ricci-flat K\"ahler manifold $X$, and $\star_8$ denotes the Hodge dual on $X$. Demanding that $A$ is a massless gauge field in the Minkowski three-space leads to the equations
\begin{equation}\label{betaeqn}
\dd\beta=0\ ,\qquad \dd(\, h\star_8\beta)=0~.
\end{equation}
In particular $\beta$ is closed and hence defines a cohomology class in $H^2(X,\R)$; we shall be interested in the case where this 
class is Poincar\'e dual to the two-cycle wrapped by the particle-like M2 brane.
In three dimensions the gauge field $A$ is dual to a periodic scalar, which can be identified as the Goldstone boson of the spontaneous symmetry breaking.
Indeed, the M2 particle is a magnetic source for this pseudoscalar. The pseudoscalar modes will therefore have unit monodromy around a circle linking the M2 particle worldline. As in \cite{Klebanov:2007cx}, this Goldstone boson is expected to appear as a phase, through the Wess-Zumino action of the Euclidean M5 brane, in the baryonic condensate. We shall see that this is indeed the case.

It thus remains to construct appropriate two-forms $\beta$ satisfying (\ref{betaeqn}) in the warped resolved backgrounds for $Q^{111}$. 
This will occupy us for the remainder of this subsection.
It will turn out to be simpler to use the metric in form given in (\ref{Q111Cvetic}), which we reproduce here for convenience:
\begin{eqnarray}
\dd s^2&=&{U}^{-1}\dd\varrho^2+{U}\varrho\Big(\dd\psi+\sum_{i=1}^3 \cos\theta_i\dd\phi_i\Big)^2+(l_2^2+\varrho)\,\Big(\dd\theta_2^2+\sin^2\theta_2\dd\phi_2^2\Big)\nonumber \\ &&+(l_3^2+\varrho)\,\Big(\dd\theta_3^2+\sin^2\theta_3\dd\phi_3^2\Big)+\varrho\, \Big(\dd\theta_1^2+\sin^2\theta_1\dd\phi_1^2\Big)~,
\end{eqnarray}
where
\begin{equation}
 U(\varrho)= \frac{3\, \varrho^3+4\, \varrho^2\, (l_2^2+l_3^2)+6\, l_2^2\, l_3^2\varrho}{6\,(l_2^2+ \varrho)\, (l_3^2+ \varrho)}~.
\end{equation}
The constants $l_2$, $l_3$ are related to the constants $a,b$ in (\ref{resolvedQ111}) via 
$a=4l_2^2$, $b=4l_3^2$.

\subsubsection{Harmonic forms: unwarped case}

As a warm-up, we begin with the unwarped case in which the warp factor $h\equiv 1$. It is convenient to introduce the following vielbein
\bea
e_{\theta_i}=\diff \theta_i ,\quad e_{\phi_i}=\sin{\theta_i}\diff \phi_i ,\quad g_5=\diff \psi + \sum_{i=1}^3 \cos{\theta_i} \diff \phi_i~.
\eea
A natural ansatz for two two-forms $\beta_i$, $i=2,3$, is then
\begin{equation}
\beta_i=e_{\theta_i}\wedge e_{\phi_i}+\dd(f_i\, g_5)~, \qquad i=2, 3~,
\end{equation}
where $f_i$ is a function of the radial coordinate $\varrho$. Without loss of generality we will focus on the case $i=2$ and, in order not to clutter notation, drop the subscript in $f_i$. After some algebra one finds the following equation for $f$
\begin{eqnarray}\label{eqnforf}
&&\frac{\varrho\, (\varrho+l_3^2)}{(\varrho+l_2^2)}(1-f)- \frac{(\varrho+l_2^2)\, (\varrho+l_3^2)}{\varrho}f- \frac{(\varrho+l_2^2)\, \varrho}{(\varrho+l_3^2)}f \nn \\
&&+\partial_{\varrho}\Big(\varrho\,(\varrho+l_2^2)\, (\varrho+l_3^2)\, \partial_{\varrho}f\Big)=0~.
\end{eqnarray}
Since the corresponding $\delta C_3$ fluctuation in (\ref{fluctuateC3}) couples to the M2 brane wrapped over the $\mathbb{CP}^1$ at $\varrho = 0$, $\beta$ should approach here the volume form of the finite sized $\mathbb{CP}^1$. For this we require the boundary condition that $f(\varrho = 0)$ vanishes. On the other hand, as we will see later, in the large $\varrho$ region $\beta$ should asymptote to a harmonic two-form $\omega_2$ on the singular cone.

It turns out that equation (\ref{eqnforf}) can be solved exactly. 
Using the above boundary conditions to fix integration constants, we have 
\begin{equation}
f=\frac{\varrho^2+\frac{3}{2}\, l_3^2\, \rho}{3\, (\varrho+l_2^2)\, (\varrho+l_3^2)}~.
\end{equation}
Given this $f$, at small $\varrho$ one can check that $\beta$ asymptotes to $e_{\theta_2}\wedge e_{\phi_2}$, which is the volume form of the submanifold that the M2 brane is wrapped on, while at large $\varrho$ instead $\beta$ asymptotes to a harmonic two-form $\omega_2$ on the singular cone which is simply 
the pull-back of a harmonic two-form on $Q^{111}$.

\subsubsection{Harmonic forms: warped case}

Let us now turn to the warped case. Since with the M2 brane stack at $\varrho=0$ the warp factor is then a function of $(\varrho, \theta_2, \theta_3)$, it is natural to consider the following ansatz for the two-form $\beta$:
\begin{equation}
\beta=e_{\theta_2}\wedge e_{\phi_2}+\dd\left(\, f_0\, g_5+\sum_{i=2}^3 f_i\, e_{\phi_i}\, \right)~,
\end{equation}
where now $f_{\mu}=f_{\mu}(\varrho,\,\theta_2,\,\theta_3)$, and $\mu=0,2$ or $3$.

The second equation in (\ref{betaeqn}) implies that the $f_{\mu}$ must satisfy
\begin{eqnarray}
&&\partial_{\rho}\Big(\frac{h\,\sqrt{\det g}\,U(\rho)}{\ell_j^2+\rho}\partial_{\rho} f_j \Big)+\sum_{i=2}^3 \frac{1}{\sin{\theta_i}} \partial_{\theta_i}\Big( \frac{h\,\sqrt{\det g}\,\sin{\theta_i}}{(\ell_j^2+\varrho)(\ell_i^2+\varrho)} \partial_{\theta_i}f_j \Big)-\\ \nn
&&-\partial_{\theta_j}\Big( \frac{h\,\sqrt{\det g}\,f_0}{(\ell_j^2+\varrho)^2}\Big)+\partial_{j}\Big( \frac{h\,\sqrt{\det g}\cot{\theta_j}}{(\ell_j^2+\varrho)^2}\Big)f_j +\partial_{\theta_j}\Big( \frac{h\,\sqrt{\det g}}{(\ell_j^2+\varrho)^2}\Big)\delta_{j,2}=0~,
\end{eqnarray}
for $\mu=j$, and
\begin{eqnarray}
&&\partial_{\varrho}\Big(h\, \sqrt{\det g}\, \partial_{\varrho}f_0\Big)+\sum_{i=1}^3\frac{1}{\sin\theta_i}\partial_{\theta_i}\Big(\frac{U^{-1}(\varrho)\, \sqrt{\det g}\, h}{(\varrho+l_i^2)}\, \sin\theta_i\,\partial_{\theta_i}\, f_0\Big) \\ \nonumber
&& +\sum_{i=2}^3\frac{\sqrt{ \det g}\, h}{(\varrho+l_i^2)^2\, \sin\theta_i}\partial_{\theta_i}\Big(\sin\theta_i\, f_i\Big)-\sum_{i=1}^3 \frac{\sqrt{\det g}\, h}{(\varrho+l_i^2)^2}f_0+\frac{h\, \sqrt{\det g}}{(\varrho+l_2^2)^2}=0~,
\end{eqnarray}
for $\mu=0$.
Note that $l_1\equiv 0$ and $\sqrt{\det g}=\varrho\, (\varrho+l_2^2)\, (\varrho+l_3^2)$.

We have three equations for three functions $f_{\mu}$, so we expect this system to contain the desired solution for $\beta$. Furthermore, if we consider the unwarped case where $h\equiv 1$ we can consistently set $f_i=0$ and assume $f_0=f_0(\varrho)$. In this case  the second equation  reduces to 
\begin{eqnarray}
&&\partial_{\varrho}\Big(\sqrt{\det g}\, \partial_{\varrho}f_0\Big) -\sum_{i=2}^3 \frac{\sqrt{\det g}}{(\varrho+l_i^2)^2}f_0+\frac{\sqrt{\det g}}{(\varrho+l_2^2)^2}=0~.
\end{eqnarray}
This is precisely the equation obtained above. Furthermore, these equations can be seen to arise from minimizing the action
\bea
\nn
&&I=\int_X \beta\wedge\star_8h\,\beta = \int_0^{\pi} \diff \theta_2 \int_0^{\pi} \diff \theta_3 \int_0^{\infty} \diff \varrho\, h \, \sqrt{\det g} \sin{\theta_2} \sin{\theta_3} \Big[(\partial_{\varrho}\,f_0)^2 + \\ \nn 
&& +\frac{(\partial_{\theta_3} f_0)^2}{U\,(l_3^2 + \varrho)} + \frac{(\partial_{\theta_2} f_0)^2}{U\,(l_2^2 + \varrho)} + \frac{(f_0 - \partial_{\theta_3} f_3- \cot{\theta_3} f_3)^2}{(l_3^2 + \varrho)^2} + \frac{(\partial_{\theta_3} f_2)^2 + (\partial_{\theta_2} f_3 )^2}{(l_2^2 + \varrho) (l_3^2 + \varrho)} + \\ \nn
&& + \frac{(-1 + f_0 - \cot{\theta_2} f_2 -\partial_{\theta_2} f_2)^2}{(l_2^2 + \varrho)^2} + \frac{U}{(l_2^2 + \varrho)} (\partial_{\varrho} f_2)^2 
+ \frac{U}{(l_3^2 + \varrho)} (\partial_{\varrho} f_3)^2 + \frac{f_0^2}{\varrho^2} \Big]~.
\eea
With the boundary conditions above, one can check that $I$ is finite. We shall give a more general argument for existence of 
this two-form later in this section.

Going back to the original physical motivation, the warped harmonic form $\beta$ allows one to construct a three-form fluctuation $\delta C=A\wedge \beta$ that satisfies the linearized SUGRA equations. Consider the Hodge dual
\begin{equation}
\delta G_{7}=\star_3 \dd A\wedge h\star_8\beta~.
\end{equation}
As noted above, the three-dimensional gauge field $A$ is dual to a periodic scalar via $\star_3\dd A=\dd p$. Thus we can write
\begin{equation}
\delta G_{7}=\dd p\,\wedge h\star_8\beta=\dd\left(p\, h\star_8\beta\right)~,
\end{equation}
thus obtaining a local form for the six-form fluctuation. At very large $\varrho$, $\beta$ can be approximated to leading order by
\begin{equation}
\beta\sim \frac{2}{3}\, e_{\theta_2}\wedge e_{\phi_2} - \frac{1}{3}\, e_{\theta_1}\wedge e_{\phi_1}- \frac{1}{3}\, e_{\theta_3}\wedge e_{\phi_3}~,
\end{equation}
so that
\begin{equation}
\delta C_{6}\supset p \,\,\frac{\dd\varrho}{\varrho^2}\wedge \dd\psi \wedge e_{\theta_1}\wedge e_{\phi_1}\wedge e_{\theta_2}\wedge e_{\phi_2}~.
\end{equation}
This form of the local potential couples to the baryonic M5 brane through the Wess-Zumino term, thus reinforcing the identification of the scalar $p$ as the phase of the baryonic condensate and Goldstone boson. Furthermore, at large $\varrho$ we can also write
\begin{equation}
\delta G_{7}\supset \dd(\, \varrho^{-1}\, \dd p\wedge \dd\psi\wedge e_{\theta_1}\wedge e_{\phi_1}\wedge e_{\theta_2}\wedge e_{\phi_2})~.
\end{equation}
Then, following \cite{Klebanov:2007cx}, we note the appearance of the volume form of the submanifold wrapped by the baryonic M5 brane, with the appropriate decay for a conserved current in three dimensions of dimension 2. This motivates the identification
\begin{equation}
\langle J^B_{\mu}\rangle \sim \partial_{\mu}p~.
\end{equation}

\subsection{General warped resolved Calabi-Yau backgrounds}\label{sec:generalwarped}

Much of our discussion of the resolved $Q^{111}$ backgrounds can be extended to general warped resolutions of 
Calabi-Yau cones. In the remainder of this section we describe what is known about such generalizations. 
In particular in the next subsection we present a novel method for computing M5 brane condensates in such 
backgrounds, or more generally the worldvolume actions of branes in warped Calabi-Yau geometries.

\subsubsection{Gravity backgrounds}

As for the $Q^{111}$ case we are interested in M-theory backgrounds of the form
\bea\label{background2} 
\dd s^2_{11} &=& h^{-2/3} \dd s^2(\R^{1,2}) + h^{1/3} \dd s^2(X)~, \\ \nn 
G &=& \dd^3 x \wedge \dd h^{-1}~,
\eea
where $X$ is a Ricci-flat K\"ahler eight-manifold that is asymptotic to a cone metric over some Sasaki-Einstein seven-manifold $Y$.  
Placing $N$ spacetime-filling M2 branes at a point $y\in X$ leads to the warp factor equation
\bea\label{green}
\Delta_x h[y] = \frac{(2\pi \ell_p)^6N}{\sqrt{\det g_X}} \delta^8(x-y)~.
\eea
Here $\Delta h = \diff^* \diff h = - \nabla^i\nabla_i h$ is the scalar Laplacian of $X$ acting on $h$. 
Thus $h[y](x)$ is simply the Green's function on $X$. More generally, one could pick different points $y_i\in X$, with $N_i$ M2 branes 
at $y_i$, such that $\sum_i N_i=N$. Then $h[\{y_i,N_i\}](x)$ will be a sum of Green's functions, weighted by $N_i$. We shall regard this as an obvious generalization. There are thus two steps involved in constructing such a solution: choose a Calabi-Yau metric 
on $X$, and then solve for the Green's function. If the latter is chosen so that it vanishes at infinity then the 
SUGRA solution (\ref{background2}) will be asymptotically $AdS_4 \times Y$ with $N$ units of $G_7$ flux through $Y$.

If $Y$ is a Sasaki-Einstein manifold then $\mathcal{C}(Y)$ defines an isolated singularity at the tip of the cone $r=0$. 
We may then take a resolution  $\pi:X\rightarrow \mathcal{C}(Y)$, which defines our manifold $X$ as a complex manifold. 
The map $\pi$ is a biholomorphism 
of complex manifolds on the complement of the singular point $\{r=0\}$, so that in $X$ the singular point is effectively replaced by a higher-dimensional 
locus, which is called the exceptional set. We require that the holomorphic $(4,0)$-form on $\mathcal{C}(Y)$ extends to a smooth holomorphic $(4,0)$-form on 
$X$. Such resolutions are said to be ``crepant'', and they are not always guaranteed to exist, even for toric singularities. 
In the latter case one can typically only partially resolve so that $X$ has at worst orbifold singularities. Having chosen such an $X$ we must then 
find a Calabi-Yau metric on $X$ that approaches the given cone metric asymptotically. Fortunately, mathematicians have very recently proved that 
you can \emph{always} find such a metric. Essentially, this is a non-compact version of Yau's theorem with a ``Dirichlet'' boundary condition, where we have a fixed Sasaki-Einstein metric 
at infinity on $\partial X=Y$ and ask to fill it in with a Ricci-flat K\"ahler metric. There are a number of papers that have developed this subject 
in recent years \cite{crep1,crep2,crep3,crep4,crep5, crep6}, but the most recent \cite{goto, vanC} prove the strongest possible result: that in each K\"ahler class in $H^2(X,\R)\cong H^{1,1}(X,\R)$ (see \cite{crep6}) there is a unique 
Calabi-Yau metric that is asymptotic to a fixed given Sasaki-Einstein metric on $Y=\partial X$. Note this result \emph{assumes} the existence of the 
Sasaki-Einstein metric -- it does not prove it.

The crepant (partial) resolutions of toric singularities are well understood, being described by toric geometry and hence fans of polyhedral cones. 
The extended K\"ahler cone for such resolutions is known as the GKZ fan, or secondary fan. The fan is a collection of polyhedral cones living in $\R^{b_2(X)}$, glued together along their boundaries, such that each cone corresponds to a particular choice of topology for $X$. Implicit here is the fact that 
$b_2(X)$ is independent of which topology for $X$ we choose.
A point inside the polyhedral cone corresponding to a given $X$ is a K\"ahler class 
on $X$. The boundaries between cones correspond to partial resolutions, where there are further residual singularities, and there is a topology change as 
one crosses a boundary from one cone into another. The GKZ fan for $Y=Q^{111}$ was described in Figure \ref{fig:GKZ1}. If we combine this description with the above 
existence results, we see that the GKZ fan is in fact classifying the space of resolved asymptotically conical Calabi-Yau metrics. 

Having chosen a particular resolution and K\"ahler class, hence metric, 
we must then find the warp factor $h$ satisfying (\ref{green}). This amounts to finding the Green's function on $X$, and this always exists 
and is unique using very general results in Riemannian geometry. A general discussion in the Type IIB context may be found in 
 \cite{Martelli:2007mk}, and the comments here apply equally to the M-theory setting. In the warped metric (\ref{background2}) the point $y\in X$ is effectively sent to infinity, and 
the geometry has two asymptotically $AdS_4$ regions: one near $r=\infty$ that is asymptotically $AdS_4\times Y$, and one near 
to the point $y$, which if $y$ is a smooth point is asymptotically $AdS_4\times S^7$. For further discussion, see 
\cite{Martelli:2007mk, Martelli:2008cm, Benishti:2009ky}.

If one places the $N$ M2 branes at the same position $y\in X$, the moduli space is naturally a copy of $X$. 
The $b_2(X)$ K\"ahler moduli are naturally complexified by noting that $H^6(X,Y,\R)\cong H_2(X,\R)\cong \R^{b_2(X)}$ by Poincar\'e 
duality, and that this group classifies the periods of $C_6$ through six-cycles in $X$, which may either be closed or have a 
boundary five-cycle on $Y=\partial X$. More precisely, taking into account large gauge transformations leads to 
the torus $H^2(X,Y,\R)/H^2(X,Y,\Z)\cong U(1)^{b_2(X)}$. Altogether this moduli space of SUGRA solutions 
should be matched to the full moduli space of the dual SCFT. At least for toric 
$X$ one can prove quite generally via an exact sequence that $b_2(X)=b_2(Y)+b_6(X)$, where $b_6(X)$ is also the number 
of irreducible exceptional divisors in the resolution. In toric language, this is the number of internal lattice points 
in the toric diagram. We shall discuss such examples in \secref{sec:5}: the presence of calibrated six-cycles 
is expected to lead to M5 brane instanton corrections in these backgrounds.

\subsubsection{Harmonic two-forms}

Recall we are also interested in fluctuations of the form
\begin{equation}
\delta C_3=A\wedge \beta~,
\end{equation}
where $A$ leads to a massless gauge field in the Minkowski three-space if
\begin{equation}\label{betaagain}
\dd\beta=0\ ,\qquad \dd(\, h\star_8\beta)=0~.
\end{equation}
For trivial warp factor $h\equiv 1$ this just says that $\beta$ is harmonic. It is a general result that 
if we also impose that $\beta$ is $L^2$ normalizable, or equivalently that $A$ has finite kinetic energy 
in three dimensions, then such forms are guaranteed to exist and are in 1-1 correspondence with 
$H^2(X,Y,\R)\cong H_6(X,\R)$ \cite{hausel}. Thus there are always $b_6(X)$ $L^2$ normalizable harmonic two-forms $\beta$ 
in the unwarped case. 

However, this case isn't the physical case for applications to $AdS/CFT$. Instead we should look for 
solutions to (\ref{betaagain}) where $h$ is the Green's function on $X$. Again, fortunately 
there are mathematical results that we may appeal to to guarantee existence of such forms. 
These are again described in the Type IIB context in \cite{Martelli:2008cm}. In the warped 
case $\beta$ is harmonic with respect to the metric $h^{1/2} \diff s^2(X)$. This manifold 
has an asymptotically conical end with boundary $S^7$ (or more generally $Y_{IR}$ if the M2 brane stack 
is placed at a singular point with horizon $Y_{IR}$), and an isolated conical singularity with 
horizon metric $\tfrac{1}{4}\diff s^2(Y)$. The number of $L^2$ harmonic two-forms on such a space 
is in fact known, and is $b_2(X)$. To see this requires combining a number of mathematical results 
that are described in \cite{Martelli:2008cm}. In particular, since $b_2(X)=b_2(Y)+b_6(X)$ there is a 
corresponding harmonic form, and hence Goldstone mode, for each of the $b_2(Y)$ baryonic $U(1)$ 
symmetries. Indeed, these $b_2(Y)$ harmonic forms can be seen to asymptote to the harmonic two-forms
on $Y$ at $r=\infty$. Thus the analysis at the end of \secref{sec:wrapped} carries over 
in much more general backgrounds.
We shall analyse the asymptotics of the $b_6(X)$ unwarped $L^2$ harmonic forms 
in more detail in \secref{sec:5}, where they will be given a very different interpretation.

\subsection{Baryonic condensates: M5 branes in general warped geometries}\label{sec:M5general}

As discussed in \secref{sec:baryons}, M5 branes wrapped on five-manifolds $\Sigma_5\subset Y$ lead, with 
appropriate choice of quantization of the gauge fields in $AdS_4$, 
to scalar operators 
in the dual SCFT that are charged under the $U(1)^{b_2(Y)}$ baryonic symmetry group. We have already described 
how to compute the VEV of such an operator in the (partially) resolved $Q^{111}$ background. More generally 
one should compute the action of a Euclidean M5 brane which is wrapped on a minimal six-submanifold $D\subset X$ with boundary $\partial D=\Sigma_5$. 
Similar computations, in some specific examples, have been performed in \cite{Baumann:2006th}. In this section we explain how 
this Euclidean action may be computed exactly, in general, in the case where $D$ is a divisor. This is essentially a technical computation that may be skipped if the reader is not interested in the details: the final formula is (\ref{thefinalcountdown}).

Let suppose that we are given a warped background (\ref{background2}), where $X$ has an asymptotically conical Calabi-Yau metric and 
the warp factor satisfying (\ref{green}) is given, with a specific choice of point $y\in X$ where the stack of $N$ M2 branes are located. 
We would like to compute 
\bea
\exp\left(-T_5\int_D \sqrt{\det g_D}\, h[y]\, \diff^6x\right)~.
\eea
Here $T_5=2\pi/(2\pi \ell_p)^6$ is the M5 brane tension, and the integrand is the worldvolume action (in the absence of a self-dual two-form). Thus 
$g_{D}$ denotes the pull-back of the unwarped metric to the worldvolume $D$. 
We also assume that $D$ is a divisor, in order to preserve supersymmetry. Then $D$ is also minimal and the integral is
\bea
\int_D \sqrt{\det g_D}\, h[y] \, \diff^6 x = \int_D \frac{\omega^{3}}{3!}\, h[y]~,
\eea
where $\omega$ denotes the K\"ahler form of the unwarped metric, pulled back to $D$. 

Before beginning our computation, we note that on a K\"ahler manifold the scalar 
Laplacian $\Delta h = \dd^* \dd h$ can be written as
\bea\label{KahlerLaplace}
\Delta h = -\omega\lrcorner \dd \dd^c h = -2\ii \omega\lrcorner\partial \bar{\partial}h~.
\eea
Here
$\dd^c \equiv I\circ \dd = \ii (\bar{\partial}-\partial)$, 
where $I$ is the complex structure tensor. The contraction 
$\omega\lrcorner \dd \dd^c h$ is then in local complex coordinates
\bea
\omega\lrcorner\dd \dd^c h = 4\omega^{i\bar{j}}\frac{\partial^2h}{\partial z^i\partial\bar{z}^{\bar{j}}} = -\Delta h~,
\eea
where 
\bea
\omega=\frac{\ii}{2}\omega_{i\bar{j}}\diff z^i\wedge \diff\bar{z}^{\bar{j}}~.
\eea

For simplicity we shall study the case in which $D$ is described globally by the equation 
$D=\{f=0\}$, where $f$ is a global holomorphic function on $X$. This means 
that the homology class of $D$ is trivial, and hence in fact the wrapped 
M5 brane carries zero charge under $U(1)^{b_2(Y)}$. The cases studied in 
\cite{Baumann:2006th} are of this form. More generally, since $D$ is a complex divisor it defines 
an associated holomorphic line bundle $\mathcal{L}_D$ over $X$. Then we may take $D$ to be the zero 
set of a holomorphic section of $\mathcal{L}_D$, with a simple zero along $D$. 
To extend the computation below to this case would require combining the argument we give here 
with the arguments in \secref{sec:5}.

Thus, suppose that $f$ is a holomorphic function with a simple zero along $D$, and introduce the
two-form
\bea
\eta_D \equiv \frac{1}{2\pi} \dd \dd^c \log |f| = -\frac{1}{2\pi \ii} \partial\bar{\partial} \log |f|^2~.
\eea
Let us examine its properties. First, note that 
\bea
\log |f|^2 = \log f + \log \bar{f}~.
\eea
Since $f$ is holomorphic, and thus $\bar{\partial}f =0$ by definition, 
this shows that away from the locus $f=0$, which is the divisor $D$, 
in fact $\eta_D=0$. On the other hand, locally along $D$ we can write 
$f = zg$ 
where $z$ is a local coordinate normal to $D$, and $g=g(z,w_1,w_2,w_3)$, where 
$w_1,w_2,w_3$ are local complex coordinates along $D$ and $g$ has no zero in this local chart. 
We may then write $z=r \me^{\ii\theta}$, and note that 
\bea
\frac{1}{2\pi}\dd \dd^c \log r = \delta^2(0) \frac{\ii}{2}\dd z\wedge\dd \bar{z}~.
\eea
This is just the elementary statement that $(1/2\pi)\log r$ is the unit 
Green's function in dimension two (the local transverse space to $D$). 
Thus we have shown that $\eta_D$ is zero away from $D$, and is  a unit 
delta function supported along $D$. 

Using these properties of $\eta_D$ we may hence write
\bea
V\equiv T_5\int_D \frac{\omega^3}{3!}\, h[y] = \int_X \frac{T_5 h[y]}{2\pi} \frac{\omega^3}{3!} \wedge \dd \dd^c \log |f|~.
\eea
Note in particular that $T_5 h[y]/2\pi$ is $N$ times the unit Green's function (with unit delta function source), {\it i.e.}
\bea
\Delta_x \left(\frac{T_5 h[y]}{2\pi}\right) = \frac{N}{\sqrt{\det g_X}}\delta^8(x-y)~.
\eea
We then integrate by parts
\bea\label{eq1}
V = \int_{\partial X=Y} \frac{T_5 h[y]}{2\pi} \frac{\omega^3}{3!} 
\wedge \dd^c \log |f| - \int_X \dd\left(\frac{T_5 h[y]}{2\pi}\right)\wedge\frac{\omega^3}{3!} \wedge \dd^c \log |f|~.
\eea
We will deal with the boundary terms later, focusing for now on the integrals over $X$. First note that
\bea
\gamma\wedge \frac{\omega^3}{3!} = - \star I(\gamma)~.
\eea
holds for any one-form $\gamma$. Using this we may write
\bea
V &=& -\int_X \dd\left(\frac{T_5 h[y]}{2\pi}\right)\wedge\frac{\omega^3}{3!} \wedge \dd^c \log |f| + \mbox{boundary term}\nn\\
  &=& \int_X \diff^c\left(\frac{T_5 h[y]}{2\pi}\right)\wedge\frac{\omega^3}{3!} \wedge \dd \log |f| + \mbox{boundary term}~.
\eea
Here we have used that the metric is of course Hermitian, so that $g_X(\gamma,\delta)=g_X(I\gamma,I\delta)$. 
We then again integrate by parts
\bea\label{eq2}
V = \int_X \dd \dd^c \left(\frac{T_5 h[y]}{2\pi}\right)\wedge\frac{\omega^3}{3!}\log |f| + \mbox{boundary terms}~,
\eea
where explicitly now
\bea
\mbox{boundary terms} = \int_{\partial X=Y} \left\{\frac{T_5 h[y]}{2\pi} \frac{\omega^3}{3!} 
\wedge \dd^c \log |f| - \dd^c \left(\frac{T_5 h[y]}{2\pi}\right) \wedge \frac{\omega^3}{3!} \log |f|\right\}~.
\eea
In fact this boundary integral is divergent -- a key physical point in interpreting it holographically. For now 
let us deal with the integral over $X$ in (\ref{eq2}). Using (\ref{KahlerLaplace}) we may write
\bea
-\dd\dd^c \left(\frac{T_5 h[y]}{2\pi}\right)\wedge \frac{\omega^3}{3!} = \Delta \left(\frac{T_5 h[y]}{2\pi}\right) \frac{\omega^4}{4!} = 
\frac{N}{\sqrt{\det g_X}}\delta^8(x-y) \frac{\omega^4}{4!}~,
\eea
and thus
\bea
V = -\int_X N \delta^8(x-y)\, \log |f|\, \diff^8 x = - N \log |f(y)|+\mbox{boundary terms}~.
\eea

Let us turn now to the boundary terms. In order to render this finite, we cut off the integral at some large $r=r_c$, and 
write the boundary integral as
\bea\label{bterms}
\mbox{boundary terms} = \int_{Y_{r_c}} \left[\frac{T_5 h[y]}{2\pi}\dd^c \log |f|  - \log |f| \dd^c \left(\frac{T_5 h[y]}{2\pi}\right)\right]\wedge \frac{\omega^3}{3!}~.
\eea
We require that $D$ is asymptotically conical, so that at large $r_c$ it approaches a cone over a compact five-manifold $\Sigma_5\subset Y$. 
In the cone geometry, a conical divisor with trivial homology class is specified as the zero set of a homogeneous function under $r\partial_r$. Thus we take
\bea
|f| = Ar^{\lambda}\left(1+\ldots\right)~,
\eea
where $A$ is homogeneous degree zero ({\it i.e.} a function on $Y$) and the $\ldots$ are terms that go to zero as $r\rightarrow\infty$. Thus $f$ has asymptotic homogeneous degree $\lambda>0$. Now, the volume form on $Y$ is 
\bea
\diff\vol(Y)=\eta\wedge \frac{(\diff\eta)^3}{2^3\cdot 3!}~,
\eea
where $\eta=\dd^c \log r$ and
$\omega_{\mathrm{cone}} = \frac{1}{2}\diff (r^2 \eta)$ is the K\"ahler form on the cone over $Y$. Asymptotically,
\bea
\omega = \omega_{\mathrm{cone}} + O(r^4)~.
\eea
The $O(r^4)$ follows since the leading correction to the cone metric is a harmonic two-form on $Y$, which is down by a factor 
of $r^{-2}$ relative to the cone metric. We also have
\bea
\frac{T_5 h[y]}{2\pi} = \frac{N}{6\vol(Y) r^6}(1 + \ldots)~.
\eea
{\it Cf.} (\ref{warping}). Thus the first term in (\ref{bterms}) is \emph{convergent} and gives
\bea
\lim_{r_c\rightarrow \infty} \int_{Y_{r_c}} \frac{T_5 h[y]}{2\pi}\dd^c \log |f| \wedge \frac{\omega^3}{3!} = \frac{N}{6\vol(Y)}\cdot \lambda \cdot \vol(Y) = \frac{N\lambda}{6}~.
\eea
Note here that the function $A$ does not contribute to the integral as it is independent of $r$, and thus $J(r\partial_r)\lrcorner \dd^c \log A=0$. On the other hand, the second term in (\ref{bterms}) is divergent, the leading divergent piece being
$N\lambda \log r_c~.$
Provided the $\ldots$ terms in $|f|$ and $h$ fall off as $o(r^{-\epsilon})$, for some $\epsilon>0$, then in fact this is the only divergence (since 
any positive power of $r$ grows faster than $\log r$). There is also a finite part, namely $N\int_Y \log A/\vol(Y)$. However, the important point is that this 
depends only on asymptotic data. 

Let us interpret this divergence. Suppose that the Sasaki-Einstein manifold $Y$ is quasi-regular, meaning that it is a $U(1)$ 
bundle over a K\"ahler-Einstein orbifold $Z$.\footnote{In fact the irregular case can be approximated arbitrarily closely by the quasi-regular case.} 
Then asymptotically $f$ is, in its dependence on $Z$, a holomorphic section of $L^k$ for some integer $k\in\Z_{>0}$, where $L=K_Z^{-1/I}$ with $I=I(Z)\in\Z_{>0}$ being the orbifold Fano index of $Z$. Here $K_Z$ denotes the orbifold canonical bundle of $Z$, and $I$ is by definition 
the largest integer so that the root $L$ is defined. It follows that
\bea
\lambda = \frac{4k}{I}~,
\eea
where $4=\dim_\C X$. The five-manifold $\Sigma_5\subset Y$ is then the total space of a $U(1)$ bundle over an orbifold surface $S\subset Z$, with 
the Poincar\'e dual to $S$ being represented by $c_1(L^k)=kc_1(Z)/I$. The K\"ahler-Einstein condition on $Z$ gives
\bea\label{KE}
8[\omega_Z] = 2\pi c_1(Z)\in H^{1,1}(Z,\R)~,
\eea
where $\omega_Z=(1/2)\diff\eta$ denotes the K\"ahler form of $Z$. 

Now, the conformal dimension of the operator dual to an M5 brane wrapped on $\Sigma_5$ is given by the general formula
\bea
\Delta(\mathcal{O}[\Sigma_5]) = \frac{\pi N \vol(\Sigma_5)}{6\vol(Y)}~.
\eea
In the quasi-regular case at hand, the length of the $U(1)$ circle cancels in the numerator and denominator and we can write this as 
\bea
\Delta(\mathcal{O}[\Sigma_5]) = \frac{\pi N \int_S \frac{\omega_Z^2}{2!}}{6 \int_Z \frac{\omega_Z^3}{3!}} = \frac{\pi N \frac{k}{I}\int_Z \frac{\omega_Z^2}{2!}\wedge 
c_1(V)}{2\int_Z \frac{\omega_Z^3}{2!}} = \frac{2Nk}{I}~.
\eea
Here in the last step we used the K\"ahler-Einstein condition (\ref{KE}). Thus we have the general result that the divergent part of the 
integral is
\bea
N\lambda \log r_c = 2\Delta(\mathcal{O}[\Sigma_5]) \log r_c = -\Delta(\mathcal{O}[\Sigma_5]) \log z~,
\eea
where we have changed to the usual $AdS_4$ coordinate $r_c^2 = 1/z$. Thus we see that the divergent part of the 
integral is precisely such that we can interpret its coefficient as the VEV of the operator $\mathcal{O}[\Sigma_5]$ in this background. 
This coefficient is, from the above computations, 
\bea\label{thefinalcountdown}
\exp(-V_{\mathrm{reg}}) = |f(y)|^N\exp\left(-\frac{N\lambda}{6}-\frac{N}{\vol(Y)}\int_Y \log (|f|/r^\lambda)\right)~.
\eea
This is an exact result for the VEV, or regularized exponential of the M5 brane action, in terms of the defining function $f$ of the divisor $D$. The integral is understood in the limit
$Y=Y_{r_c}$ as $r=r_c\rightarrow \infty$, which is convergent as the integrand is independent of $r$ in this limit. Notice, in particular, that 
if one multiplies $f$ by a constant, this constant drops out of the formula, as it should. 

\section{Six-cycles and non-perturbative superpotentials} \label{sec:5}

So far our discussion has mainly focused on the $Q^{111}$ theory and its warped resolved supergravity solutions. However, this 
solution is somewhat special in that the resolved Calabi-Yau manifolds are ``small resolutions'' -- that is, there are no 
exceptional divisors, or six-cycles, in the resolved solution. 
An isolated toric Calabi-Yau four-fold singularity will typically have exceptional divisors when it is resolved.\footnote{As already 
mentioned, generically one can at best partially resolve such singularities so that the remaining singularities are all of orbifold type.} 
The irreducible components of these divisors are in 1-1 correspondence with the internal lattice points in the toric diagram, and 
a simple homology calculation shows that these generate the group $H_6(X,\R)$ of six-cycles. This immediately 
raises the question of what is the $AdS/CFT$ interpretation of such six-cycles, as we mentioned briefly at the end of 
section \ref{sec:generalwarped}. Again, in order to make our discussion concrete we will begin by focusing on a simple example, namely the cone over $Q^{222}$. This is a $\mathbb{Z}_2$ orbifold of $Q^{111}$ in which the free $\Z_2$ quotient is along the R-symmetry $U(1)$ fibre. Further details about the 
geometry of this manifold are contained in appendix \ref{sec:Q222-geometry}.

\subsection{The $\mathcal{C}(Q^{222})$ theory}

A candidate dual field theory to $\mathcal{C}(Q^{222})$ was proposed in \cite{Franco:2009sp}. In fact there are (at least) two possible toric phases for this theory. The quivers of the two phases are

\begin{center}
\includegraphics[scale=.8]{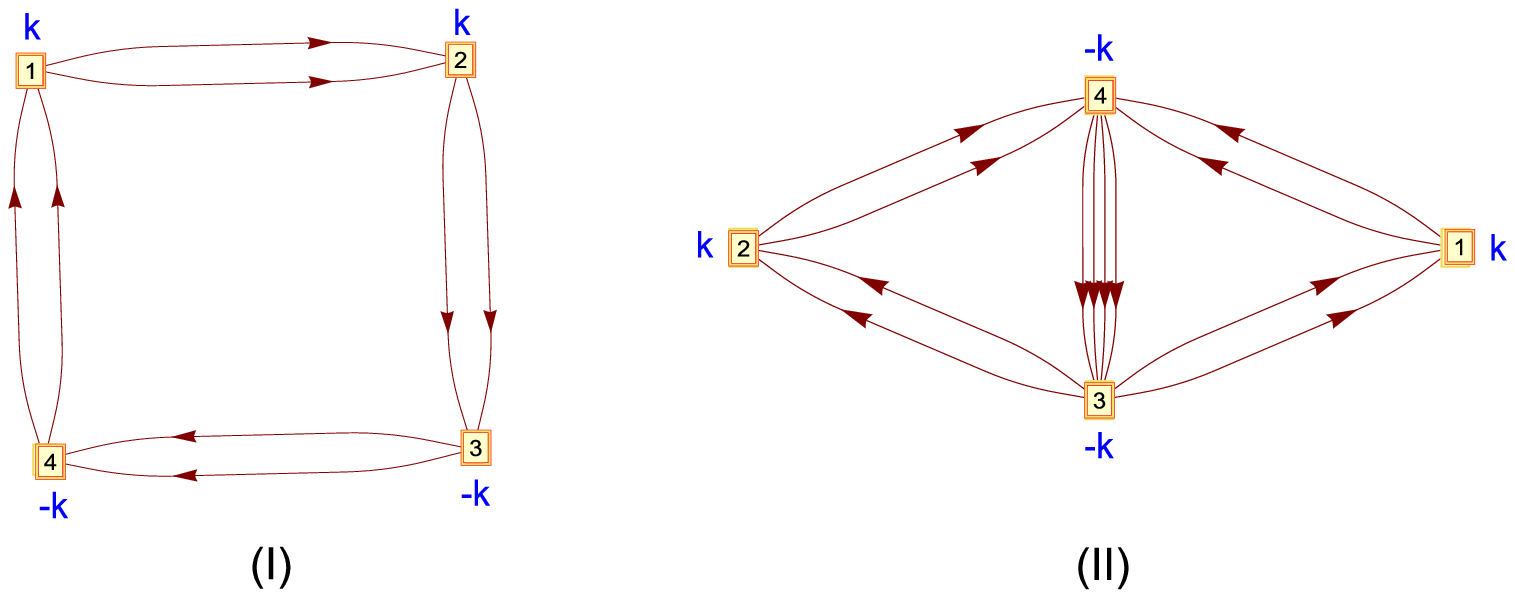}
\end{center}
with superpotentials
\bea
\nn
&&W_I\,=\,{\rm Tr}\, \epsilon_{ij}\, \epsilon_{mn}\, X_{12}^i\, X_{23}^m\, X_{34}^j\, X_{41}^n~, \\
&&W_{II}\, =\, {\rm Tr}\,\Big(\, \epsilon_{ij}\, \epsilon_{mn}\, X_{32}^i\, X_{24}^m\, X_{43}^{jn}-\epsilon_{ij}\, \epsilon_{mn}\, X_{31}^m\, X_{14}^i\, X_{43}^{jn}\,\Big)~.
\eea

Following the same prescription for $\mathcal{C}(Q^{111})$, we will consider quantizations of the gauge fields in $AdS_4$ such that 
the gauge groups are $U(1)_G\times U(1)_{G-1}\times SU(N)^4$. As shown in \cite{Franco:2009sp}, the moduli spaces of both phases give the desired cone over $Q^{222}$. Furthermore, with $SU(N)$ gauge groups we have two global $U(1)$ baryonic symmetries, which precisely match the $b_2(Q^{222})=2$ gauge fields we find in $AdS_4$ from the gravitational point of view. Thus, as far as the moduli space and baryonic symmetries are concerned, the $\mathcal{C}(Q^{222})$ case closely resembles its $\mathcal{C}(Q^{111})$ cousin. 

\begin{figure}[ht]
\begin{center}
\includegraphics[scale=.7]{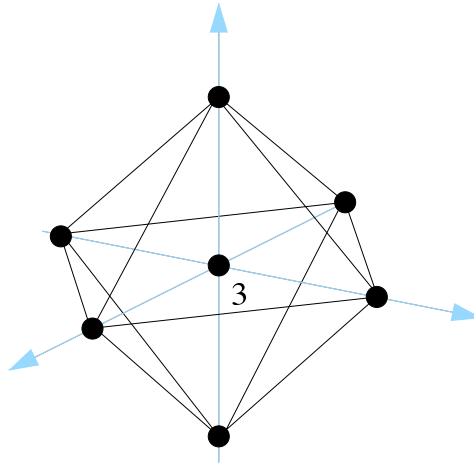}
\end{center}
\caption{The toric diagram for $\mathcal{C}(Q^{222})$.}
\label{fig:toricdiagramQ222}
\end{figure}

However, inspection of the $\mathcal{C}(Q^{222})$ toric diagram in Figure \ref{fig:toricdiagramQ222}
shows an interior lattice point, signaling the possibility of blowing up a six-cycle. We shall discuss the geometry of such resolutions 
in more detail later in this section. However, this immediately raises the question of the field theory interpretation of this six-cycle. As we discuss in the next subsection, such six-cycles have been shown to be responsible for non-perturbative superpotentials in Calabi-Yau compactifications via wrapped Euclidean M5 branes. We are interested in such contributions to a non-perturbative superpotential 
in warped Calabi-Yau backgrounds. The warping here is induced by the back-reaction of point-like M2 brane sources on the Calabi-Yau.\footnote{It might be possible to generalize this to the case with SUSY $G$-flux, in which the flux also sources the warp factor.} 

In previous sections we have argued that, for toric singularities with no vanishing six-cycles, we can choose a set of boundary conditions which amounts to ungauging all but the $\mathcal{B}_{G-1},\, \mathcal{B}_G$ $U(1)$ symmetries. This gives rise to a field theory moduli space which is roughly a $(\C^*)^{b_2(Y)}$ fibration over the mesonic space. Here the fibres are the global $U(1)$s we do not quotient by, and their corresponding D-terms that we do not impose. On the SUGRA side, the mesonic part of the moduli space arises from the $N$ M2 branes that are free to move in the geometry. The $b_2(Y)$ $\C^*$s are instead naturally the $b_2(Y)=b_2(X)$ K\"ahler classes plus the corresponding periodic scalars coming from modes of $C_3$ on associated two-cycles. More generally we have $b_2(X)=b_2(Y)+b_6(X)$, where $X$ is a Calabi-Yau resolution of $Y$. We can at this point consider the same operation for geometries with six-cycles, so that $b_6(X)\neq 0$. However, in this case the field theory moduli space seems to be too small. Let us consider the $\mathcal{C}(Q^{222})$ example. In field theory we still have a $(\C^*)^2$ fibration over Sym$^N$ $\mathcal{C}(Q^{222})$ as moduli space, since there are again only 4 nodes in the quiver. On the other hand, the gravity side has 3 K\"ahler classes and corresponding periodic scalars coming from modes of $C_3$ on the two-cycles (see \secref{s:GKZ}). This immediately implies that the classical field theory moduli space cannot possibly match all the gravity solutions. However, as argued in the next subsections, geometries with exceptional six-cycles may have non-perturbative corrections which could potentially fix this mismatch. 

\subsection{Non-perturbative superpotentials from six-cycles}

The presence of exceptional six-cycles in warped resolved Calabi-Yau backgrounds raises the very interesting possibility that Euclidean M5 branes (EM5) may wrap such cycles. Indeed, the toric geometries under consideration do not have three-cycles, either in the boundary Sasaki-Einstein manifold 
$Y$, or in the Calabi-Yau resolution $X$ of $\mathcal{C}(Y)$. Thus there are no cycles on which to wrap Euclidean M2 branes, leaving 
only EM5 branes as possible instantonic branes in these backgrounds. More precisely, such EM5 branes may be wrapped on the irreducible 
components of the exceptional divisor in the Calabi-Yau resolution. Such cycles, being complex submanifolds, are automatically 
supersymmetric. Moreover, as already mentioned, the irreducible components are in 1-1 
 correspondence with the generating homology classes in $H_{6}(X,\R)$. 

A very similar situation was considered in \cite{Witten:1996bn}, where compactifications of M-theory to three dimensions on a Calabi-Yau four-fold were discussed. In that reference the r\^ole of Euclidean instantonic M5 branes, and their possible contribution to non-perturbative superpotentials, was studied in detail. In order to generate such contributions, the number of zero modes, which includes the Goldstinos of the SUSYs broken by the brane, must be appropriate to saturate the superspace measure. In particular, such instantons must wrap cycles without infinitesimal holomorphic deformations, since the superpartners of the deformation moduli would provide additional fermionic zero modes. In \cite{Witten:1996bn} it was shown that the appropriate zero mode counting in the case of an M5 brane wrapping a divisor $D$ in this set-up requires the necessary condition that
\bea
\label{Todd}
\chi(D,\mathcal{O}_D)\equiv \sum_{i=0}^3 (-1)^i \dim H^i(D,\mathcal{O}_D)=1~.
\label{arithmetic-genus}
\eea
This is the \emph{arithmetic genus} of $D$. 
Assuming this necessary condition is satisfied, the structure of the non-perturbative superpotential generated by an EM5 brane also requires one to understand its dependence on the K\"ahler moduli. As explained in \cite{Witten:1996bn}, the dependence on these K\"ahler moduli is known exactly, being encoded entirely in the semi-classical term $\me^{-V + \ii\phi}$. Here $V$ denotes the volume of the six-cycle, while $\phi$ is the expected linear multiplet superpartner to $V$ and is given by the period of $C_6$ through $D$. 
This latter structure is determined from holomorphy of the superpotential.

It is well-known that for \emph{any} smooth compact toric manifold the arithmetic genus in (\ref{Todd}) is indeed equal to 1.
This suggests that EM5 branes would typically generate non-perturbative contributions to the superpotential in such cases. However, at this point we should recall that the situation in \cite{Witten:1996bn} is slightly different from the one at hand. Firstly, our Calabi-Yau four-fold is 
non-compact, so that gravity is decoupled from the point of view of reduction to Minkowski three-space. Secondly, 
 our set-up contains also point-like M2 branes, and moreover in the warped solutions the back-reaction of these M2 branes is also included. This leads to the asymptotically 
$AdS_4$ backgrounds discussed in section \ref{sec:FR}. It should however be possible to extend the analysis of  \cite{Witten:1996bn} to such warped cases. 

Let us first briefly discuss a similar situation in the more controlled Type IIB scenario.  In that case one can consider a Calabi-Yau three-fold singularity with colour and fractional D branes wrapping the collapsed cycles at the singularity, leading 
to a four-dimensional $\mathcal{N}=1$ SUSY field theory at the singularity.
In addition one can consider instantonic Euclidean Ep branes. The various types of strings stretching between these branes can give rise to non-perturbative contributions to the superpotential 
of the field theory at the Calabi-Yau singularity.
Exactly as for the M-theory case, in order for this non-perturbative superpotential to be generated at all the right number of zero modes must be present. An important remark here is that the Ep-Ep sector sees the full $\mathcal{N}=2$ Calabi-Yau three-fold background, thus generically leading to too many zero modes to saturate the $\mathcal{N}=1$ superspace measure. Therefore in order for a non-perturbative superpotential to be generated, some method of eliminating these extra zero modes is required. On the other hand, the situation in M-theory is very different since the colour M2 branes do not break 
any further the SUSYs of the Calabi-Yau four-fold background. From this point of view, we then expect EM5 instantons to generically contribute to non-perturbative corrections to the superpotential in three 
dimensions.

Nevertheless, the structure and interpretation of such corrections is far from clear. 
We can think of the gravitational background as a warped Calabi-Yau four-fold compactification, albeit one which is asymptotically an $AdS_4$ background. As such, we expect to be able to promote all moduli, both K\"ahler and those related to the positions of the M2 branes in the colour stack, to  spacetime fields in the Minkowski three-space. These will be dynamical fields provided the fluctuations are normalizable in the warped metric; 
at least for the K\"ahler moduli this is expected to be the case, as discussed in the Type IIB context in \cite{Martelli:2008cm}.
However, finding an ansatz for such a reduction is far from trivial, and at the time of writing there is no complete proposal for such an ansatz which would allow one 
to compute the precise form of the non-pertubative superpotential for these modes. The most recent paper on this subject is \cite{Frey:2008xw}, 
where the authors consider only the universal K\"ahler modulus in a warped compactification. 
 On the other hand, following the more controlled Type IIB case, one might expect that computing the warped volume of the Euclidean brane is the dual ``closed membrane channel'' of the picture above, described in terms of M2 branes in the blown up Calabi-Yau four-fold in the presence of the EM5 branes. In the IIB case it has been explicitly checked \cite{Baumann:2006th} in some simple situations how the computation of open string diagrams in the relevant sector \cite{Berg:2004ek, Berg:2005ja} can be reproduced though the computation of warped volumes, which are then interpreted as a resummation of such open string diagrams. Of course, it should be stressed that in the M-theory scenario at hand this can be taken only as a heuristic picture. In any case, one would expect that a general \emph{holographic} interpretation of the superpotential should also be available, since the warped 
background is asymptotically $AdS_4$. One natural suggestion is that this might come from considering the boundary behaviour of the  six-form fluctuation 
sourced by the EM5 branes. We leave a more complete investigation of these issues for further work. Instead in this paper we focus 
on computing the warped volumes of the EM5 branes as a function of the moduli. Understanding precisely how this is related to non-perturbative 
corrections in these warped resolved geometries will require further work.


\subsection{EM5 instantons in the resolved $Q^{222}$ background} \label{s:GKZ}

\begin{figure}[ht]
\begin{center}
\includegraphics[scale=.5]{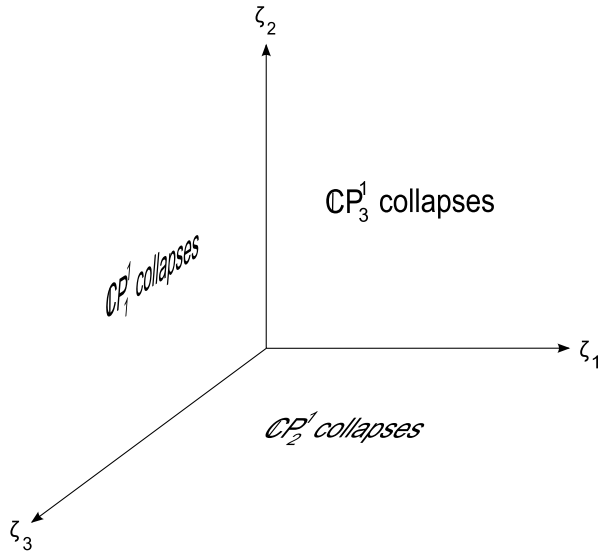}
\end{center}
\caption{The GKZ fan for $Q^{222}$ is $(\R_{\geq 0})^3$. The axes $\zeta_a$, $a=1,2,3$, may be identified with the K\"ahler classes 
of each factor in $\mathbb{CP}^1\times\mathbb{CP}^1\times\mathbb{CP}^1$, or equivalently the FI parameters in the GLSM.}
\label{fig:GKZ2}
\end{figure}

There is a unique resolution of $\mathcal{C}(Q^{222})$ where one blows up the six-cycle corresponding to the internal lattice point in 
Figure \ref{fig:toricdiagramQ222}. This gives a Calabi-Yau four-fold $X$ which is the total space of the canonical bundle 
$\mathcal{O}(-2,-2,-2)\rightarrow\mathbb{CP}^1\times\mathbb{CP}^1\times\mathbb{CP}^1$. There is a K\"ahler class 
for each factor in the zero section exceptional divisor $(\mathbb{CP}^1)^3$, leading to a GKZ fan which is 
$(\R_{\geq 0})^3$, as shown in Figure \ref{fig:GKZ2}. 
By the general theorem mentioned earlier we know that there will be a unique Ricci-flat K\"ahler metric, which is asymptotic to the 
cone metric over $Q^{222}$, for each choice of K\"ahler class. Again, in this case one can write these metrics explicitly:
\begin{eqnarray}
\diff s^2&=&\kappa(r)^{-1}\diff r^2+\kappa(r)\frac{r^2}{16}\Big(\diff \psi+\sum_{i=1}^3 \cos\theta_i \diff \phi_i\Big)^2+\frac{(2a+r^2)}{8}\Big(\diff \theta_2^2+\sin^2\theta_2 \diff \phi_2^2\Big)\nonumber \\ &&+\frac{(2b+r^2)}{8}\Big(\diff \theta_3^2+\sin^2\theta_3 \diff \phi_3^2\Big)+\frac{r^2}{8}\Big(\diff \theta_1^2+\sin^2\theta_1 \diff \phi_1^2\Big)~,
\end{eqnarray}
where now
\begin{equation}
\kappa(r)=\frac{r^8+\frac{8}{3}\, (a+b)\, r^6+8\, a\,b \,r^4-16\, k}{r^4\, (2\,a+r^2)\,(2\,b+r^2)} \ .
\end{equation}
Here $a$, $b$ and $k$ are arbitrary constants, and correspond to the choice of K\"ahler classes $\zeta_a$, $a=1,2,3$. 
In particular, setting $a=b=0$ implies that all three $\mathbb{CP}^1$s have the same volume, and the metric 
simplifies considerably. In this case it is convenient to define $r_\star^8=16k$, so that the exceptional 
divisor is at the radial position $r=r_\star$. For further details about these metrics we refer the reader to appendix \ref{sec:D}.

Again, it is also possible to solve explicitly for the warp factor for these metrics. In the simplified case 
with $a=b=0$, one can place the stack of $N$ M2 branes at an arbitrary radial position $r=r_0\geq r_\star$ and solve 
for the Green's function. Again, we refer the reader to appendix \ref{sec:D} for details of this warp factor. 
From now on we focus exclusively on the case $a=b=0$, where $\zeta_1=\zeta_2=\zeta_3$ is parametrized by the radius $r_\star>0$.

\subsubsection{Warped volumes}

We are interested in computing the warped volume of the compact exceptional divisor $(\mathbb{CP}^1)^3$ at $r=r_{\star}$, with the stack of $N$ M2 branes at the position $y=(r_0,\xi_0)$. Here $\xi_0$ denotes the point in the copy of $Q^{222}$ at radius $r=r_0$. Thus we define
\begin{equation}
S=T_5\, \int_D\, \sqrt{\det g_D}\, h\, \diff^6 x
\end{equation}
where $D$ is the exceptional divisor and $h$ is the (pull back of the) warp factor. Here the latter is given by the expression (\ref{general-warp}) 
in terms of the mode expansion discussed in appendix \ref{sec:D}. The determinant of the metric pulled back to the divisor is simply
\begin{equation}
\sqrt{\det g_D}=\frac{r_{\star}^6}{8^3}\, \sin\theta_1\, \sin\theta_2\, \sin\theta_3~.
\end{equation}
After subtituting these results into the worldvolume action one obtains
\begin{equation}
S=\frac{{T_5}\,r_{\star}^6}{8^3}\, \sum_I\,  Y_I(\xi_0)^*\,\psi_I(r_{\star})\,\int_D\, \sin\theta_1\, \sin\theta_2\, \sin\theta_3 \, Y_I(\xi)\, \diff^6 x~.
\end{equation}
Explicitly, the integral reads
\begin{eqnarray}
\int_D\, \sin\theta_1\, \sin\theta_2\, \sin\theta_3 \, Y_I(\xi)\, \diff^6 x=\mathcal{C}_{I}\,\prod_{i=1}^3  \int_{\theta_i=0}^\pi\int_{\phi_i=0}^{2\pi} \sin\theta_i\, \me^{\ii\, m_i\,\phi_i}J_{0,\, l_i,\, m_i}(\theta_i)\, \diff\theta_i\, \diff\phi_i~.
\end{eqnarray}
Here $\mathcal{C}_I$ is a normalization constant that ensures the mode $Y_I$ has unit norm.
The $\phi_i$ integrals vanish unless $m_i=0$. Then $J_{0,\,l_i,\, 0}(\theta_i)$ reduces to a Legendre polynomial ${\rm P}_{l_i}(\cos\theta_i)$, so that
\begin{equation}
\int_{\theta_i=0}^\pi \sin\theta_i\, {\rm P}_{l_i}(\cos\theta_i)\, \diff\theta_i=2\, \delta_{l_i,\,0}~.
\end{equation}
Therefore
\begin{equation}
\int_D\, \sin\theta_1\, \sin\theta_2\, \sin\theta_3 \, Y_I(\xi)\, \diff^6 x=2^6\, \pi^3\, \mathcal{C}_{0}\,\prod_{i=1}^3 \delta_{l_i,\,0}\,\delta_{m_i,\,0}~.
\end{equation}
Subtituting this back into the expression for $S$ one finds
\begin{equation}
S=\frac{2^6\, \pi^3\,{T}_5\,r_\star^6}{8^3}\, |\mathcal{C}_0|^2 \,\psi_{0,0,0}(r_{\star})~.
\end{equation}
From \eqref{warp-Q222-I}, (\ref{warp-Q222-II}) and (\ref{warp-Q222-III}) we have
\begin{equation}
\psi_{0,0,0}(r_{\star})=\frac{1}{r_0^2}\, _2F_1\Big(\frac{3}{4},\, 1,\, \frac{7}{4},\Big(\frac{r_{\star}^2}{r_0^2}\Big)^4\Big)\, , \quad\, |\mathcal{C}_0|^2=\frac{2^9\, \pi^2\, N\, \ell_p^6}{3}~.
\end{equation}
Finally, substituting the explicit M5 brane tension results in the warped volume
\begin{equation}
S=\frac{2\,N}{3}\,\frac{r_{\star}^6}{r_0^6}\, _2F_1\Big(\frac{3}{4},\, 1,\, \frac{7}{4},\Big(\frac{r_{\star}^2}{r_0^2}\Big)^4\Big)
\label{Q222-action}~.
\end{equation}
Note that for $r_0 \rightarrow r_{\star}$ we get $S\sim -\frac{N}{2}\,\log(r_0-r_{\star})=-\log\,\rho^N$, where for the last equality we used (\ref{gamma-equ}) and (\ref{gamma-equ-II}).
\subsubsection{The $L^2$ harmonic two-form}

The key observation in this subsection is that the warped volume (\ref{Q222-action}) of the exceptional divisor 
is closely related to the $L^2$ normalizable two-form $\beta$ which is Poincar\'e dual to the six-cycle. 
The claim is that the precise relation between the two is (for $N=1$)
\bea
\beta=\frac{1}{\pi \, \ii}\,\partial \bar{\partial} S~.
\label{beta-S}
\eea
Here the derivatives are regarded  as acting on the coordinates of the point $y=(r_0,\xi_0)$, which recall is the location of the stack 
of M2 branes. 
We shall prove this claim in full generality in \secref{s:general-V}. Here we first prove it in the current explicit example, where it 
is convenient to use the coordinate system in section \ref{coordinates}. From \eqref{gamma-equ} we see that $r_0^8=r_{\star}^8+\rho^4$, so in the new coordinate system we can write (setting $N=1$)
\begin{equation}
S=\frac{2}{3}\,\frac{r_{\star}^6}{(\rho^4+r_{\star}^8)^{\frac{3}{4}}}\, _2F_1\Big(\frac{3}{4},\, 1,\, \frac{7}{4},\frac{r_{\star}^8}{\rho^4+r_{\star}^8}\Big)~.
\end{equation}
After some algebra it can be shown that 
\begin{equation}
(\bar{\partial}-\partial)S=\ii \,\frac{r_{\star}^6}{(\rho^4+r_{\star}^8)^{{3}/{4}}}\,g_5~,
\end{equation}
where
\bea
g_5=\dd\psi+\sum_{i=1}^3\cos\theta_i\, \diff\phi_i~,
\eea
and therefore (\ref{beta-S}) reads
\begin{equation}
\beta=\partial \bar{\partial}\,\left(\frac{1}{\pi \, \ii}\,S\right)=\frac{1}{2}\dd(\bar{\partial}-\partial)\left(\frac{1}{\pi \, \ii}\,S\right)=\frac{1}{2 \pi}\dd\left(\frac{r_{\star}^6}{(\rho^4+r_{\star}^8)^{{3}/{4}}}\,g_5\right)~.
\end{equation}
Going back to the original coordinates we then have 
\begin{equation}
\beta=\frac{1}{2 \pi}\,\dd\left(\frac{r_{\star}^6}{r_0^6}\,g_5\right)~.
\end{equation}
This is easily checked to be a harmonic two-form with respect to the unwarped metric, and also $L^2$ with respect to this metric. To see that $\beta$ is indeed Poincare dual to the six-cycle, we choose the following closed form on $X=\mathcal{O}(-2,-2,-2)\rightarrow (\mathbb{CP}^1)^3$
\begin{equation}
\mu = \frac{\sin\,\theta_1\,\sin\,\theta_2\,\sin\,\theta_3}{2^6 \, \pi^3} \dd \theta_1 \wedge \dd \phi_1 \wedge \dd \theta_2 \wedge \dd \phi_2 \wedge \dd \theta_3 \wedge \dd \phi_3~.
\end{equation}
It is easy to see that 
\begin{equation}
\int_D\,\mu = 1 \, , \quad \, \int_X\,\beta\,\wedge\,\mu = 1~.
\end{equation}
Hence $\beta$ is $L^2$ normalizable and Poincar\'e dual to the divisor, as claimed.

\subsubsection{Critical points}

Formally, the superpotential that is induced by the instanton action that we have calculated in the previous subsections is given by
\bea\label{supQ222}
W=\me^{-S}=\exp\left[\frac{2\,N}{3}\,\frac{r_{\star}^6}{r_0^6}\, _2F_1\Big(\frac{3}{4},\, 1,\, \frac{7}{4},\Big(\frac{r_{\star}}{r_0}\Big)^8\Big)\right]~,
\eea
where $S$ is given in (\ref{Q222-action}). This is something of a formal statement, since in reality what we have computed is the 
on-shell Euclidean action of the wrapped M5 brane as a function of the moduli of the SUGRA background. Here essentially $r_\star$ is a K\"ahler 
modulus, while $r_0$ is a modulus associated to the position of the stack of M2 branes. On the other hand, the superpotential 
should be a function of the corresponding spacetime fields in Minkowski space, obtained by promoting these moduli to dynamical fields. 
In the unwarped case there is essentially no distinction between the two, but in the warped case it is not known how to do this at present, and the situation is much less clear. Note also that we have only computed the real part of $S$, and hence absolute value of $W$. Thus the best we can do is to examine the critical points of $\me^{-S}$ interpreted directly as a 
superpotential on the SUGRA moduli space. 
It is straightforward to compute
\bea
\partial_{r_0}\,S=\frac{4\,N\,r_{\star}^6\,r_0}{r_{\star}^8-r_0^8}~.
\eea
For $r_{\star}>0$ there are no critical points of $S$. In order for $\diff W=0$ we must then necessarily have $r_0=r_{\star}$, which gives $S=+\infty$ and $W=0$ on this locus. In this case the branes move only on the six-cycle. Clearly, this is always a solution since the ``superpotential'' (\ref{supQ222}) is identically zero if the branes are moving on the divisor $D$. For $r_{\star}=0$ there are formally no contributions of instantons to the superpotential. However, notice this is a singular limit of the SUGRA solution in which the six-cycle is blown down. 
In the absence of instantons of the course M2 branes are free to propagate on the cone $\mathcal{C}(Q^{222})$ in which the six-cycle is blown down. 

In \secref{sec:EH} we have performed the same computation of the warped volume in the much simpler toy example of the Eguchi-Hanson manifold. Interestingly, the same qualitative results on the behaviour of the warped volume (basically the non-perturbative superpotential) are obtained.

\subsection{EM5 brane instantons: general discussion} \label{s:general-V}

In this section we describe how the above calculations for $Q^{222}$ generalize to more general Calabi-Yaus. 

\subsubsection{Geometric set-up}

Throughout this section we assume we are given an asymptotically conical K\"ahler manifold $X$ of complex dimension $n$, with metric $g=g_X$ and K\"ahler two-form $\omega$. This means that the manifold $X$ is non-compact, and that the metric $g$ asymptotically approaches the cone metric $\diff r^2 + r^2 g_Y$,
where $Y=\partial X$ is the compact base of the cone. The metric $g_Y$ is then Sasakian, by definition. 
In fact this is slightly too general for the situation we are interested in. More precisely, we want $X$ to be a \emph{resolution} of the singularity at $r=0$ of the cone $\mathcal{C}(Y)\cong (\R_{\geq 0}\times Y)\cup \{r=0\}$. This means that there is a proper birational map $\pi:X\rightarrow \mathcal{C}(Y)$ which is a biholomorphism on the restriction $X\setminus E\rightarrow \{r>0\}\subset \mathcal{C}(Y)$. Less formally, the resolution $X$ replaces the singular point $\{r=0\}$ of the cone by the \emph{exceptional set} $E=\pi^{-1}(r=0)$. 
For physical applications we require the metric $g$ on $X$ to be Ricci-flat and hence Calabi-Yau. Then by definition $X$ above is a \emph{crepant} resolution of the cone singularity. In fact this won't really affect the computations that follow.

Of particular interest for us in this section are the exceptional divisors in 
$X$. These are the irreducible (prime) divisors of $E=\pi^{-1}(r=0)$. Call these irreducible components $E_i$. Since $X$ is contractable onto $E$ we have
\bea
H_{2n-2}(X,\R)\cong H_{2n-2}(E,\R)\cong \bigoplus_i H_{2n-2}(E_i,\R)\cong \R^{b_{2n-2}(X)}~,
\eea
so that $i=1,\ldots,b_{2n-2}(X)=\dim H_{2n-2}(X,\R)$. Thus the 
exceptional divisors $E_i$ generate the homology of $X$ in codimension two. Notice that if $b_{2n-2}(X)=0$ then the resolution has no exceptional divisors and the resolution is said to be \emph{small}. For example, the resolved conifold is a small resolution of the cone over $T^{1,1}$, since the exceptional set is $E\cong \mathbb{CP}^1$; similarly, the resolutions of $Q^{111}$ considered in \secref{sec:4} are small.

\subsubsection{$L^2$ harmonic two-forms}

Another key result for us is that in the above situation
\bea
\mathcal{H}^2_{L^2}(X,g)\cong H^2(X,Y,\R)\cong H_{2n-2}(X,\R)~,
\eea
for $n>2$.
Here $\mathcal{H}^2_{L^2}(X,g)$ denotes the $L^2$ normalizable harmonic two-forms on $X$ (which thus depends on the metric $g$). 
That is, the codimension two cycles in 
$X$ are 1-1 with the prime exceptional divisors, and these are also 1-1 with the $L^2$ normalizable harmonic two-forms on $X$, as long as $n>2$. This result about $L^2$ harmonic forms holds in general for complete asymptotically conical manifolds, and was proven in 
\cite{hausel}. In dimension $n=2$, instead $\mathcal{H}^2_{L^2}(X) \cong \mathrm{Im}\left[H^2(X,Y,\R)\rightarrow 
H^2(X,\R)\right]$. For example, for the  Eguchi-Hanson manifold the map $H^2(X,Y,\Z)\rightarrow H^2(X,\Z)$ is multiplication by 2, so that 
there is a unique $L^2$ harmonic two-form, up to scale.

We shall need some more information about these harmonic forms. 
First, we note that the harmonic two-forms are of Hodge type $(1,1)$. This is because they are Poincar\'e dual to divisors in $X$, so if there 
was a $(0,2)$ part of the harmonic form it would be cohomologically trivial and hence\footnote{Notice this result definitely fails for spaces that are not asymptotically conical. A good example is the Taub-NUT space, which has no two-cycles but does have an $L^2$ harmonic two-form.} identically zero by the result of \cite{hausel}.

Pick a particular $D=E_i$ and normalize the associated $L^2$ harmonic two-form so that it is Poincar\'e dual to $D$. 
We may then 
think of the harmonic form as the curvature of a Hermitian line bundle $\mathcal{L}=\mathcal{L}_D$ -- the divisor bundle for 
$D$. If $s$ is a \emph{local} nowhere zero holomorphic section of $\mathcal{L}$ over an open set $U\subset X$, and $H$ is the Hermitian metric 
on $\mathcal{L}$, then we may write the harmonic form as
\bea\label{betaform}
\beta\mid_U\, = \frac{1}{2\pi\ii}\, \partial\bar{\partial} \log H(s,s)~.
\eea
This is a standard result. Notice that $\beta$ here does not depend on the choice of local holomorphic section $s$. 
We will be interested in the special case where we take $U$ as large as possible, which is $U = X\setminus D$. 
By construction, the line bundle $\mathcal{L}_D$ is trivial over $U$, with trivializing nowhere zero holomorphic section $s$. 
We pick an $s$, and write $H=H(s,s)$ for the corresponding real function on $X\setminus D$.

We next note that $\log H$ is itself a harmonic function on $X\setminus D$. 
To see this, suppose generally we have a harmonic $(1,1)$-form $\beta$. Because $(X,g)$ is complete, this means 
$\beta$ is both closed and co-closed. The co-closed condition involves computing the Hodge dual of $\beta$, which is
\bea
\star\beta = -\beta\wedge \omega^{n-2} + (\omega\lrcorner\beta)\omega^{n-1}~.
\eea
Here we have simply used that $\vol = \omega^n/n!$ and that $\beta$ is type $(1,1)$. Thus 
$\beta^{ij}\omega_{im}\omega_{jn}=\beta_{mn}$. By definition, $\omega\lrcorner\beta=(1/2!)\omega_{ij}\beta^{ij}$. 
Thus if $\beta$ is closed, then $\star\beta$ is co-closed if and only if $\omega\lrcorner\beta$ is closed, {\it i.e.} 
constant. 

Thus we learn that for a harmonic $(1,1)$-form $\beta$, $\omega\lrcorner\beta$ is in fact constant. 
Now, for an $L^2$ form on an asymptotically conical manifold this constant is in fact necessarily zero. 
To see why, we must look at the asymptotics of $\beta$. This was studied in appendix A of 
\cite{Martelli:2008cm}. Here we have a two-form, so $p=2$ in Table 4 of that reference, and we are interested in 
$L^2_{\infty}$, so that the form is normalizable at infinity. For $n>2$, $p<n$ so the normalizable modes 
are of type II and type III$^-$. However, the type II modes are constructed from harmonic one-forms on the 
base $Y$, and there are not any of these as $(Y,g_Y)$ is a positive curvature Einstein manifold so $b_1(Y)=0$ by Myers' theorem.
Thus asymptotically, the two-form $\beta$ 
is to leading order of the form
\bea
\beta \sim \diff (r^{2-n-\nu}\beta_\mu)~,
\eea
where $\beta_\mu$ is a massive co-closed one-form along $Y$
\bea
\Delta_Y \beta_\mu = \mu\beta~,
\eea
and 
\bea
\nu = \sqrt{(n-2)^2+\mu}~.
\eea
Consider now $\omega\lrcorner\beta$. Since asymptotically $\omega\sim (1/2)\diff (r^2\eta)$, where $\eta=\ii (\bar\partial-\partial)\log r$ is the contact one-form of the Sasakian manifold $(Y,g_Y)$, we have
\bea
\omega\lrcorner\beta \sim r^{-n-\nu}\left[(2-n-\nu)\beta_\mu\lrcorner\eta + \frac{1}{2}\diff\beta_\mu\lrcorner\diff\eta\right]~.
\eea
Since $\nu>0$ it follows that $\omega\lrcorner\beta\rightarrow 0$ at infinity. Since we have also shown that $\omega\lrcorner\beta$ is constant, it follows that 
this constant is zero.
If we write 
$\beta$ as in (\ref{betaform}) on $X\setminus D$, then $\omega\lrcorner\beta=0$ is equivalent to saying that $\log H$ is harmonic. This just follows from the form 
of the scalar Laplacian on a K\"ahler manifold: $\Delta f = -2\omega\lrcorner\ii\partial\bar{\partial} f$. 

As an aside comment that will be important below, the first non-zero eigenvalue $\mu$ is bounded below by $4(n-1)$. In fact, this is saturated precisely by \emph{Killing} one-forms -- 
see, for example, \cite{DNP}. Thus $\mu\geq 4(n-1)$ and correspondingly $\nu \geq \sqrt{(n-2)^2+4(n-1)} = n$.

We thus conclude that $\log H$ is a harmonic function. It will be crucial in what follows that $\log H$ is not in fact defined everywhere on $X$. 
By construction, it is defined only on $X\setminus D$. Along $D$ in fact $\log H$ is singular. This is simply because $D$ is the zero set of $s$, which has a simple zero along $D$ by assumption. Thus if $z=\rho\, \me^{\ii\theta}$ is a local complex coordinate normal to $D$, with $D$ at $\rho=0$, then $\log H$ blows up near to $D$ like $\log \rho^2 = 2 \log \rho$. 

\subsubsection{The instanton action}

An instantonic brane wrapped on an exceptional divisor $D=E_i$ is calibrated and supersymmetric -- for example, a D3 brane 
for $n=3$ in Type IIB string theory or an M5 brane for $n=4$ in M-theory. These are in 1-1 correspondence with the homology classes $H_{2n-2}(X,\R)$, and moreover there is a unique $L^2$ harmonic two-form associated to each irreducible exceptional divisor, which is Poincar\'e dual to the divisor, as discussed above. 

Let $G[y](x)$ denote the Green's function on $X$, with a fixed (Ricci-flat) K\"ahler metric, normalized so that 
\bea
\Delta_x G[y](x) = 2\pi \frac{1}{\sqrt{\det g_X}}\, \delta^{2n}(x-y)~. 
\eea
Consider the on-shell action of an instantonic brane, given by the following Green's function weighted volume of $D$:
\bea
V = \int_D G[y]\sqrt{\det g_D}\, \diff^{2n-2}x = \int_D G[y]\frac{\omega^{n-1}}{(n-1)!}~.
\eea
This is the relevant formula both for D3 branes and M5 branes, where the warp factor is $h= NG/T$, with $N$ the number of 
spacetime filling branes and $T$ the tension of the wrapped instanton brane.
The warped volume $V$ depends on the source point $y\in X$, so $V=V(y)$. Of course, it also depends on the choice of K\"ahler metric. If we consider the Calabi-Yau case, we know that there is a unique metric in each K\"ahler class, so we may think of $V$ also as a function of the K\"ahler class: $V=V(y;[\omega])$. Then we claim that
\bea
V(y) = -\frac{1}{2}\log H(y)~.
\eea
Of course, $\log H$ depends implicitly on the K\"ahler metric since the associated harmonic form does also. 
Notice this result actually provides a \emph{formula} for the $L^2$ harmonic two-forms on an asymptotically conical 
K\"ahler manifold, in terms of the Green's function on $X$. This is a new mathematical result, as far as we are aware.\footnote{We thank Tamas Hausel 
for discussions on this.}

The strategy for proving this claim involves three steps: (i) show that $V(y)$ is a harmonic function on $X\setminus D$, (ii) show that $V$ diverges as $-\log \rho$ along $D$, (iii) show that the two-form $\ii\partial\bar{\partial} V$ is $L^2$. These steps show that the latter two-form is an $L^2$ harmonic form that is Poincar\'e dual to $D$. We may then appeal to the uniqueness of such a form.

Step (i). This is straightforward. We want 
to compute $\Delta_y$ acting on $V$. Using that 
the Green's function $G[y](x)$ is symmetric in its arguments, so $G[y](x)=G[x](y)$, then 
provided $y\notin D$ 
\bea
\Delta_y V= \int_{D}(\Delta_y G[y](x))\sqrt{\det g_D}\, \diff^{2n-2}x = 0~.
\eea
The last step follows since $y\notin D$.
This shows that $V(y)$, interpreted as a function on $X\setminus D$, is indeed harmonic. 

Step (ii). Near to the source point $y$ we have
\bea\label{Gnearsource}
G[y](x) = \frac{2\pi}{(2n-2)\vol(S^{2n-1})\rho^{2n-2}} (1+{o}(1))~,
\eea
where $\rho$ denotes geodesic distance from $y$, where we regard the latter as fixed. Here $\vol (S^{2n-1})$ is the volume of a unit $(2n-1)$-sphere, which appears 
in this computation as a small sphere around the point $y$. 
The divergence is a local question, so this is really a question in Euclidean space. Let us take the local model $\C^{n-1}\times \C$, with complex coordinate $z$ on $\C$. We suppose that the divisor $D$ is locally $z=0$ in this patch, and that the point $y$ is $(0,z)$ in this coordinate system. The integral over $D$ is finite outside a ball $B_\delta$ of fixed radius $\delta>0$ in $\C^{n-1}\times\{0\}$, so we would like to analyse the divergence in the integral
\bea
\frac{2\pi}{(2n-2)\vol(S^{2n-1})}\int_{B_\delta} \frac{1}{(R^2 + |z|^2)^{n-1}}R^{2n-3}\diff R \, \diff\mathrm{Vol}(S^{2n-3})~,
\eea
as $|z|\rightarrow 0$. This is easily seen to be 
\bea
\left(-\frac{1}{2}\log |z|^2\right) \cdot \frac{2\pi\vol(S^{2n-3})}{(2n-2)\vol(S^{2n-1})}~.
\eea
We next need the ratio of volumes of spheres:
\bea
\vol(S^{2n-1})=\frac{2\pi^n}{(n-1)!}~,
\eea
so that the divergence is
\bea
\left(-\frac{1}{2}\log |z^2|\right)\cdot \frac{2\pi \cdot 2\pi^{n-1}(n-1)!}{(2n-2)\cdot 2\pi^n (n-2)!} = -\frac{1}{2}\log |z|^2 = - \log |z|~.
\eea
This shows that $V$ diverges as $-\log |z|=-\log \rho$ near to $D$, as claimed.

Step (iii). We consider the two-form $\diff \diff^c V(y)$ as the point $y$ tends to infinity. Since we have already shown this two-form is harmonic, it will have an asymptotic expansion as in Appendix A of \cite{Martelli:2008cm}. There are three types of modes, I, II and III. 
As already mentioned, there are in fact no modes of type II since $Y$ has no harmonic one-forms. The modes of type I are pull-backs of harmonic two-forms on $Y$, which are not $L^2$. The modes of type III$^{\pm}$ are of the form
\bea
\diff\left(r^{2-n\pm\nu}\beta_\mu\right)~,
\eea
with $\nu=\sqrt{(n-2)^2+\mu^2}$ as above. The III$^{-}$ modes are $L^2$, while III$^{+}$ are not. Thus we must show that $\diff\diff^c V$ has leading term of type III$^{-}$, so that it is normalizable at infinity and hence normalizable.

Since $V$ is globally defined on $X\setminus D$, we immediately see that there can be no mode of type I since near infinity $\diff\diff^c V$ is exact. Thus we are reduced to analysing the asymptotic $r$-dependence of the one-form $\diff^c V$. 

If we regard the point $x$ as fixed, then as $y$ tends to infinity we have
\bea
G[y](x) = \frac{2\pi}{(2n-2)\vol(Y)r^{2n-2}} (1+o(1)).
\eea
Then
\bea
\beta = \frac{1}{2\pi}\diff\diff^c \int_D G[y]\frac{\omega^{n-1}}{(n-1)!}~,
\eea
gives to leading order
\bea\label{asym}
\beta \sim -\frac{\mathrm{vol}(D)}{\mathrm{vol}(Y)}\diff\left(r^{2-2n}\eta\right)~.
\eea
We thus conclude that 
\bea
2-n\pm\nu = 2-2n~.
\eea
From the comment above, this means that we indeed have a normalizable mode III$^{-}$, and moveover
that $\nu=n$ and hence $\mu=4(n-1)$ saturates the lower bound on the eigenvalue. The Killing one-form $\eta$ is of course dual to the R-symmetry.

This completes our proof.

As a final check on the last asymptotic formula, we can compare with the Eguchi-Hanson result (\ref{EHasym}). These indeed agree, noting that $\mathrm{vol}(D)=\pi c^2$ and $\vol(Y)=\pi^2$ in this case.

\subsubsection{The superpotential} \label{s:W}

To conclude our results thus far, we have proven in general that
\bea
\me^{-V}(y,[\omega]) = \sqrt{H(y)}~,
\eea
where $(1/2\pi \ii) \partial\bar{\partial} \log H$ is the unique $L^2$ harmonic two-form that is Poincar\'e dual to the divisor $D$ wrapped by the 
instanton. The on-shell action of this instanton is $V$. Notice that $\sqrt{H}$ has a simple zero along $D$, as expected on general grounds. In fact locally $H=H(s,s)=\me^{2g}|s|^2$, where $g$ is function and $s$ is a holomorphic section of the divisor bundle $\mathcal{L}_D$. Again, this was expected from arguments in \cite{Ganor:1996pe}, where the phase that pairs with $V$, coming from the Wess-Zumino term in the action, was studied. Thus the result 
presented here is rather complimentary to the discussion in reference \cite{Ganor:1996pe}.
Restoring the factor of $N$, our computation hence shows that, formally at least, we have the superpotential
\bea
W = \me^{-NV}(y;[\omega]) = \sqrt{H^N} = \me^{Ng} |s|^N~.
\eea
This is interpreted as a function of both the K\"ahler class, and also the position of the stack of M2 branes $y\in X$, and generalizes 
the result (\ref{supQ222}) we derived explicitly for $Q^{222}$.
A critical point of this $W$  requires either a critical point of $V$,  or else $V=\infty$. Since $V=-\frac{1}{2}\log H$, the first case requires a critical point of the harmonic function $\log H$. By the maximum principle, notice that such a critical point cannot be either a local maximum or a local minimum of $\log H$. 

\section{Conclusions}\label{sec:6}

In this paper we  set out to study abelian symmetries in the context of the $AdS_4/CFT_3$ correspondence. In particular, we considered gauge fields in $AdS_4$ arising from KK reduction of the SUGRA potentials over the $b_2(Y)$ topologically non-trivial cycles (sometimes called Betti multiplets in the literature). In contrast to its better-understood $AdS_5/CFT_4$ relative, the case at hand displays many more subtleties. The key difference resides in the fact that gauge fields in $AdS_4$ admit, in a consistent manner, quantizations with either of two possible fall-offs at the boundary, implying that the gauge field can be dual to either a global symmetry or to a dynamical gauge field in the boundary CFT. From the bulk perspective, electric-magnetic duality  in the four-dimensional electromagnetic theory in $AdS_4$ amounts to exchanging these two boundary fall-offs. In addition, from the bulk perspective one can shift the $\theta$-angle by $2\pi$. Following \cite{Witten:2003ya}, these two actions translate into particular operations on the boundary theory, the $\mathcal{T}$ and $\mathcal{S}$ operations reviewed in \secref{sec:baryons}, which then generate the group $SL(2,\mathbb{Z})$. As stressed in the main text, these actions exchange different boundary conditions for the gauge field in $AdS_4$. Correspondingly, the dual boundary CFTs are different. Indeed, the whole of $SL(2,\mathbb{Z})$ acts on the boundary conditions for the bulk gauge fields, leading in general to an infinite orbit of CFTs for each $U(1)$ gauge symmetry in $AdS_4$. Understanding the structure of such orbits is a very interesting problem which we postpone for further work. 

In this paper we have contented ourselves with studying the particular case of M2 branes moving in $\mathcal{C}(Q^{111})$.\footnote{Although as described in \secref{sec:gen}, and further elaborated in \secref{sec:classification}, we expect similar results to hold for other toric isolated four-fold singularities with no exceptional six-cycles.} In \cite{Franco:2008um, Franco:2009sp} a $U(N)^4$ dual theory was proposed and further studied. In \secref{sec:Q111} we proposed a choice of quantization for the abelian vector fields in the Betti multiplets  such that precisely two $U(1)$s are ungauged, leading to the gauge group  $U(1)^2\times SU(N)^4$. This leaves precisely two global symmetries that may be identified with the two gauge fields coming from KK reduction of the SUGRA six-form potential over five-cycles in $Q^{111}$. A key point in that identification is that the corresponding boundary conditions in the bulk $AdS_4$ allow for electric wrapped M5 brane states. These M5 branes can be easily identified in terms of the toric geometry of the variety. Since the field theory realizes the minimal GLSM, it is then straightforward to identify the relevant $U(1)$ symmetries in the QCS theory. In turn, this allows one to construct dual baryonic operators to such M5 branes. It is then natural to consider the spontaneous breaking of such symmetries, where the operator dual to such an M5 brane acquires a VEV.
We analyzed in detail such SSB in \secref{sec:4}. In particular, we have been able to compute the VEV of the baryonic condensate, with precise agreement with field theory expectations. We stress that this is a non-trivial check of the dual theory, as this suggests that it does admit an IR superconformal fixed point with the correct properties (specifically, R-charges) to be dual to M2 branes moving in $\mathcal{C}(Q^{111})$. Along the lines of \cite{Klebanov:2007us, Klebanov:2007cx}, we have also been able to identify the Goldstone boson of this SSB.  However, a comprehensive understanding of these resolutions in the context of the actions on the boundary conditions is still lacking. We postpone this for further work. An interesting by-product of our computation is the finding of general expressions for warped volumes in Calabi-Yau backgrounds, which are potentially of interest for other, similar computations.

It is natural to extend our analysis and consider backgrounds with exceptional six-cycles, as we briefly considered in \secref{sec:5}.  
Upon resolution, Euclidean M5 branes can be wrapped on these exceptional divisors. As opposed to the Type IIB counterpart case for four-cycles, the  M2 branes sourcing the background do not break SUSY any futher than that preserved by the Euclidean brane in the resolution of the cone. Thus, in very much the same spirit as in \cite{Witten:1996bn}, it is natural to expect that these Euclidean branes contribute as non-perturbative effects to the superpotential, even in the warped case. Nevertheless, a comprehensive understanding of these issues is lacking. In this paper we have however taken some first steps towards understanding this by computing the warped volume of such branes. In particular we studied in detail the example $\mathcal{C}(Q^{222})$, which is a certain $\mathbb{Z}_2$ orbifold of $\mathcal{C}(Q^{111})$, as well as the Eguchi-Hanson manifold. In extending our findings to more general geometries we have found expressions which might be of relevance in other contexts.

It is fair to say that the $AdS_4/CFT_3$ correspondence still hides many mysteries. In this paper we have scratched the surface of a few of them. We hope to be able to report on further progress in the near future.


\subsection*{Acknowledgments}
\noindent N.~B.~would like to thank the ISEF for their support; this work was completed with the support of a University of Oxford Clarendon Fund Scholarship.
D. R-G. acknowledges financial support from the European Commission through Marie Curie OIF grant contract No. MOIF-CT-2006-38381, Spanish Ministry of Science through the research grant No. FPA2009-07122 and Spanish Consolider-Ingenio 2010 Programme CPAN (CSD2007-00042).
J. F. S. is supported by a Royal Society University Research Fellowship. 

\begin{appendix}

\section{Isolated toric Calabi-Yau four-fold singularities with no vanishing six-cycles} \label{sec:classification}

In \cite{Benishti:2009ky} it was shown that the toric diagram of an isolated  Calabi-Yau four-fold singularity should satisfy the following conditions:
\begin{enumerate}
 \item All the faces of the polytope should be triangular.
 \item No lattice point should appear on faces or edges of the polytope.
\end{enumerate}
Recall that, as explained in \secref{s:bc}, the $v_{\alpha}$ vectors define an affine toric Calabi-Yau four-fold. This definition is unique up to a unimodular transformation $\mathcal{R}$, where $\mathcal{R} \in GL(4,\Z)$ and $\det \mathcal{R}=\pm 1$. The vectors $v_{\alpha}$ may be written as $v_{\alpha} = (1,w_{\alpha})$ for an appropriate choice of basis, where $w_{\alpha} \in \Z^3$ are the vertices of the three-dimensional toric diagram. 
We will be interested in singularities with no vanishing six-cycles. We will therefore demand, in addition, that no lattice points appear inside the polytope. These toric diagrams are known as \textit{lattice-free} polytopes. Such lattice-free polytopes in three dimensions are characterized by the fact that they have \textit{width} one (see for example \cite{kantor-1997} and references therein). This is sometimes referred to as Howe's theorem, and is translated into the fact that the vertices of any lattice-free polytope are sitting in adjacent planes, {\it i.e} two lattice planes with no lattice points inbetween. These planes can be chosen to be $\{z=0\}$ and $\{z=1\}$.\footnote{If there are more vertices in one plane we choose it to be the $\{z=0\}$ plane without loss of generality.} We want to start by showing that any three-dimensional lattice-free polytope with more than four vertices describes a cone over a seven-dimensional simply-connected Sasakian manifold.

From \cite{Lerman2} we learn that the first and second homotopy groups of a toric Sasakian manifold $Y$ can be read straightforwardly from the toric diagram. The results for Calabi-Yau four-folds are
\bea
\pi_1(Y)\cong \Z^4/L \quad , \quad \pi_2(Y)\cong\Z^{d-4} \ ,
\label{homotopy}
\eea
where $L=\mathrm{span}_\Z\{v_\alpha\}$ is the span over $\Z$ of the space of $d$ external vertices of the toric diagram. According to the Hurewicz Theorem $H_2(Y)\cong\pi_2(Y)$ whenever $H_1(Y)\cong\pi_1(Y)/[\pi_1(Y),\pi_1(Y)]$ is trivial. Therefore we see from \eqref{homotopy} that for simply-connected Sasakian manifolds $b_2(Y)=d-4$, which is also the number of gauge groups in the minimal GLSM describing this geometry. This  immediately suggests that the number of gauge nodes of the corresponding field theory, in the case that the latter is identified with the minimal GLSM, is $b_2(Y)+2$. The two additional gauge nodes correspond to the $U(1)$s which are not quotiented by in forming the moduli space. Note from \eqref{homotopy} that $Y$ is simply-connected if and only if the external vertices span $\Z^4$. 

For polytopes with more than four vertices, three of the vertices must be co-planar. Thus the matrix that describes four of the vertices can be written as follows 
\bea
A=
\left(
\begin{array}{cccc}
 1 & 1 & 1 & 1  \\
 0 & x_1 & x_2 & x_3 \\
 0 & y_1 & y_2 & y_3 \\ 
 0 & 0 & 0 & 1
\end{array}\right) \ ,
\eea
where each column corresponds to a vertex (the $x$, $y$ and $z$ coordinates correspond to the 2nd, 3rd and 4th rows, respectively). It is easy to see that by an $\mathcal{R}$ transformation this can be brought into the form
\bea
B=
\left(
\begin{array}{cccc}
 1 & 1 & 1 & 1  \\
 0 & 1 & 0 & 0 \\
 0 & 0 & 1 & 0 \\ 
 0 & 0 & 0 & 1
\end{array}\right) \ .
\eea
To see that the matrices $A$ and $B$ are related by an $\mathcal{R}$ transformation first note that $|\det (A^{-1}\,B)|=1/|(x_1\,y_2-x_2\,y_1)|$. The denominator of the latter is just the area of the parallelogram made up of two identical triangles defined by the first three columns in $A$. The area of this triangle is $1/2$, as this is the condition for a lattice-free triangle in two dimensions.

The four vertices described by $B$ span $\Z^4$. Therefore any three-dimensional lattice-free polytope with more than four vertices corresponds to a simply-connected Sasakian  seven-manifold. This not always true for diagrams with four vertices, since in this case each plane can contain two points. 

To start our analysis we note that lattice-free polytopes with 4 vertices correspond to a type of orbifold singularity that have been discussed intensively in the literature (see {\it e.g.} Section 3.1 in \cite{Morrison:1998cs}). This is a supersymmetric $\C^4/\Z_k$ orbifold corresponding to an isolated singularity that cannot be resolved, with the orbifold weights in this case being $(1,-1,q,-q)$ with $\gcd \{k,q\}=1$. As already noted in \cite{Jafferis:2009th}, if $q>1$, for any choice of $U(1)$ isometry to reduce on, one can show that $Y$ reduces in Type IIA to a space with orbifold singularities. Therefore, it seems that these $AdS_4 \times Y$ solutions are dual to field theories with no Lagrangian description. Thus we are left with the ABJM orbifolds, obtained by taking $q=1$, for which there are of course already field theory candidates.

We continue with the classification of diagrams with five vertices, where recall that we have shown that the first four vertices are described by $B$. Since the fifth vertex should be in the $\{z=1\}$ plane (to prevent a face with 4 vertices) it can be written as $(1,x,y,1)$ with $x,y>0$.\footnote{As can be easily seen from (\ref{Qt-matrix}), toric diagrams with triangular faces obtained by picking other values of $x$ and $y$ are related by $\mathcal{R}$ transformations.} The only way to break the lattice-free condition would be if there were points between $(1,x,y,1)$ and $(1,0,0,1)$. Thus we have to require $\gcd\{x,y\}=1$. This concludes the classification of polytopes with five vertices. There are no additional $\mathcal{R}$ transformations that connect between diagrams in this set; as we show later, the corresponding GLSM charge matrix that describes this toric diagram is unique for any choice of $x$ and $y$. 

Toric diagrams with 6 vertices are potentially more complicated, although we note that 
these include the $Q^{111}$ example studied in detail in this paper. 
6 vertices is also the maximal number since, otherwise, it is not possible to arrange the vertices in two adjacent planes with the constraint that all faces are triangular. 

\begin{figure}[ht]
\begin{center}
\includegraphics[scale=.8]{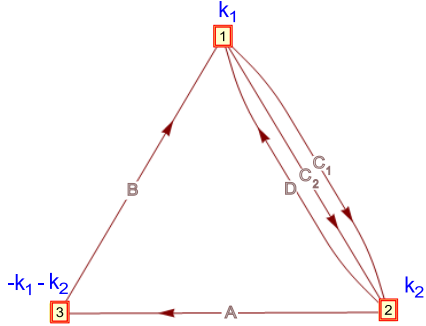}
\end{center}
\caption{The unique quiver that contains 3 nodes, 5 fields and no adjoints.}
\label{General_quiver}
\end{figure}

We continue now with a discussion of the geometries that correspond to the polytopes obtained above. Recall that, given a toric diagram, one can recover the corresponding Calabi-Yau four-fold via Delzant's construction. In physics terms, this would be called a GLSM description of the four-fold. Let us discuss the toric diagrams with five vertices described above. The GLSM charge matrix can be computed by taking the null-space of the $G$-matrix, obtaining
\bea
\label{Qt-matrix}
Q_t=\left(x+y , -x , -y , -1 , 1 \right) \ .
\eea
Since the GLSM charge matrix contains one gauge group, we find that the corresponding quiver should have 3 nodes. However, it is not possible to find a QCS field theory for every value of $x$ and $y$. First, note that there are no zero entries in $Q_t$, therefore there should be no adjoint fields in the quiver. The most general way to construct a quiver with 3 nodes and 5 fields with no adjoints, such that there is an equal number of in-going and out-going arrows at each node, is given in Figure~\ref{General_quiver}.
This quiver was also discussed in \cite{Hanany:2008gx}. Since we are interested in field theories which reproduce the minimal GLSM, we must have a toric superpotential which vanishes in the abelian case. The natural candidate is
\begin{equation}
 W = \mathrm{Tr}\, \epsilon^{ij} C_iDC_jAB~,
 \label{WW}
\end{equation}
where $\epsilon^{ij}$ is the usual alternating symbol.

In this theory the only contribution to the GLSM matrix comes from the D-term, which in this case reduces to 
\bea
Q_{\mathrm{quiver}}=\left(-k_1 - k_2, -k_1 - k_2, k_1 + k_2, k_2, k_1 \right) \ ,
\eea
Note that two of the entries are equal while in general there are no equal entries in $Q_t$. Therefore the only hope to reproduce $Q_t$ (up to an overall minus sign) is to choose $k_1+k_2=\pm\,1$. Substituting this back into $Q_{\mathrm{quiver}}$ we obtain
\bea
Q_{\mathrm{quiver}}=\left(\mp\,1, \mp\,1, \pm\,1, \pm\,1-k_1, k_1 \right) \ .
\eea
Obviously we can reproduce $Q_t$ only for $x=1$ or $y=1$. For other values, the geometries are not captured by the quiver that we have written. Indeed, it seems that the $AdS_4 \times Y$ spaces, where $Y$ is the base of the corresponding isolated Calabi-Yau singularity, reduce in Type IIA to singular spaces, for any choice of $U(1)$ isometry on $Y$. Thus the corresponding M2 brane theories apparently do not admit a Lagrangian description, according to 
\cite{Jafferis:2009th}. 


\section{Geometry of $\mathcal{C}(Q^{111})$ and its resolutions} \label{sec:Q111-geometry}


The cone over $Q^{111}$ is a non-complete intersection defined by 8 $w_i\in \mathbb{C}$ such that
\begin{eqnarray}
\nonumber
&& w_1w_2-w_3w_4=w_1w_2-w_5w_8=w_1w_2-w_6w_7=0\ ,\\ \nonumber
&& w_1w_3-w_5w_7=w_1w_6-w_4w_5=w_1w_8-w_4w_7=0\ ,\\ \nonumber
&& w_2w_4-w_6w_8=w_2w_5-w_3w_6=w_2w_7-w_3w_8=0\ .\nonumber
\end{eqnarray}

One can check that these equations can be solved in general by taking 
\begin{equation}
\begin{tabular}{l l}
$w_1=\rho \me^{\frac{\ii}{2}(\psi+\phi_1+\phi_2+\phi_3)}\cos\frac{\theta_1}{2}\cos\frac{\theta_2}{2}\cos\frac{\theta_3}{2}\ ,$ & $w_2=\rho   \me^{\frac{\ii}{2}(\psi-\phi_1-\phi_2-\phi_3)}\sin\frac{\theta_1}{2}\sin\frac{\theta_2}{2}\sin\frac{\theta_3}{2}\ ,$\\
$w_3=\rho  \me^{\frac{\ii}{2}(\psi+\phi_1-\phi_2-\phi_3)}\cos\frac{\theta_1}{2}\sin\frac{\theta_2}{2}\sin\frac{\theta_3}{2}\ ,$ & $w_4=\rho   \me^{\frac{\ii}{2}(\psi-\phi_1+\phi_2+\phi_3)}\sin\frac{\theta_1}{2}\cos\frac{\theta_2}{2}\cos\frac{\theta_3}{2}\ ,$\\
$w_5=\rho   \me^{\frac{\ii}{2}(\psi+\phi_1+\phi_2-\phi_3)}\cos\frac{\theta_1}{2}\cos\frac{\theta_2}{2}\sin\frac{\theta_3}{2}\ ,$  &$w_6=\rho \me^{\frac{\ii}{2}(\psi-\phi_1+\phi_2-\phi_3)}\sin\frac{\theta_1}{2}\cos\frac{\theta_2}{2}\sin\frac{\theta_3}{2}\ ,$\\
$w_7=\rho   \me^{\frac{\ii}{2}(\psi+\phi_1-\phi_2+\phi_3)}\cos\frac{\theta_1}{2}\sin\frac{\theta_2}{2}\cos\frac{\theta_3}{2}\ ,$ & $w_8=\rho  \me^{\frac{\ii}{2}(\psi-\phi_1-\phi_2+\phi_3)}\sin\frac{\theta_1}{2}\sin\frac{\theta_2}{2}\cos\frac{\theta_3}{2}\ .$
\end{tabular}
\label{coordinates}
\end{equation}
Since the cone over $Q^{111}$ is both Ricci-flat and K\"ahler the metric can be written as $g_{m\bar{n}} = \partial_m\,\partial_{\bar{n}}K$, where $K$ is the K\"ahler potential. In the singular case, the most general such K\"ahler potential which one could write, compatible with the 
$SU(2)^3\times U(1)_R$ symmetry, is
\begin{equation}
K=F(\rho^2)\ .
\end{equation}
We may resolve $\mathcal{C}(Q^{111})$ by blowing up a copy of $\mathbb{CP}^1\times\mathbb{CP}^1$, as explained in the main text. We take the
corresponding K\"ahler potential to be
\begin{equation}
K=F(\rho^2)+a\log(1+|\lambda_1|^2)+b\log(1+|\lambda_2|^2)\ .
\end{equation}
Here $a, b$  are the resolution parameters of the two $\mathbb{CP}^1$s, coordinatized respectively by
\begin{equation}
\lambda_1=\frac{w_2}{w_6}=\me^{-\ii\phi_2}\tan\frac{\theta_2}{2}\ ,\quad \lambda_1=\frac{w_5}{w_1}=\me^{-\ii\phi_3}\tan\frac{\theta_3}{2}\ .
\end{equation}
The Ricci tensor for a K\"ahler manifold is related to the determinant of the metric as $R_{\bar{a}b}=-\bar{\partial}_{\bar{a}}\partial_b\log{\rm det}\, g$. For the case at hand, the determinant of the metric reads
\begin{equation}
{\rm det}\, g= (F'+\rho^2F'')(a+\rho^2 F')(b+\rho^2F')F' \ ,
\end{equation}
where $'\equiv\frac{\diff }{\diff \rho^2}$. It is useful to define $\gamma=\rho^2\, F'$. Then Ricci flatness implies
\begin{equation}
\gamma'\gamma(a+\gamma)(b+\gamma)=\frac{\rho^2}{32}\ .
\end{equation}
Integrating this expression and setting the integration constant to zero we obtain
\begin{equation}
\gamma^4+\frac{4}{3}(a+b)\gamma^3+2ab\gamma^2=\frac{\rho^4}{16} \ .
\end{equation}
In terms of $\gamma$ the metric then becomes
\begin{eqnarray}
\diff s^2&=&\gamma'\, \diff \rho^2+\frac{\rho^2 \gamma'}{4}\Big(\diff \psi+\sum_{i=1}^3 \cos\theta_i \diff \phi_i\Big)^2+\frac{(a+\gamma)}{4}\Big(\diff \theta_2^2+\sin^2\theta_2 \diff \phi_2^2\Big)\nonumber \\ &&+\frac{(b+\gamma)}{4}\Big(\diff \theta_3^2+\sin^2\theta_3 \diff \phi_3^2\Big)+\frac{\gamma}{4}\Big(\diff \theta_1^2+\sin^2\theta_1 \diff \phi_1^2\Big)\ .
\end{eqnarray}
It is now convenient to introduce a new radial coordinate $r$ as $r^2=2\gamma$. In terms of this $r$ the metric becomes
\begin{eqnarray}
\label{Q111}
\diff s^2&=&\kappa^{-1}\diff r^2+\kappa\frac{r^2}{16}\Big(\diff \psi+\sum_{i=1}^3 \cos\theta_i \diff \phi_i\Big)^2+\frac{(2a+r^2)}{8}\Big(\diff \theta_2^2+\sin^2\theta_2 \diff \phi_2^2\Big)\nonumber \\ &&+\frac{(2b+r^2)}{8}\Big(\diff \theta_3^2+\sin^2\theta_3 \diff \phi_3^2\Big)+\frac{r^2}{8}\Big(\diff \theta_1^2+\sin^2\theta_1 \diff \phi_1^2\Big)\ ,
\label{Q111-resolved-metric}
\end{eqnarray}
with
\begin{equation}
\kappa=\frac{2\rho^2\gamma'}{\gamma}\ .
\end{equation}
After some algebra, this is
\begin{equation}
\label{kappaQ111}
\kappa=\frac{(2A_-\,+r^2)(2A_+\,+r^2)}{(2a+r^2)(2b+r^2)}\ ,
\end{equation}
where
\begin{equation}
A_{\pm}=\frac{1}{3}\Big(2a+2b \pm \sqrt{4a^2-10ab+4b^2}\Big)\ .
\end{equation}

This metric has appeared in the literature in a slightly different form \cite{Cvetic:2000db, Cvetic:2001ma}.\footnote{This form can be recovered by defining $r=\sqrt{2}\rho$, $a=l_2^2$ and $b=l_3^2$.} In the main text we make use of yet another form which can be obtained by redefining $r=\sqrt{8\,\varrho}$, $a=4\,l_2^2$ and $b=4\,l_3^2$. After these redefinitions the metric becomes
\begin{eqnarray}
\label{Q111Cvetic}
\diff s^2&=&{U}^{-1}\diff \varrho^2+{U}\varrho\Big(\diff \psi+\sum_{i=1}^3 \cos\theta_i \diff \phi_i\Big)^2+(l_2^2+\varrho)\,\Big(\diff \theta_2^2+\sin^2\theta_2 \diff \phi_2^2\Big)\nonumber \\ &&+(l_3^2+\varrho)\,\Big(\diff \theta_3^2+\sin^2\theta_3 \diff \phi_3^2\Big)+\varrho\, \Big(\diff \theta_1^2+\sin^2\theta_1 \diff \phi_1^2\Big)\ ;
\end{eqnarray}
where

\begin{equation}
{U}=\frac{3\, \varrho^3+4\, \varrho^2\, (l_2^2+l_3^2)+6\, l_2^2\, l_3^2\varrho}{6\,(l_2^2+ \varrho)\, (l_3^2+ \varrho)} \ .
\end{equation}

\section{Geometry of $\mathcal{C}(Q^{222})$ and its resolutions} \label{sec:Q222-geometry}

The cone $\mathcal{C}(Q^{222})$ is a $\mathbb{Z}_2$ orbifold of $\mathcal{C}(Q^{111})$, where $\Z_2\subset U(1)_R$ acts on the fibre coordinate $\psi$. Thus we can construct the variety by starting with the $w_i$ holomorphic coordinates and taking the desired orbifold. In particular, it follows that the metric is just that of $\mathcal{C}(Q^{111})$ with $\psi\in[0,2\pi]$.

We are interested in resolving this conical singularity. The computation of the metric is formally similar to that for $\mathcal{C}(Q^{111})$. Thus, after defining $\gamma=\rho^2\, F'$, we have the equation
\begin{equation}
\gamma'\, \gamma\, (a+\gamma)\, (b+\gamma)=\frac{\rho^2}{32} \ .
\end{equation}
This equation is integrated into
\begin{equation}
\gamma^4+\frac{4}{3}(a+b)\, \gamma^3+2\,a\,b\,\gamma^2-k=\frac{\rho^4}{16} \ ,
\label{gamma-equ}
\end{equation}
where we have left a non-zero integration constant $k$. If $k=0$ we reduce to the $\mathcal{C}(Q^{111})$ case (locally). Following the same steps as for $\mathcal{C}(Q^{111})$, we introduce a new radial variable 
\bea
r^2=2\,\gamma \ ,
\label{gamma-equ-II}
\eea
such that the metric reduces to
\begin{eqnarray}
\label{Q222}
\diff s^2&=&\kappa^{-1}\diff r^2+\kappa\frac{r^2}{16}\Big(\diff \psi+\sum_{i=1}^3 \cos\theta_i \diff \phi_i\Big)^2+\frac{(2a+r^2)}{8}\Big(\diff \theta_2^2+\sin^2\theta_2 \diff \phi_2^2\Big)\nonumber \\ &&+\frac{(2b+r^2)}{8}\Big(\diff \theta_3^2+\sin^2\theta_3 \diff \phi_3^2\Big)+\frac{r^2}{8}\Big(\diff \theta_1^2+\sin^2\theta_1 \diff \phi_1^2\Big)\ ,
\end{eqnarray}
where now
\begin{equation}
\label{kappaQ222}
\kappa=\frac{r^8+\frac{8}{3}\, (a+b)\, r^6+8\, a\,b \,r^4-16\, k}{r^4\, (2\,a+r^2)\,(2\,b+r^2)} \ .
\end{equation}

Let us consider the case in which $a=b=0$ and $k\ne 0$. Then
\begin{equation}
\kappa=1-\frac{16\,k}{r^8} \ .
\end{equation}
Defining $r_{\star}^8=16\, k$, we have that close to $r_{\star}$, the metric approaches
\begin{equation}
\diff s^2=\frac{r_{\star}}{8\, (r-r_{\star})}\, \diff r^2+\frac{r_{\star}}{2}\, (r-r_{\star})\, \Big(\diff \psi+\sum_{i=1}^3 \cos\theta_i\, \diff \phi_i\Big)^2+\frac{r_{\star}^2}{8}\, \sum_{i=1}^3 \diff \theta_i^2+\sin^2\theta_i\,\diff \phi_i^2 \ .
\end{equation}
Introducing a new radial variable $u=\sqrt{\frac{r_{\star}}{2}\, (r-r_{\star})}$ we have
\begin{equation}
\diff s^2=\diff u^2+u^2\, \Big(\diff \psi+\sum_{i=1}^3 \cos\theta_i\, \diff \phi_i\Big)^2+\frac{r_{\star}^2}{8}\, \sum_{i=1}^3 \diff \theta_i^2+\sin^2\theta_i\,\diff \phi_i^2 \ .
\end{equation}
Here in order to avoid a conical singularity the period of $\psi$ must be $2\pi$. Thus the metric (\ref{Q222}) is the resolution of $\mathcal{C}(Q^{222})$. Indeed, by redefinition of variables we can recover the metric for the resolved $\mathcal{C}(Q^{222})$ that appeared in \cite{Cvetic:2000db, Cvetic:2001ma}.\footnote{This form can be recovered by starting with (\ref{Q222}) and redefining $r=\sqrt{8\, \rho+r_{\star}}\ , l_1^2=\frac{2\,a+r_{\star}}{8}\ ,  l_2^2=\frac{2\,b+r_{\star}}{8}\ , l_3^2=\frac{r_{\star}}{8} \ .$}
\comment{
Plugging these back to the metric it turns out that the metric reduces to
\begin{equation}
\diff s^2=U^{-1}\, \diff \rho^2+U\, \Big(\diff \psi+\sum_{i=1}^3 \cos\theta_i\, \diff \phi_i\Big)^2+\sum(\rho+l_i^2)\, \Big(\diff \theta_i^2+\sin^2\theta_i\,\diff \phi_i^2\Big)
\end{equation}
with
\begin{equation}
U= \frac{\frac{\rho}{6}\,\Big\{\rho\, \Big(6\, l_s^2\, l_3^2+4\,\rho\, (l_2^2+l_3^2)+3\, \rho^2\Big)+2\, l_1^2\, \Big(3\, l_2^2\, (2\, l_3^2+\rho)+\rho\, (3\, l_3^2+2\, \rho)\Big)\Big\}}{(\rho+l_1^2)\, (\rho+l_2^2)\, (\rho+l_3^2)}
\end{equation}
which is precisely the form obtained in \cite{Cvetic:2000db, Cvetic:2001ma}.
}

\section{The warp factor for resolutions of $\mathcal{C}(Q^{111})$ and $\mathcal{C}(Q^{222})$} \label{sec:D}
We are interested in studying supergravity backgrounds corresponding to M2 branes localized on the space $X$, which will be the resolution of either $\mathcal{C}(Q^{111})$ or $\mathcal{C}(Q^{111})$ described in \secref{sec:Q111-geometry} and  \secref{sec:Q222-geometry} respectively. After placing $N$ spacetime-filling M2 branes at a point $y$ in the resolved space $X$ we must solve the Green's equation
\bea
\Delta_x h[y] = \frac{(2\pi \ell_p)^6N}{\sqrt{\det g_X}} \delta^8(x-y) \ ,
\eea
for the warp factor $h=h[y]$. Here $\Delta$ is the scalar Laplacian on $X$ and $g_X$ is the metric on the resolved cone. Using the explicit form of the Laplacian we can write
\begin{equation}
\frac{1}{\sqrt{\det g}}\, 
\partial_i\Big(\sqrt{\det g}\, g^{ij}\partial_j h\Big)=-\frac{(2\pi \ell_p)^6N}{\sqrt{\det g}} \delta^8(x-y) \ .
\end{equation}

Since we can choose coordinates such that the metrics we are considering are formally identical
\begin{eqnarray}
\diff s^2&=&\kappa^{-1}\diff r^2+\kappa\frac{r^2}{16}\Big(\diff \psi+\sum_{i=1}^3 \cos\theta_i \diff \phi_i\Big)^2+\frac{(2a+r^2)}{8}\Big(\diff \theta_2^2+\sin^2\theta_2 \diff \phi_2^2\Big)\nonumber \\ &&+\frac{(2b+r^2)}{8}\Big(\diff \theta_3^2+\sin^2\theta_3 \diff \phi_3^2\Big)+\frac{r^2}{8}\Big(\diff \theta_1^2+\sin^2\theta_1 \diff \phi_1^2\Big)\ ,
\end{eqnarray}
we have that the Laplacian in both cases can be written as
\begin{equation}
\frac{1}{\sqrt{\det g}}\,\partial_i\Big(\sqrt{\det g}\, g^{ij}\partial_j h\Big)=\frac{\partial_r\Big(r^3(2a+r^2)(2b+r^2)\kappa\,\partial_rh\Big)}{r^3(2a+r^2)(2b+r^2)}+\mathbf{A}h \ ,
\end{equation}
where the angular Laplacian $\mathbf{A}$ is
\begin{equation}
\label{AngLap}
\mathbf{A}h=\frac{8}{r^2}\Delta_1h+\frac{8}{2a+r^2}\Delta_2 h+\frac{8}{2b+r^2}\Delta_3h+\frac{16}{r^2\kappa}\partial^2_{\psi}h \ ,
\end{equation}
with
\begin{equation}
\Delta_i=\frac{1}{\sin\theta_i}\partial_{\theta_i}(\sin\theta_i\partial_{\theta_i})+\Big(\frac{1}{\sin\theta_i}\partial_{\phi_i}-\cot\theta_i\partial_{\psi}\Big)^2 \ .
\end{equation}
As we show in (\ref{delta}) in the next subsection, we can expand the delta function in terms of eigenfunctions of the Laplacian such that
\begin{equation}
\frac{1}{\sqrt{\det g}} \delta^8(x-y)=\frac{1}{r^3(r^2+2a)(r^2+2b)}\delta(r-r_0)\sum_I\, Y_I(\xi_0)^*Y_I(\xi) \ ,
\end{equation}
where we denote collectively the angular coordinates as $\xi$, and define $x=(r,\xi)$ and $y=(r_0,\xi_0)$. Then, the equation for the warp factor reads
\begin{equation}
\frac{1}{f}\partial_r\Big(f\,\partial_rh\Big)+\kappa^{-1}\,\mathbf{A}h=- \frac{(2\pi \ell_p)^6N}{f}\delta(r-r_0)\sum_I\, Y_I(\xi_0)^*Y_I(\xi) \ ,
\end{equation}
where we have defined for simplicity
\begin{equation}
\label{f-def}
f=r^3(2a+r^2)(2b+r^2)\kappa \ .
\end{equation}
We can now expand $h$ in eigenfunctions of the angular Laplacian. Since $\xi_0$ is just a point (not a variable), we can write
\begin{equation}
h=\sum_I \psi_I(r)\, Y_I(\xi_0)^*\, Y_I(\xi) \ .
\label{general-warp}
\end{equation}
Then the radial equation we have to solve reduces to
\begin{equation}
\frac{1}{f}\partial_r\Big(f\,\partial_r\psi_I\Big)-\kappa^{-1}\,E_I\,\psi_I=- \frac{(2\pi \ell_p)^6N}{f}\delta(r-r_0) \ ,
\label{radial-to-solve}
\end{equation}
where $E_I$ is the angular eigenvalue of $Y_I$, to which we now turn.

\subsection{Angular eigenfunctions in $Q^{111}$}

We want to consider (\ref{AngLap}) with fixed $r$ and construct eigenfunctions of such an operator. For this we first concentrate on each of the $\Delta_i$ operators. For each $\theta$, these look like
\begin{equation}
\Delta_{i} =\frac{1}{\sin\theta }\partial_{\theta}(\sin\theta\partial_{\theta})+\Big(\frac{1}{\sin\theta}\partial_{\phi}-\cot\theta\partial_{\psi}\Big)^2 \ .
\end{equation}
Note that these angular Laplacians are the same as those for the conifold. As such, many technical details and results can be borrowed from \cite{Klebanov:2007us}.

We consider the following function
\begin{equation}
Y=J(\theta)\, \me^{\ii\, m\, \phi}\, \me^{\ii\, {R\psi}/{2}} \ .
\end{equation}
It is obvious that
\begin{equation}
\Delta_{i} Y=\Big\{\frac{1}{\sin\theta}\partial_{\theta}(\sin\theta\partial_{\theta} J)-\Big(\frac{m}{\sin\theta}-\cot\theta\frac{R}{2}\Big)^2 J\Big\}J^{-1} Y \ .
\end{equation}
Therefore it is interesting to consider the following eigenfunctions
\begin{equation}
\frac{1}{\sin\theta}\partial_{\theta}(\sin\theta\partial_{\theta} J)-\Big(\frac{m}{\sin\theta}-\cot\theta\frac{R}{2}\Big)^2 J=-E J \ .
\end{equation}
This equation has two solutions, given in terms of hypergeometric functions
\begin{equation}
\label{sol_A}
J^A_{l,m,R}=\sin^m\theta\, \cot^{\frac{R}{2}}\frac{\theta}{2}\, \,_2F_1\Big(-l+m,1+l+m;1+m-\frac{R}{2};\sin^2\frac{\theta}{2}\Big) ~,
\end{equation}
and
\begin{equation}
\label{sol_B}
J^B_{l,m,R}=\sin^{\frac{R}{2}}\theta\, \cot^{m}\frac{\theta}{2}\, \,_2F_1\Big(-l+\frac{R}{2},1+l+\frac{R}{2};1-m+\frac{R}{2};\sin^2\frac{\theta}{2}\Big) \ ,
\end{equation}
where we have introduced a labelling for the quantum numbers distinguishing the eigenfunctions. If $m\ge \frac{R}{2}$ solution (\ref{sol_A}) is non-singular while if $m\le \frac{R}{2}$ it is (\ref{sol_B}) that is the non-singular solution. Both have eigenvalue under each $\Delta_i$ operator given by
\begin{equation}
E=l(l+1)-\frac{R^2}{4} \ .
\end{equation}
Given these results, we can consider the functions
\begin{equation}
\label{Y-function}
Y_{I}=\mathcal{C}_I\,J_{l_1,m_1,R}(\theta_1)J_{l_2,m_2,R}(\theta_2)J_{l_3,m_3,R}(\theta_3)\, \me^{\ii\, (m_1\phi_1+m_2\phi_2+m_3\phi_3)}\, \me^{\ii\, {R\psi}/{2}} \ ,
\end{equation}
where the multi-index $I$ stands for $\{(l_1,m_1),(l_2,m_2),(l_3,m_3),R\}$ and where $\mathcal{C}_I$ is just a normalization factor such that the norm of $Y_I$ is one.  It is now clear that
\begin{equation}
\mathbf{A}Y_I=-E_I Y_I \ ,
\end{equation}
with
\begin{equation}
E_I=\frac{8\,l_1(l_1+1)}{r^2}+\frac{8\,l_2(l_2+1)}{r^2+2a}+\frac{8\,l_3(l_3+1)}{r^2+2b}+2R^2\Big(\frac{2}{r^2\kappa}-\frac{1}{r^2}-\frac{1}{r^2+2a}-\frac{1}{r^2+2b}\Big) \ .
\end{equation}
We now note that the $Y_I$ are also eigenfunctions of the singular cone. Indeed, we can consider the Laplacian on the unit $Q^{111}$, namely $\mathbf{\tilde{A}}|_{a=0=b;r=1}$. Then
\begin{equation}
\mathbf{\tilde{A}}Y_I=-\tilde{E}_I Y_I \ ,
\end{equation}
with
\begin{equation}
\tilde{E}_I=8\,l_1(l_1+1)+8\,l_2(l_2+1)+8\,l_3(l_3+1)-2R^2 \ .
\end{equation}
Therefore, the $Y_I$ are also normalized eigenfunctions for the $\mathbf{\tilde{A}}$ operator. Being eigenfunctions of a Hermitian operator, these satisfy
\begin{equation}
\int \dd^7\xi\, \sqrt{\det \tilde{g}}\, Y_I(\xi)^*Y_J(\xi)=\delta_{I-J} \ ,
\end{equation}
and therefore
\begin{equation}
\sum_I\, Y_I(\xi_1)^*Y_I(\xi_2)=\frac{1}{\sqrt{\det \tilde{g}}}\, \delta^7(\xi_1-\xi_2) \ ,
\end{equation}
where we use $\xi$ to generically parametrize the angular coordinates and $\tilde{g}$ stands for the angular part of the metric. One can check very easily that
\begin{equation}
\sqrt{\det g}=r^3(r^2+2a)(r^2+2b)\, \sqrt{\det \tilde{g}} \ .
\end{equation}
Therefore, if we denote $x=(r,\xi)$ and $y=(r_0,\xi_0)$ we have
\begin{equation}
\frac{1}{\sqrt{\det g}}\, \delta^8(x-y)=\frac{1}{r^3(r^2+2a)(r^2+2b)}\delta(r-r_0)\frac{1}{\sqrt{\det \tilde{g}}}\, \delta^7(\xi-\xi_0) \ .
\end{equation}
Using the completeness relation above we may hence write
\begin{equation}
\label{delta}
\frac{1}{\sqrt{\det g}}\, \delta^8(x-y)=\frac{1}{r^3(r^2+2a)(r^2+2b)}\delta(r-r_0)\sum_I\, Y_I(\xi_0)^*Y_I(\xi) \ .
\end{equation}
\subsection{Angular eigenfunctions in $Q^{222}$}

Since $\mathcal{C}(Q^{222})$ is a $\mathbb{Z}_2$ orbifold of $\mathcal{C}(Q^{111})$ along $\psi$ it is clear that the local computation of the previous subsection will not be changed. Thus, we just have to take care of global issues. Recall that the wavefunctions in $\mathcal{C}(Q^{111})$ are
\begin{equation}
Y_{I}=\mathcal{C}_I\,J_{l_1,m_1,R}(\theta_1)J_{l_2,m_2,R}(\theta_2)J_{l_3,m_3,R}(\theta_3)\, \me^{\ii\, (m_1\phi_1+m_2\phi_2+m_3\phi_3)}\, \me^{\ii\, {R\psi}/{2}} \ .
\end{equation}
Since now $\psi\in [0,2\pi]$, it is clear that the well-behaved $Y_I$ will be those for which $R$ is even; that is, $R=2\, \tilde{R}$. Therefore, dropping the tilde, the angular wavefunctions in $\mathcal{C}(Q^{222})$ are
\begin{equation}
Y_{I}=\mathcal{C}_I\,J_{l_1,m_1,R}(\theta_1)J_{l_2,m_2,R}(\theta_2)J_{l_3,m_3,R}(\theta_3)\, \me^{\ii\, (m_1\phi_1+m_2\phi_2+m_3\phi_3)}\, \me^{\ii\, R\psi} \ ,
\end{equation}
such that
\begin{equation}
\mathbf{A}Y_I=-E_I Y_I \ ,
\end{equation}
with
\begin{equation}
E_I=\frac{8\,l_1(l_1+1)}{r^2}+\frac{8\,l_2(l_2+1)}{r^2+2a}+\frac{8\,l_3(l_3+1)}{r^2+2b}+8R^2\Big(\frac{2}{r^2\kappa}-\frac{1}{r^2}-\frac{1}{r^2+2a}-\frac{1}{r^2+2b}\Big) \ .
\end{equation}

\subsection{The warp factor for $Q^{111}$} \label{sec:Q111-warp}

We now want to use the results derived so far to compute explicitly the warp factor for the resolution of the $\mathcal{C}(Q^{111})$ space. We will consider the stack of branes to be sitting on the exceptional locus, where both the $U(1)$ fibre and the $(\theta_1,\phi_1)$ sphere shrink to zero size. This means that $h=h(r,\theta_2,\theta_3)$, which in turn implies that $R$ and $l_1$ in (\ref{Y-function}) vanish. Then, under these assumptions, the multi-index $I$ takes the values $I=\{(l_2,m_2),(l_3,m_3)\}$. Indeed, we will assume the branes are located at the north pole of each of the two two-spheres. As such, we should consider also $m_2=m_3=0$. Therefore, for such cases the angular eigenfunctions $J^A$ and $J^B$ coincide and reduce, for each sphere, to Legendre polynomials
\begin{equation}
J_{l,0,0}=\,_2F_1(-l,1+l;1;\sin^2\tfrac{\theta}{2})=P_l(\cos\theta) \ ,
\end{equation}
such that for the case at hand where only $l_2,l_3\ne 0$
\begin{equation}
Y_{l_2,l_3}=\mathcal{C}_{l_2,l_3}\, P_{l_2}(\cos\theta_2)\, P_{l_3}(\cos\theta_3) \ ,
\end{equation}
and 
\begin{equation}
E_I=\frac{8\,l_2(l_2+1)}{r^2+2a}+\frac{8\,l_3(l_3+1)}{r^2+2b} \ .
\end{equation}
Thus, from (\ref{radial-to-solve}) we see that the equation to solve reads
\begin{equation}
\frac{1}{f}\partial_r\Big(f\,\partial_r\psi_I\Big)-\Big(\frac{8\,l_2(l_2+1)}{r^2+2a}+\frac{8\,l_3(l_3+1)}{r^2+2b}\Big)\kappa^{-1}\,\psi_I=-\frac{(2\pi \ell_p)^6N}{f}\, \delta(r) \ .
\end{equation}
We are interested in the simplified case in which, say, only $b\ne 0$. Under such assumption, also the $(\theta_2,\phi_2)$ sphere shrinks to zero, so that we can also consider $l_2=0$. Then the corresponding angular wavefunctions are
\begin{equation}
Y_{l_3}=\mathcal{C}_{l_3}\, P_{l_3}(\cos\theta_3) \ .
\end{equation}
Also from (\ref{kappaQ111}) and (\ref{f-def}) we see that
\begin{equation}
\kappa=\frac{r^2+\frac{8b}{3}}{r^2+2b}\ ,\qquad f=r^5\, (r^2+\frac{8b}{3}) \ .
\end{equation}
Then the equation to solve reduces to
\begin{equation}
\partial_r\Big[r^5\, (r^2+\frac{8b}{3})\, \partial_r\psi_I\Big]-8\,r^5\, l_3(l_3+1)\,\psi_I+(2\pi \ell_p)^6N\delta(r)=0 \ .
\end{equation}
The two solutions are
\begin{eqnarray}
\psi_I^{(1)}& \sim &\Big(\frac{8b}{3r^2}\Big)^{\frac{3}{2}(1-\beta)}\, _2F_1(-\frac{1}{2}-\frac{3}{2}\beta,\frac{3}{2}-\frac{3}{2}\beta,1-3\beta,-\frac{8b}{3r^2})\ , \\ \nonumber
\psi_I^{(2)}& \sim &\Big(\frac{8b}{3r^2}\Big)^{\frac{3}{2}(1+\beta)}\, _2F_1(-\frac{1}{2}+\frac{3}{2}\beta,\frac{3}{2}+\frac{3}{2}\beta,1+3\beta,-\frac{8b}{3r^2}) \ ,
\end{eqnarray}
with
\begin{equation}
\beta=\sqrt{1+\frac{8}{9}l_3(l_3+1)} \ .
\end{equation}
Since $\beta\ge 1$, for large $r$ only the $\psi_I^{(2)}$ solutions decay at infinity, and these are therefore the solutions of interest.

We can now state the result for the warp factor, which turns out to be
\begin{equation}
h=\sum_{l_3}\, \mathcal{C}_{l_3}\, \Big(\frac{8b}{3r^2}\Big)^{\frac{3}{2}(1+\beta)}\, _2F_1\Big(-\frac{1}{2}+\frac{3}{2}\beta,\frac{3}{2}+\frac{3}{2}\beta,1+3\beta,-\frac{8b}{3r^2}\Big)\, P_{l_3}(\cos\theta_3) \ ,
\end{equation}
where we collect all normalization factors in $\mathcal{C}_{l_3}$.

\subsection{The warp factor for $Q^{222}$} \label{s:warp_factor-Q222}

We will compute the warp factor for $N$ M2 branes in arbitrary location. As was shown above, the equation to solve is (\ref{radial-to-solve}) where now we should use (\ref{kappaQ222}) for $\kappa$. We will be interested in the simpler case in which $a=b=0$. Moreover, as we explain in the main text, the interesting contribution is that coming from $R=l_i=0$. Under this simplification, the equation to solve now reads
\begin{equation}
\partial_r\Big(r^7\, (1-\frac{r_{\star}^8}{r^8})\, \partial_r\psi_I\Big)=-(2\pi \ell_p)^6N\, \delta(r-r_0) \ .
\end{equation}
Solving for $r>r_0$ we obtain
\begin{equation}
\psi_>=\frac{1}{r^6}\, _2F_1\Big(\frac{6}{8},\, 1,\, \frac{7}{4},\Big(\frac{r_{\star}^2}{r^2}\Big)^4\Big) \ ,
\end{equation}
and for $r<r_0$ instead
\begin{equation}
\psi_<=\frac{1}{r_0^6}\, _2F_1\Big(\frac{3}{4},\, 1,\, \frac{7}{4},\Big(\frac{r_{\star}^2}{r_0^2}\Big)^4\Big) \, _2F_1\Big(0,\, \frac{6}{8},\, \frac{3}{4},\Big(\frac{r^2}{r_{\star}^2}\Big)^4\Big) \ .
\label{warp-Q222-I}
\end{equation}
Indeed, the leading term for large $r$ corresponds to $l_i=0$, and in this limit 
\begin{equation}
h\sim \frac{ |\mathcal{C}_0|^2}{r^6} \equiv \frac{R^6}{r^6} \ .
\label{warp-Q222-II}
\end{equation}
Recall that for $\mathcal{C}(Q^{222})$
\begin{equation}
R^6= \frac{2^9\, \pi^2\, N\, l_p^6}{3} \ .
\label{warp-Q222-III}
\end{equation}

\section{The Eguchi-Hanson manifold as a toy model}\label{sec:EH}

In this appendix we illustrate, with the aid of a simple toy model, some of the computations derived in \secref{s:general-V}.

\subsection{Harmonic forms}

In this subsection we derive the normalizable harmonic two-form in the Eguchi-Hanson manifold. This two-form will be important later on when we show how the warped volume of the exceptional divisor can be inferred from it. The Eguchi-Hanson metric is
\begin{equation}
\diff s^2=\frac{\diff r^2}{1-\frac{c^4}{r^4}}+\frac{r^2}{4}\, \Big(1-\frac{c^4}{r^4}\Big)\, \Big(\diff\psi+\cos\theta \, \diff\phi\Big)^2+\frac{r^2}{4}\,\Big(\diff\theta^2+\sin^2\theta\, \diff\phi^2\Big) \ .
\end{equation}
To avoid a conical singularity at $r=c$ when $c>0$ we have to take $\psi\in[0,2\pi]$. It is natural to define the one-form vielbein
\begin{equation}
g_5=\diff\psi+\cos\theta\, \diff\phi\ , \qquad e_{\theta}=\diff\theta\ , \qquad e_{\phi}=\sin\theta\, \diff\phi \ .
\end{equation}
Then a natural ansatz for a closed and co-closed two-form is
\begin{equation}
\tilde\beta=e_{\theta}\wedge e_{\phi}+\diff(f\, g_5)\ , \qquad f=f(r) \ .
\end{equation}
It is immediate that this form is closed. The Hodge dual is
\begin{equation}
\star\tilde\beta=\frac{2}{r}\, (1-f)\, \diff r\wedge g_5+\frac{r}{2}\, \frac{\diff f}{\diff r}\, e_{\theta}\wedge e_{\phi} \ .
\end{equation}
Thus co-closedness implies
\begin{equation}
\frac{\diff}{\diff r}\Big(r\, \frac{\diff f}{\diff r}\Big)+\frac{4}{r}(1-f)=0 \ .
\end{equation}
Choosing the solution decaying at infinity, this can be integrated into
\begin{equation}
f=1+\frac{A}{r^2} \ ,
\end{equation}
$A$ being an integration constant which we will fix to $A=-1$. Thus the self-dual harmonic two-form is
\begin{equation}
\tilde\beta=\frac{1}{r^2}\, \sin\theta\, \diff \theta\wedge \diff \phi+\frac{2}{r^3}\, \diff r\wedge g_5 \ .
\end{equation}
We may define a normalized two-form $\hat{\beta}$ as $\hat{\beta}=\frac{ c^2}{2\sqrt{2}\,\pi}\, \tilde\beta$. Changing now to the following coordinate system
\begin{equation}
z_1=(r^4-c^4)^{1/4}\, \me^{\frac{\ii}{2}\, (\psi+\phi)}\, \sin\frac{\theta}{2}\qquad z_2=(r^4-c^4)^{1/4}\, \me^{\frac{\ii}{2}\, (\psi-\phi)}\, \cos\frac{\theta}{2} \ ,
\end{equation}
the normalized two-form now reads
\begin{equation}
\hat{\beta}= \partial_{z_i}\partial_{\bar{z}_j}\, S_{\beta}\,\diff z_i\wedge \diff \bar{z}_j\ ,\qquad S_{\beta}=\ii\frac{1}{\pi\, \sqrt{2}}\, {\rm arctanh}\frac{c^2}{r^2} \ .
\end{equation}
It is then easy to see that
\begin{equation}
\int_{X_{EH}} \hat{\beta}\wedge \star \hat{\beta}=1 \ .
\end{equation}

\subsection{Laplace operator}

Prior to the computation of the warped volume, in this subsection we derive the warp factor for the Eguchi-Hanson manifold. Thus, we are interested in solving the Green's equation
\begin{equation}
\Box_{EH}\, h=\frac{1}{\sqrt{\det g}}\, \partial_i\Big(\sqrt{\det g}\, g^{ij}\, \partial_j\, h\Big)=\mathcal{N}\, \frac{\delta^4(x-p)}{\sqrt{\det g}} \ ,
\label{laplace}
\end{equation}
where $p$ is some arbitrary point in the Eguchi-Hanson manifold and $\mathcal{N}$ is a normalization factor. After some algebra one can see that
\begin{equation}
\Box_{EH}\, h=\frac{1}{r^3}\, \partial_r\Big(r^3\, f\, \partial_r\, h\Big)+\frac{4}{r^2}\,\Delta\, h+\frac{4}{r^2\, f}\, \partial_{\psi}^2\, h \ ,
\end{equation}
where we have introduced the operator
\begin{equation}
\Delta=\frac{1}{\sin\theta}\partial_{\theta}\Big(\sin\theta\,\partial_{\theta}\Big)+\Big(\frac{\partial_{\phi}}{\sin\theta}-\cot\theta\, \partial_{\psi}\Big)^2 \ ,
\end{equation}
and defined for simplicity
\begin{equation}
f=1-\frac{c^4}{r^4} \ .
\end{equation}
Coming back to the Green's equation, denoting collectively the angular coordinates by $\xi$ and  $x=(r,\xi)$ and  $p=(r_p,\xi_p)$, we can write (\ref{laplace}) as
\begin{equation}
\Box_{EH}\, h=\frac{8\,\mathcal{N}}{r^3}\, \delta(r-r_p)\, \frac{\delta(\psi-\psi_p)\, \delta(\theta-\theta_p)\, \delta(\phi-\phi_p)}{\sin\theta} \ .
\label{laplace2}
\end{equation}

It is now useful to think of the Eguchi-Hanson manifold as a resolution of a $\mathbb{Z}_2$ orbifold of $\mathbb{C}^2$ which has been resolved with the parameter $c$. From this point of view, the metric on the singular cone would be simply
\begin{equation}
\diff s^2=\diff r^2+r^2\Big\{ \frac{1}{4}\Big(\diff \psi+\cos\theta \, \diff \phi\Big)^2+\frac{1}{4}\,\Big(\diff \theta^2+\sin^2\theta\, \diff \phi^2\Big)\Big\} \ ,
\end{equation}
where $\psi\in [0,2\pi]$. The Laplacian in this cone is 
\begin{equation}
\Box_{C}\, h=\frac{1}{r^3}\, \partial_r\Big(r^3\, \partial_r\, h\Big)+\frac{1}{r^2}\, \mathbf{A}\, h \ ,
\end{equation}
where we have introduced the angular Laplacian $\mathbf{A}$ 
\begin{equation}
\mathbf{A}=4\Delta +4\, \partial_{\psi}^2 \ .
\end{equation}
As usual, with the normalized eigenfunctions of $\mathcal{A}$
\begin{equation}
\mathcal{A}\, Y_{R,\,l,\,m}=-E_{R,\, l,\,m}\, Y_{R,\,l,\,m} \ ,
\end{equation}
we can construct a representation of the delta function on the base of the cone. This delta function is precisely the part appearing in the original Green's problem we are interested in. Thus
\begin{equation}
\sum_{R,\, l,\, m}\, Y_{R,\,l,\,m}(\xi_p)^*\, Y_{R,\,l,\,m}(\xi)=\frac{\delta(\psi-\psi_p)\, \delta(\theta-\theta_p)\, \delta(\phi-\phi_p)}{\sin\theta} \ .
\label{deltaEH}
\end{equation}

In order to find the explicit form for the $Y_{R,\,l,\,m}(\xi)$ eigenfunctions we start by writing
\begin{equation}
Y_{R,\,l,\,m}(\xi)=\me^{\ii\,R\,\psi}\, \me^{\ii\, m\, \phi}\, J_{l,\,m}(\theta) \ ,
\end{equation}
where $R,\, m\in \mathbb{Z}$. Then, the $J_{l,\,m}$ functions satisfy
\begin{equation}
\frac{1}{\sin\theta}\partial_{\theta}\Big(\sin\theta\, \partial_{\theta} J_{l,\,m}\Big)-\Big(\frac{m}{\sin\theta}-R\, \cot\theta\Big)^2\,  J_{l,\,m}=-\frac{E_{R,\,l,\,m}}{4}\,  J_{l,\,m} \ .
\end{equation}
One can then verify that the solutions to this equation are
\begin{equation}
 J_{l,\,m}^A(\theta)=\sin^m\theta\, \cot^R\frac{\theta}{2}\, _2F_1\Big(-l+m,\, 1+l+m,\, 1+m-R,\, \sin^2\frac{\theta}{2}\Big)~,
 \end{equation}
 and
\begin{equation}
 J_{l,\,m}^B(\theta)=\sin^R\theta\, \cot^m\frac{\theta}{2}\, _2F_1\Big(-l+R,\, 1+l+R,\, 1-m+R,\, \sin^2\frac{\theta}{2}\Big) \ ,
 \end{equation}
 and that for both
 \begin{equation}
 E_{R,\,l,\,m}=4\,\Big(l(l+1)-R^2\Big) \ .
 \end{equation}
If $m\ge R$ the solution $J^A_{l,\, m}$ is singular, while if $m\le R$ it is the solution $J^B_{l,\,m}$ that becomes singular. Of course, for $R=m$ both solutions coincide. Because of this, depending on $R$ we should use one or the other. 
 
Finally, the normalized eigenfunctions which we are after are
\begin{equation}
Y_{R,\,l,\,m}(\xi)=\mathcal{A}_{R,\,l,\,m}\, \me^{\ii\,R\,\psi}\, \me^{\ii\, m\, \phi}\, J_{l,\,m}(\theta) \ ,
\end{equation}
where $\mathcal{A}_{R,\,l,\,m}$ encodes the normalization. It is now clear that we should expand $h$ as
\begin{equation}
h=\sum_{R,\, l,\,m}\, \psi_{R,\, l}(r)\, Y_{R,\,l,\,m}(\xi_p)^*\, Y_{R,\,l,\,m}(\xi) \ .
\label{warp_factor}
\end{equation}
After substituting this and (\ref{deltaEH}) into (\ref{laplace2}), a straightforward computation shows that $ \psi_{R,\, l}$ satisfies
\begin{equation}
\frac{1}{r^3}\,\partial_r\Big(r^3\, f\, \partial_r\,  \psi_{R,\, l}\Big)-\Big\{\frac{4\,\Big(l(l+1)-R^2\Big)}{r^2}+\frac{4\,R^2}{r^2\, f}\Big\} \psi_{R,\, l}=\frac{8\, \mathcal{N}}{r^3}\, \delta(r-r_p) \ .
\end{equation}
The two solutions of this equation are
\begin{equation}
\psi^{(1)} \sim \Big(1-\frac{c^4}{r^4}\Big)^{\frac{|R|}{2}}\, \Big(\frac{r}{c}\Big)^{2\,l}\, _2F_1\Big(\frac{-l+|R|}{2},\frac{1-l+|R|}{2},\frac{1}{2}-l,\frac{c^4}{r^4}\Big)~,
\end{equation}
and
\begin{equation}
\label{warp1}
\psi^{(2)} \sim \Big(1-\frac{c^4}{r^4}\Big)^{\frac{|R|}{2}}\, \Big(\frac{c}{r}\Big)^{2+2\,l}\, _2F_1\Big(\frac{1+l+|R|}{2},\frac{2+l+|R|}{2},\frac{3}{2}+l,\frac{c^4}{r^4}\Big) \ .
\end{equation}
First we want to check which is the regular solution for $r>r_p$. For that we check the large $r$ limit
\begin{equation}
\psi^{(1)}\rightarrow \Big(\frac{r}{c}\Big)^{2\,l}\ ,\qquad \psi^{(2)}\rightarrow \Big(\frac{c}{r}\Big)^{2+2\,l} \ .
\end{equation}
Thus we conclude that the solution to use for $r> r_p$ is $\psi^{(2)}$. 

In order to find the $r<r_p$ solution, it is better to re-write the solutions of the above equation as
\begin{equation}
\tilde{\psi}^{(1)} \sim \Big(\frac{r^4}{c^4}-1\Big)^{\frac{|R|}{2}}\, _2F_1\Big(\frac{-l+|R|}{2},\, \frac{1+l+|R|}{2},\, \frac{1}{2},\, \frac{r^4}{c^4}\Big)~,
\end{equation}
and
\begin{equation}
\tilde{\psi}^{(2)} \sim \frac{r^2}{c^2}\, \Big(\frac{r^4}{c^4}-1\Big)^{\frac{|R|}{2}}\, _2F_1\Big(\frac{1-l+|R|}{2},\, \frac{2+l+|R|}{2},\, \frac{3}{2},\, \frac{r^4}{c^4}\Big) \ .
\end{equation}
The regular solution appears as a linear combination of these two solutions. To see this, let us define
\begin{eqnarray}
\label{combained}
\psi&=&\Big(\frac{r^4}{c^4}-1\Big)^{\frac{|R|}{2}}\, \Big[A_1\, _2F_1\Big(\frac{-l+|R|}{2},\, \frac{1+l+|R|}{2},\, \frac{1}{2},\, \frac{r^4}{c^4}\Big)+\\ \nonumber && A_2\, \frac{r^2}{c^2}\, _2F_1\Big(\frac{1-l+|R|}{2},\, \frac{2+l+|R|}{2},\, \frac{3}{2},\, \frac{r^4}{c^4}\Big)\Big] \ .
\end{eqnarray}
In order to ensure finiteness as $r$ tends to $c$, we have to set
\begin{equation}
A_2=-A_1\,\frac{2\,\Gamma\Big(\frac{1-l+|R|}{2}\Big)\, \Gamma\Big(\frac{2+l+|R|}{2}\Big)}{\Gamma\Big(\frac{-l+|R|}{2}\Big)\, \Gamma\Big(\frac{1+l+|R|}{2}\Big)} \ ,
\end{equation}
for modes where $|R| \ \geqslant l$ and the parity of $R$ and $l$ is the same. The other modes should be set to zero.
Subtituting this result into (\ref{combained}) and suppressing the overall factor $A_1$, we obtain
\begin{eqnarray}
\label{warp2}
\psi& \sim &\Big(\frac{r^4}{c^4}-1\Big)^{\frac{|R|}{2}}\, \Big[_2F_1\Big(\frac{-l+|R|}{2},\, \frac{1+l+|R|}{2},\, \frac{1}{2},\, \frac{r^4}{c^4}\Big)\\ \nonumber && -\frac{2\,\Gamma\Big(\frac{1-l+|R|}{2}\Big)\, \Gamma\Big(\frac{2+l+|R|}{2}\Big)}{\Gamma\Big(\frac{-l+|R|}{2}\Big)\, \Gamma\Big(\frac{1+l+|R|}{2}\Big)}\, \frac{r^2}{c^2}\, _2F_1\Big(\frac{1-l+|R|}{2},\, \frac{2+l+|R|}{2},\, \frac{3}{2},\, \frac{r^4}{c^4}\Big)\Big]  \ ,
\end{eqnarray}
which is the well-behaved solution for $r<r_p$.

In the following subsections it will become clear that the interesting mode for us is the one with $R=l=0$. With this choice (\ref{warp1}) and (\ref{warp2}) become
\begin{eqnarray}
\psi_> &=& A\,\Big( \frac{c}{r}\Big)^2\,_2F_1\Big(\frac{1}{2},\, 1,\, \frac{3}{2},\, \frac{c^4}{r^4}\Big)~,
\end{eqnarray}
and
\begin{eqnarray}
\label{EH-sol}
\psi_< &=& A\,\Big( \frac{c}{r_p}\Big)^2\,_2F_1\Big(\frac{1}{2},\, 1,\, \frac{3}{2},\, \frac{c^4}{r_p^4}\Big) \,_2F_1\Big(0,\, \frac{1}{2},\, \frac{1}{2},\, \frac{r^4}{c^4}\Big) \ ,
\end{eqnarray}
where $A$ is a normalization constant. To normalize these solutions we consider the $r \rightarrow \infty$ limit in (\ref{warp_factor}). We see that in this limit $h \simeq \, A\, |\mathcal{A}_{0,\,0,\,0}|^2 \,c^2/r^2$. Thus, if we integrate (\ref{laplace}) we get 
\begin{equation}
A\, |\mathcal{A}_{0,\,0,\,0}|^2\,c^2=\frac{\mathcal{N}}{2\,\vol(\Omega_{EH})} \ ,
\end{equation}
where $\Omega_{EH}\cong S^3/\Z_2$ is the the base of the Eguchi-Hanson manifold at infinity. Explicitly this volume is
\begin{equation} 
\vol(\Omega_{EH})=\int_{\Omega_{EH}} \sqrt{\det g}=\frac{1}{8}\int_0^{2\,\pi}\diff \,\psi\,\int_0^{2\,\pi}\,\diff \,\phi\,\int_0^{\pi}\,\sin\,\theta\,\diff \,\theta=\pi^2 \ .
\end{equation}
Thus, in the convention in which $\mathcal{N}=2\,\pi$ we have
\begin{equation}
\label{EH-sol-norm}
A=\frac{1}{\pi\,c^2\, |\mathcal{A}_{0,\,0,\,0}|^2} \ .
\end{equation}
\subsection{Warped volumes} \label{s:EH-V}

We are now interested in computing the warped volume of the blown-up divisor in the Eguchi-Hanson manifold. This reads
\begin{equation}
S=\int_D\, \sqrt{\det g_D}\, h \, \diff^2x~,
\end{equation}
where $D$ is the blown-up $S^2=\mathbb{CP}^1$ at $r=c$, and $g_D$ and $h$ are the pull-backs of the metric and warp factor to $D$, respectively. After substituting (\ref{warp_factor}) and
\begin{equation}
\sqrt{\det g_D}=\frac{c^2}{4} \sin\theta \ ,
\end{equation}
into this expression one obtains
\begin{equation}
S=\frac{c^2}{4}\,\sum_{l,\,m}\,  \psi_l(c)\,\int_{D}\, Y_{0,\,l,\,m}(\xi_p)^*\, Y_{0,\,l,\,m}(\xi)\, \diff^2\xi \ .
\end{equation}
We now proceed by evaluating the integral 
\begin{equation}
\int_{D}\, Y_{0,\,l,\,m}(\xi_p)^*\, Y_{0,\,l,\,m}(\xi)\, \diff^2\xi=|\mathcal{A}_{0,\,l,\,m}|^2\, \me^{-\ii\,m\,\phi_p}\,J_{l,\,m}(\theta_p)\,\int\, \diff \theta\, \diff \phi\, \sin\theta\, \me^{\ii\,m\,\phi}\,  J_{l,\,m}(\theta) \ .
\end{equation}
The $\phi$ integral forces that only the $m=0$ term contributes. Thus
\begin{equation}
\int_{D}\, Y_{0,\,l,\,m}(\xi_p)^*\, Y_{0,\,l,\,m}(\xi)\, \diff^2 \xi=2\pi\,\delta_{m,0}\,|\mathcal{A}_{0,\,l,\,0}|^2\,J_{l,\,0}(\theta_p)\,\int_0^\pi\, \diff \theta\, \sin\theta\,  J_{l,\,0}(\theta) \ .
\end{equation}
Furthermore, the $J_{l,0}(\theta)$ are just Legendre polynomials in $\cos\theta$. One can then see that all integrals are zero except for the $l=0$ mode. Thus
\begin{equation}
\int_{D}\,Y_{0,\,l,\,m}(\xi_p)^*\, Y_{0,\,l,\,m}(\xi)\, \diff^2 \xi=4\pi\,\delta_{m,0}\, \delta_{l,0}\,|\mathcal{A}_{0,\,0,\,0}|^2 \ .
\end{equation}
So finally
\begin{equation}
S=\pi\,c^2\,|\mathcal{A}_{0,\,0,\,0}|^2\,  \psi_<(c) \ .
\end{equation}
$\psi_<(c)$ can be read from (\ref{EH-sol}) together with the normalization at (\ref{EH-sol-norm}). Neglecting the label $p$ in $r_p$ we obtain our final extremely simple result:
\begin{equation}
S={\rm arctanh} \frac{c^2}{r^2} \ .
\end{equation}

\subsection{Harmonic forms from the warped volume}

In this subsection we  show how to rederive the harmonic two-form, that was derived in the first subsection, using the warped volume  just computed.  As shown in the first subsection, the two-form
\bea
\hat{\beta} = \partial\bar{\partial}S_\beta
\eea
is harmonic, where 
\bea
S_\beta = \ii\frac{1}{\pi\sqrt{2}}\mathrm{arctanh}\, \frac{c^2}{r^2}~.
\eea
Recall here that $r\geq c$, with the exceptional divisor $D=\{r=c\}$ being a copy of $\mathbb{CP}^1$. Moreover, $\hat{\beta}$ is normalized so that
\bea
\int_{X_{EH}} \hat{\beta}\wedge \star\hat{\beta} = 1~,
\eea
where $X_{EH}=\mathcal{O}(-2)\rightarrow\mathbb{CP}^1$ is the Eguchi-Hanson manifold.
The first claim is that the correctly normalized Poincar\'e dual to $D=\mathbb{CP}^1$ is 
\bea
\beta = \sqrt{2}\hat{\beta}~.
\eea
Indeed, then
\bea
\int_{X_{EH}} \beta\wedge \star\beta = -\int_{X_{EH}} \beta\wedge \beta = 2~.
\eea
We require the 2 here since this is the Euler number of the normal bundle to the exceptional $\mathbb{CP}^1$. Thus, in our notation in the main text,
\bea
\log H = (2\pi\ii)\sqrt{2}S_\beta = -2\pi\cdot\frac{1}{\pi}\mathrm{arctanh}\, \frac{c^2}{r^2} = \log \frac{r^2-c^2}{r^2+c^2}~.
\eea
The radial coordinate near to $D=\mathbb{CP}^1$ is $\rho=\sqrt{r-c}$, so that $D$ is at $\rho=0$. Thus we see that, near to $D$, $\log H$ blows up as $\log \rho^2$, precisely as claimed in the main text.

Also notice that $\beta$ may be written
\bea\label{EHasym}
\beta = -\frac{c^2}{\pi}\diff\left(r^{-2}\eta\right)~.
\eea
Here $\eta=\tfrac{1}{2}\left(\diff\psi-\cos\theta\diff\phi\right)$. Thus the mode $\beta_\mu=\eta$ in this case, which in particular is a Killing one-form. Thus $\mu=4$ and hence $2-n-\nu=-\nu=-\sqrt{\mu}=-2$, which is the power of $r$ above. This confirms the claims about the asymptotic expansion in this case.

\end{appendix}


\begin{thebibliography}{99}

\bibitem{Gustavsson:2007vu}
  A.~Gustavsson,
  ``Algebraic structures on parallel M2 branes,''
  Nucl.\ Phys.\  B {\bf 811} (2009) 66
  [arXiv:0709.1260 [hep-th]].

\bibitem{Bagger:2007vi}
  J.~Bagger and N.~Lambert,
  ``Comments On Multiple M2 branes,''
  JHEP {\bf 0802} (2008) 105
  [arXiv:0712.3738 [hep-th]].

\bibitem{Aharony:2008ug}
  O.~Aharony, O.~Bergman, D.~L.~Jafferis and J.~Maldacena,
  ``N=6 superconformal Chern-Simons-matter theories, M2 branes and their
  gravity duals,''
  JHEP {\bf 0810}, 091 (2008)
  [arXiv:0806.1218 [hep-th]].

\bibitem{Benna:2008zy}
  M.~Benna, I.~Klebanov, T.~Klose and M.~Smedback,
  ``Superconformal Chern-Simons Theories and AdS$_4$/CFT$_3$ Correspondence,''
  JHEP {\bf 0809} (2008) 072
  [arXiv:0806.1519 [hep-th]].
  
\bibitem{Benna:2009xd}
  M.~K.~Benna, I.~R.~Klebanov and T.~Klose,
  ``Charges of Monopole Operators in Chern-Simons Yang-Mills Theory,''
  JHEP {\bf 1001} (2010) 110
  [arXiv:0906.3008 [hep-th]].
  
\bibitem{Gaiotto:2009mv}
  D.~Gaiotto and A.~Tomasiello,
  ``The gauge dual of Romans mass,''
  JHEP {\bf 1001} (2010) 015
  [arXiv:0901.0969 [hep-th]].

\bibitem{Gaiotto:2009yz}
  D.~Gaiotto and A.~Tomasiello,
  ``Perturbing gauge/gravity duals by a Romans mass,''
  J.\ Phys.\ A  {\bf 42} (2009) 465205
  [arXiv:0904.3959 [hep-th]].

\bibitem{Jafferis:2008qz}
  D.~L.~Jafferis and A.~Tomasiello,
  ``A simple class of N=3 gauge/gravity duals,''
  JHEP {\bf 0810} (2008) 101
  [arXiv:0808.0864 [hep-th]].

\bibitem{Martelli:2008si}
  D.~Martelli and J.~Sparks,
  ``Moduli spaces of Chern-Simons quiver gauge theories and AdS(4)/CFT(3),''
  Phys.\ Rev.\  D {\bf 78} (2008) 126005
  [arXiv:0808.0912 [hep-th]].

\bibitem{Hanany:2008cd}
  A.~Hanany and A.~Zaffaroni,
  ``Tilings, Chern-Simons Theories and M2 Branes,''
  JHEP {\bf 0810} (2008) 111
  [arXiv:0808.1244 [hep-th]].
  
\bibitem{Benini:2009qs}
  F.~Benini, C.~Closset and S.~Cremonesi,
  ``Chiral flavors and M2 branes at toric CY4 singularities,''
  arXiv:0911.4127 [hep-th].
  
\bibitem{Jafferis:2009th}
  D.~L.~Jafferis,
  ``Quantum corrections to N=2 Chern-Simons theories with flavor and their AdS4
  duals,''
  arXiv:0911.4324 [hep-th].
  
\bibitem{Intriligator:2003jj}
  K.~A.~Intriligator and B.~Wecht,
  ``The exact superconformal R-symmetry maximizes a,''
  Nucl.\ Phys.\  B {\bf 667}, 183 (2003)
  [arXiv:hep-th/0304128].
  
\bibitem{D'Auria:1984vv}
  R.~D'Auria and P.~Fre,
  ``On The Spectrum Of The N=2 SU(3) X SU(2) X U(1) Gauge Theory From D = 11 Supergravity,''
  Class.\ Quant.\ Grav.\  {\bf 1} (1984) 447.
  
\bibitem{D'Auria:1984vy}
  R.~D'Auria and P.~Fre,
  ``Universal Bose-Fermi Mass Relations In Kaluza-Klein Supergravity And Harmonic Analysis On Coset Manifolds With Killing Spinors,''
  Annals Phys.\  {\bf 162} (1985) 372.
   
\bibitem{Fabbri:1999hw}
  D.~Fabbri, P.~Fre', L.~Gualtieri, C.~Reina, A.~Tomasiello, A.~Zaffaroni and A.~Zampa,
  ``3D superconformal theories from Sasakian seven-manifolds: New  nontrivial evidences for AdS(4)/CFT(3),''
  Nucl.\ Phys.\  B {\bf 577} (2000) 547
  [arXiv:hep-th/9907219].
  
\bibitem{Merlatti:2000ed}
  P.~Merlatti,
  ``M-theory on AdS(4) x Q(111): The complete Osp(2|4) x SU(2) x SU(2) x  SU(2) spectrum from harmonic analysis,''
  Class.\ Quant.\ Grav.\  {\bf 18} (2001) 2797
  [arXiv:hep-th/0012159].
  
\bibitem{Klebanov:1999tb}
  I.~R.~Klebanov and E.~Witten,
  ``AdS/CFT correspondence and symmetry breaking,''
  Nucl.\ Phys.\  B {\bf 556} (1999) 89
  [arXiv:hep-th/9905104].
  
\bibitem{Franco:2005sm}
  S.~Franco, A.~Hanany, D.~Martelli, J.~Sparks, D.~Vegh and B.~Wecht,
  ``Gauge theories from toric geometry and brane tilings,''
  JHEP {\bf 0601} (2006) 128
  [arXiv:hep-th/0505211].
  
\bibitem{Martelli:2007mk}
  D.~Martelli and J.~Sparks,
  ``Baryonic branches and resolutions of Ricci-flat Kahler cones,''
  JHEP {\bf 0804}, 067 (2008)
  [arXiv:0709.2894 [hep-th]].

\bibitem{Martelli:2008cm}
  D.~Martelli and J.~Sparks,
  ``Symmetry-breaking vacua and baryon condensates in AdS/CFT,''
  Phys.\ Rev.\  D {\bf 79}, 065009 (2009)
  [arXiv:0804.3999 [hep-th]].
  
\bibitem{Witten:2003ya}
  E.~Witten,
  ``SL(2,Z) action on three-dimensional conformal field theories with Abelian
  symmetry,''
  arXiv:hep-th/0307041.
  
\bibitem{Marolf:2006nd}
  D.~Marolf and S.~F.~Ross,
  ``Boundary conditions and new dualities: Vector fields in AdS/CFT,''
  JHEP {\bf 0611} (2006) 085
  [arXiv:hep-th/0606113].
  
\bibitem{Imamura:2008nn}
  Y.~Imamura and K.~Kimura,
  ``On the moduli space of elliptic Maxwell-Chern-Simons theories,''
  Prog.\ Theor.\ Phys.\  {\bf 120} (2008) 509
  [arXiv:0806.3727 [hep-th]].
  
\bibitem{Lambert:2010ji}
  N.~Lambert and C.~Papageorgakis,
  ``Relating U(N)xU(N) to SU(N)xSU(N) Chern-Simons Membrane theories,''
  arXiv:1001.4779 [hep-th].

\bibitem{Franco:2008um}
  S.~Franco, A.~Hanany, J.~Park and D.~Rodriguez-Gomez,
  ``Towards M2 brane Theories for Generic Toric Singularities,''
  JHEP {\bf 0812}, 110 (2008)
  [arXiv:0809.3237 [hep-th]].

\bibitem{Franco:2009sp}
  S.~Franco, I.~R.~Klebanov and D.~Rodriguez-Gomez,
  ``M2 branes on Orbifolds of the Cone over $Q^{1,1,1}$,''
  JHEP {\bf 0908}, 033 (2009)
  [arXiv:0903.3231 [hep-th]].
  
\bibitem{Imamura:2008ji}
  Y.~Imamura and S.~Yokoyama,
  ``N=4 Chern-Simons theories and wrapped M5-branes in their gravity duals,''
  Prog.\ Theor.\ Phys.\  {\bf 121} (2009) 915
  [arXiv:0812.1331 [hep-th]].

\bibitem{Klebanov:2007us}
  I.~R.~Klebanov and A.~Murugan,
  ``Gauge/Gravity Duality and Warped Resolved Conifold,''
  JHEP {\bf 0703}, 042 (2007)
  [arXiv:hep-th/0701064].

\bibitem{Klebanov:2007cx}
  I.~R.~Klebanov, A.~Murugan, D.~Rodriguez-Gomez and J.~Ward,
  ``Goldstone Bosons and Global Strings in a Warped Resolved Conifold,''
  JHEP {\bf 0805}, 090 (2008)
  [arXiv:0712.2224 [hep-th]].

\bibitem{Amariti:2009rb}
  A.~Amariti, D.~Forcella, L.~Girardello and A.~Mariotti,
  ``3D Seiberg-like Dualities and M2 Branes,''
  arXiv:0903.3222 [hep-th].
  
\bibitem{Davey:2009sr}
  J.~Davey, A.~Hanany, N.~Mekareeya and G.~Torri,
  ``Phases of M2 brane Theories,''
  JHEP {\bf 0906} (2009) 025
  [arXiv:0903.3234 [hep-th]].

\bibitem{Witten:1996bn}
  E.~Witten,
  ``Non-Perturbative Superpotentials In String Theory,''
  Nucl.\ Phys.\  B {\bf 474}, 343 (1996)
  [arXiv:hep-th/9604030].

\bibitem{Acharya:1998db}
  B.~S.~Acharya, J.~M.~Figueroa-O'Farrill, C.~M.~Hull and B.~J.~Spence,
 ``Branes at conical singularities and holography,''
  Adv.\ Theor.\ Math.\ Phys.\  {\bf 2}, 1249 (1999)
  [arXiv:hep-th/9808014].

\bibitem{BG} C.~P.~Boyer and K.~Galicki, ``Sasakian Geometry, Hypersurface Singularities, and Einstein Metrics,'' Supplemento ai Rendiconti del Circolo Matematico di Palermo Serie II. Suppl {\bf 75} (2005), 57-87 [arXiv:math/0405256].

\bibitem{Gauntlett:2004hh}
  J.~P.~Gauntlett, D.~Martelli, J.~F.~Sparks and D.~Waldram,
  ``A new infinite class of Sasaki-Einstein manifolds,''
  Adv.\ Theor.\ Math.\ Phys.\  {\bf 8}, 987 (2006)
  [arXiv:hep-th/0403038].

\bibitem{Res}  D.~Martelli and J.~Sparks, ``Resolutions of non-regular Ricci-flat Kahler cones,'' J. Geom. Phys. {\bf 59}, 1175-1195  (2009), [arXiv:0707.1674 [math.DG]].

\bibitem{Martelli:2008rt}
  D.~Martelli and J.~Sparks,
  ``Notes on toric Sasaki-Einstein seven-manifolds and AdS$_4$/CFT$_3$,''
  JHEP {\bf 0811}, 016 (2008)
  [arXiv:0808.0904 [hep-th]].

\bibitem{FOW} A.~Futaki, H.~Ono, G.~Wang, ``Transverse K\"ahler geometry of Sasaki manifolds and toric Sasaki-Einstein manifolds,'' 
arXiv:math/0607586. 

\bibitem{Martelli:2009ga}
  D.~Martelli and J.~Sparks,
  ``AdS$_4$/CFT$_3$ duals from M2 branes at hypersurface singularities and their
  deformations,''
  arXiv:0909.2036 [hep-th].

\bibitem{Becker:1996gj}
  K.~Becker and M.~Becker,
  ``M-Theory on Eight-Manifolds,''
  Nucl.\ Phys.\  B {\bf 477}, 155 (1996)
  [arXiv:hep-th/9605053].

\bibitem{Witten:1996md}
  E.~Witten,
  ``On flux quantization in M-theory and the effective action,''
  J.\ Geom.\ Phys.\  {\bf 22}, 1 (1997)
  [arXiv:hep-th/9609122].

\bibitem{Aharony:2008gk}
  O.~Aharony, O.~Bergman and D.~L.~Jafferis,
  ``Fractional M2 branes,''
  JHEP {\bf 0811}, 043 (2008)
  [arXiv:0807.4924 [hep-th]].

\bibitem{Boyer:1998sf}
  C.~P.~Boyer and K.~Galicki,
  ``3-Sasakian Manifolds,''
  Surveys Diff.\ Geom.\  {\bf 7}, 123 (1999)
  [arXiv:hep-th/9810250].

\bibitem{Gubser:1998fp}
  S.~S.~Gubser and I.~R.~Klebanov,
  ``Baryons and domain walls in an N = 1 superconformal gauge theory,''
  Phys.\ Rev.\  D {\bf 58}, 125025 (1998)
  [arXiv:hep-th/9808075].

\bibitem{Hanany:2008fj}
  A.~Hanany, D.~Vegh and A.~Zaffaroni,
 ``Brane Tilings and M2 Branes,''
  JHEP {\bf 0903} (2009) 012
  [arXiv:0809.1440 [hep-th]].
  
  \bibitem{Lerman2} E.~Lerman, ``Homotopy groups of K-contact toric manifold,'' arXiv:math/0204064 [math.SG].

\bibitem{Benishti:2009ky}
  N.~Benishti, Y.~H.~He and J.~Sparks,
  ``(Un)Higgsing the M2 brane,'' 
  arXiv: 0909.4557 [hep-th].

\bibitem{Benvenuti:2005qb}
  S.~Benvenuti, M.~Mahato, L.~A.~Pando Zayas and Y.~Tachikawa,
  ``The gauge / gravity theory of blown up four cycles,''
  arXiv:hep-th/0512061.
  
\bibitem{Baumann:2006th}
  D.~Baumann, A.~Dymarsky, I.~R.~Klebanov, J.~M.~Maldacena, L.~P.~McAllister and A.~Murugan,
  ``On D3-brane potentials in compactifications with fluxes and wrapped
  D-branes,''
  JHEP {\bf 0611}, 031 (2006)
  [arXiv:hep-th/0607050].
  
\bibitem{Davey:2009qx}
  J.~Davey, A.~Hanany, N.~Mekareeya and G.~Torri,
  ``Higgsing M2 brane Theories,''
  JHEP {\bf 0911} (2009) 028
  [arXiv:0908.4033 [hep-th]].
  
  \bibitem{crep1} G.~Tian and S.-T.~Yau, ``Complete K\"ahler manifolds with zero Ricci
curvature II,'' Invent. Math. {\bf 106}, 27-60 (1991).

 \bibitem{crep2} S.~Bando and R.~Kobayashi, ``Ricci-flat K\"ahler metrics on affine algebraic
manifolds,'' II, Math. Ann. {\bf 287} (1990), 175-180.

 \bibitem{crep3} D.~Joyce, ``Asymptotically Locally Euclidean metrics with holonomy $SU(m)$,''
Annals of Global Analysis and Geometry {\bf 19} (2001), 55-73 [arXiv:math/9905041].

 \bibitem{crep4} C.~van Coevering, ``A Construction of Complete Ricci-flat Kaehler Manifolds,'' arXiv:0803.0112 [math.DG].

 \bibitem{crep5} C.~van~Coevering, ``Ricci-flat K\"ahler metrics on crepant resolutions of K\"ahler cones,'' 
arXiv:0806.3728 [math.DG].

\bibitem{crep6} C.~van~Coevering, ``Examples of asymptotically conical Ricci-flat K\"ahler manifolds,'' arXiv:0812.4745 [math.DG]. 

\bibitem{goto} R.~Goto, ``Calabi-Yau structures and Einstein-Sasakian structures on crepant resolutions of isolated singularities,'' arXiv:0906.5191v2 [math.DG].

\bibitem{vanC} C.~van~Coevering, ``Regularity of asymptotically conical Ricci-flat K\"ahler metrics,'' arXiv:0912.3946 [math-DG].

\bibitem{hausel}  T. Hausel, E. Hunsicker, R. Mazzeo, ``Hodge cohomology of gravitational instantons,'' [arXiv:hep-th/0207169].

\bibitem{DNP} M. J. Duff, B. E. W. Nilsson and C. N. Pope, ``Kaluza-Klein Supergravity,'' Phys.Rept. 130, 1 (1986).
  
\bibitem{Frey:2008xw}
  A.~R.~Frey, G.~Torroba, B.~Underwood and M.~R.~Douglas,
  ``The Universal Kaehler Modulus in Warped Compactifications,''
  JHEP {\bf 0901} (2009) 036
  [arXiv:0810.5768 [hep-th]].
  
\bibitem{Berg:2004ek}
  M.~Berg, M.~Haack and B.~Kors,
  ``Loop corrections to volume moduli and inflation in string theory,''
  Phys.\ Rev.\  D {\bf 71} (2005) 026005
  [arXiv:hep-th/0404087].
  
\bibitem{Berg:2005ja}
  M.~Berg, M.~Haack and B.~Kors,
  ``String loop corrections to Kaehler potentials in orientifolds,''
  JHEP {\bf 0511} (2005) 030
  [arXiv:hep-th/0508043].
  
\bibitem{Ganor:1996pe}
  O.~J.~Ganor,
  ``A note on zeroes of superpotentials in F-theory,''
  Nucl.\ Phys.\  B {\bf 499} (1997) 55
  [arXiv:hep-th/9612077].
  
  \bibitem{kantor-1997}
  Jean-Michel Kantor, "On the width of lattice-free simplices," arXiv:9709026 [alg-geom] (1997).
  
\bibitem{Hanany:2008gx}
  A.~Hanany and Y.~H.~He,
  ``M2-Branes and Quiver Chern-Simons: A Taxonomic Study,''
  arXiv:0811.4044 [hep-th].
  
\bibitem{Cvetic:2000db}
  M.~Cvetic, G.~W.~Gibbons, H.~Lu and C.~N.~Pope,
  ``Ricci-flat metrics, harmonic forms and brane resolutions,''
  Commun.\ Math.\ Phys.\  {\bf 232}, 457 (2003)
  [arXiv:hep-th/0012011].

\bibitem{Cvetic:2001ma}
  M.~Cvetic, G.~W.~Gibbons, H.~Lu and C.~N.~Pope,
  ``Supersymmetric non-singular fractional D2-branes and NS-NS 2-branes,''
  Nucl.\ Phys.\  B {\bf 606}, 18 (2001)
  [arXiv:hep-th/0101096].
%
\bibitem{Klebanov:2010tj}
  I.~R.~Klebanov, S.~S.~Pufu and T.~Tesileanu,
  ``Membranes with Topological Charge and AdS4/CFT3 Correspondence,''
  arXiv:1004.0413 [hep-th].
  
\bibitem{Morrison:1998cs}
  D.~R.~Morrison and M.~R.~Plesser,
  ``Non-spherical horizons. I,''
  Adv.\ Theor.\ Math.\ Phys.\  {\bf 3}, 1 (1999)
  [arXiv:hep-th/9810201].

\end{thebibliography}
\end{document}